\documentstyle[eqsecnum,aps]{revtex}
\begin{document}
\draft
\title{Theory of a  dilute low-temperature
trapped Bose  condensate}
\author{Alexander L.~Fetter}
\address{Departments of Physics and Applied
Physics, Stanford University, Stanford, CA
94305-4060}
\maketitle
\begin{abstract}
This set of four lectures reviews various aspects of the theory of a dilute
low-temperature trapped Bose gas, starting with (I) a review of the Bogoliubov
description of the elementary excitations in a uniform system.  The
treatment is
then generalized (II) to include the new physical effects of a confining
harmonic trap potential on  the condensate and its normal modes.  An equivalent
hydrodynamic description (III) focuses directly on the density and velocity
fluctuations. The physics of vortices (IV) in an incompressible fluid is
summarized
and extended to the case of a trapped Bose condensate.
\end{abstract}

\section{Uniform dilute Bose gas}\label{sec:I}

Bogoliubov's \cite{Bog} seminal
1947 paper originated the modern theoretical description of a dilute
interacting
low-temperature Bose gas. He considered how  a weak repulsive
interparticle potential affects  the ground state and low-lying excited
states of
a uniform condensed Bose gas, showing that the interactions
qualitatively alter the dispersion relation at long wavelengths from
the familiar quadratic free-particle form $ p^2/2M$
to a linear ``phonon''-like structure $ sp$  with   speed
of sound $s$.

\subsection{Brief review of a uniform ideal Bose gas}

The standard textbook example \cite{LP} of an ideal Bose gas treats a
uniform system of $N$ particles  in a
cubical box of volume
$V = L^3$ with periodic boundary conditions and number density $n = N/V$.  The
relevant single-particle states are plane waves
$V^{-1/2}
\exp(i{\bf k\cdot r})$, with
${\bf k} = (2\pi/L)(n_1, n_2, n_3)$, where $n_i$ is any integer;
the energy of such a state is $\epsilon_k^0 =
\hbar^2k^2/2M$, where $M$ is the particle's mass.  In the classical
limit,  the thermal de
Broglie wavelength
$\Lambda_T = (2\pi\hbar^2/Mk_BT)^{1/2}$ is much smaller than the
interparticle spacing $l \approx n^{-1/3}$, so that diffraction
effects are negligible.  The inequality $\Lambda_T\ll l$ necessarily fails
as the
temperature is reduced (or the number density increases),  and  quantum
diffraction
becomes important at a temperature $T_c$ determined by the approximate
condition
$\Lambda_{T_c} \approx l$ (or, equivalently,  $n\Lambda_{T_c}^3 \approx 1$).  A
detailed calculation yields the exact expression

\begin{equation}n\Lambda_{T_c}^3= \zeta(\case{3}{2})
\approx 2.612,\label{eq:Tc1}\end{equation}
 where $\zeta(\frac{3}{2})$ is a  Riemann zeta function. The transition at
$T_c$
represents the onset of the special form of quantum degeneracy known as
Bose-Einstein condensation.  Alternatively, the transition temperature
$T_c$ is given by the approximate relation

\begin{equation}k_BT_c \approx
\frac{\hbar^2n^{2/3}}{M}\label{eq:Tc}
\end{equation}
 that  characterizes the onset of degeneracy in
any ideal quantum gas, providing a qualitative description of electrons in
metals and white dwarf stars, nucleons in nuclear matter and neutron stars,
and the
fermionic isotope
${}^3$He, as well as  the bosonic isotope
${}^4$He.

For $T< T_c$, a uniform ideal Bose gas has a macroscopic number
of particles $N_0(T)$ in the
lowest single-particle state (here, that with  ${\bf k = 0}$), given by

\begin{equation}
\frac{N_0(T)}{N} = 1 -
\left(\frac{T}{T_c}\right)^{\!\!3/2}\quad\hbox{for
$T\le T_c$}.\label{eq:cond1}
\end{equation}
Thus a finite fraction of all the particles in a condensed Bose
gas occupies a single quantum-mechanical state.  In the
limit $T=0$, this fraction is unity;  all the particles in a  uniform ideal
Bose
gas at zero temperature and fixed volume $V$  have zero momentum (and therefore
exert no pressure on the confining walls).

\subsection{Effect of weak repulsive interactions}

Bogoliubov \cite{Bog} introduced the microscopic treatment of a
``weakly interacting'' or ``nearly ideal'' Bose gas, starting from the
second-quantized hamiltonian

\begin{equation}
\hat H = \sum_{\bf k} \epsilon_k^0 a_{\bf k}^\dagger a_{\bf k} +
\frac{1}{2V}\sum_{{\bf k}_1,{\bf k}_2,{\bf k}_3,{\bf
k}_4}\,\tilde V_{{\bf k}_1-{\bf k}_3}\,a_{{\bf k}_1}^\dagger a_{{\bf
k}_2}^\dagger a_{{\bf k}_3}a_{{\bf k}_4}\,\delta_{{\bf k}_1+{\bf
k}_2,{\bf k}_3+{\bf k}_4},\label{eq:ham1}
\end{equation}
where the Kronecker $\delta$ ensures momentum conservation, and
$\tilde V_{\bf k} = \int d^3r\,\exp(-i{\bf k\cdot r} ) V({\bf r}) $ is the
Fourier transform of the interparticle potential.  The operators
$a_{\bf k}^\dagger$ and $a_{\bf k} $ obey bosonic commutation
relations

\begin{equation}
[a_{\bf k}, a_{\bf k'}^\dagger] = \delta_{\bf k, k'},\quad [a_{\bf k},
a_{\bf k'}] = [a_{\bf k}^\dagger, a_{\bf k'}^\dagger] =
0.\label{eq:comm1}
\end{equation}

Bogoliubov's basic idea is very simple and elegant.  If the
interacting Bose gas is   sufficiently dilute, it should differ only
slightly from an ideal gas, so that most of the particles in the true
interacting ground state should have  zero momentum.  Furthermore,
only two-body collisions with small momentum transfer are significant,
and these may be characterized by a single parameter known as the
$s$-wave scattering length $a$ (here generally taken as positive,
corresponding to a repulsive interaction).  For a strong repulsive
potential, $a$
is simply the range, and the more general case is treated below.
In the dilute limit
$a\ll l$ (or, equivalently,  $na^3\ll 1$), the interparticle potential may be
approximated by a ``pseudopotential''  $V({\bf r}) \approx
g\delta^{(3)}({\bf r}) $, with a constant Fourier transform $\tilde V_{\bf k}
= g$, and the original hamiltonian in eq.\ (\ref{eq:ham1}) becomes
\begin{equation}
\hat H = \sum_{\bf k} \epsilon_k^0 a_{\bf k}^\dagger a_{\bf k} +
\frac{g}{2V}\sum_{{\bf k}_1,{\bf k}_2,{\bf k}_3,{\bf
k}_4}\,a_{{\bf k}_1}^\dagger a_{{\bf
k}_2}^\dagger a_{{\bf k}_3}a_{{\bf k}_4}\,\delta_{{\bf k}_1+{\bf
k}_2,{\bf k}_3+{\bf k}_4}.\label{eq:ham2}
\end{equation}

\subsection{Review of scattering theory}

The model interaction strength $g$ can be related to the  more
physical parameter $a$ by requiring that the many-body hamiltonian
(\ref{eq:ham2}) reproduce the correct two-body scattering in vacuum
\cite{FW}.  The  Schr\"odinger equation for the scattering wave function
$\psi_{\bf k} ({\bf r})$ of two particles with mass
$M$ and initial relative wave vector $\bf k$ can be rewritten in terms
of  the relative separation
${\bf r } = {\bf r}_1-{\bf r}_2$ and the reduced mass $\frac{1}{2}M$
\begin{equation}
-\frac{\hbar^2}{M} \nabla^2\psi_{\bf k} ({\bf r}) + V({\bf r}
)\psi_{\bf k} ({\bf r}) = E\psi_{\bf k} ({\bf r}),\label{eq:Sch1}
\end{equation}
where $E = 2\epsilon^0_k = \hbar^2k^2/M$.
With the outgoing-wave Green's function of the Helmholtz
equation \cite{FW1}, this partial differential equation can be recast
as an integral equation

\begin{equation}
\psi_{\bf k} ({\bf r}) = e^{i{\bf k\cdot r}} -\frac{M}{4\pi \hbar^2}
\, \int  d^3 r' \,\frac{e^{ik|{\bf r-r'}|}}{|{\bf r-r'}|}\,V({\bf
r'})\,
\psi_{\bf k} ({\bf r'})\label{eq:int1}.
\end{equation}
For simplicity, assume that the potential has a finite range (and no
bound states),  so that  eq.~(\ref{eq:int1}) has the asymptotic form for
$|{\bf r} | = r \to \infty$

\begin{equation}
\psi_{\bf k} ({\bf r}) \sim e^{i{\bf k\cdot r}}  + f({\bf k'},{\bf k})
\,\frac{e^{ikr}}{r}.\label{eq:out}
\end{equation}
This expression is the sum of  an incident plane wave $\exp(i{\bf k\cdot
r}) $ with
wave vector
$\bf k$, and an  outgoing scattered spherical wave $r^{-1}\exp(ikr)$ with a
proportionality factor

\begin{equation}
f({\bf k',k})\equiv -\frac{M}{4\pi\hbar^2}\int d^3r'\,e^{-i{\bf k'\cdot
r'}}\,V({\bf r'})\,\psi_{\bf k}({\bf r'})\label{eq:scatt}
\end{equation}
known as  the {\it scattering amplitude\/}
for a transition from the initial relative wave vector $\bf k$ to a
final relative wave vector $\bf k'$.  Note that this definition
remains valid even for a singular potential because it
involves the true wave function $\psi_{\bf k}$ that vanishes wherever
$V({\bf r}) $ diverges.  In contrast,  the Born approximation takes
$\psi_{\bf k}({\bf r}) \approx \psi^B_{\bf k}({\bf r}) = \exp(i{\bf
k\cdot r})$, and the resulting Born approximation to the scattering
amplitude

\begin{equation}
f_B({\bf k', k}) = -\frac{M}{4\pi \hbar^2}\,\tilde V_{\bf k' - k}
\label{eq;scattB}
\end{equation}
clearly  requires that $V({\bf r})$ be integrable and hence
non-singular.

For a spherically symmetric potential with $k' = k$ (so that energy
is conserved), the scattering amplitude depends  on $k$ and the
scattering angle $\theta$ (defined by $\cos\theta = \hat k\cdot \hat k'$)

\begin{equation}
f(k,\theta) = \sum_{l=0}^\infty \frac{2l+1}{k}\,e^{i\delta_l}\,\sin
\delta_l\,P_l(\cos\theta),
\end{equation}
where $\delta_l$ is the energy-dependent phase shift for the $l$th partial
wave.
In the low-energy limit ($k\to 0$), the $s$-wave phase shift reduces to
$\delta_0\approx -ka$, which defines the $s$-wave scattering length $a$, and
higher-order phase shifts are of order  $k^{2l+1}$ for $l\ge
1$. Thus, only  the
$s$-wave term ($l=0$) contributes as $k\to 0$,  and the scattering
amplitude takes
the simple form

\begin{equation}\lim_{|{\bf k}| = |{\bf k'}|\to 0} f({\bf k', k})
\approx -a.\label{eq:scatt0}
\end{equation}

For many purposes, it is convenient to introduce a Fourier
decomposition of the scattering wave function, with

\begin{equation}
\psi_{\bf k} ({\bf r}) = \int \frac{d^3p}{(2\pi)^3} \, e^{i{\bf
p\cdot r}}\,\tilde \psi_{\bf k}({\bf p}),
\end{equation}
and  the exact scattering amplitude in eq.\ (\ref{eq:scatt}) can be
rewritten as a
convolution  of the Fourier transforms of the interaction potential and the
wave function

\begin{equation}
f({\bf k',k}) = -\frac{M}{4\pi\hbar^2}\int
\frac{d^3p}{(2\pi)^3}\,\tilde V_{\bf k'-p}\,\tilde\psi_{\bf k}({\bf
p}).\label{eq:scattmom}
\end{equation}
In addition, the outgoing-wave Green's function has  the
Fourier representation~\cite{FW1}

\begin{equation}
-\frac{e^{ik|{\bf r-r'}|}}{|{\bf r-r'}|}= \int \frac{d^3
q}{(2\pi)^3}\,\frac{e^{i\bf q\cdot(r-r')}}{k^2-q^2+i\eta},
\end{equation}
where $\eta\to 0^+$ defines the appropriate contour in the complex $q$
plane. A combination with the  integral equation (\ref{eq:int1})
then yields an exact expression for the Fourier transform $\tilde \psi_{\bf
k}({\bf p})$

\begin{equation}
\tilde\psi_{\bf k}({\bf p}) = \underbrace{(2\pi)^3 \,\delta^{(3)}({\bf
k-p})}_{\rm incident\>wave} -\underbrace{\frac {4\pi f({\bf p,
k})}{k^2-p^2+i\eta}}_{\rm scattered\>wave},\label{eq:psimom}
\end{equation}
again expressed as the sum of an incident wave and an outgoing scattered
wave.

As a final step, a combination of eqs.~(\ref{eq:scattmom})
and (\ref{eq:psimom}) gives a corresponding   integral equation for the
scattering amplitude

\begin{equation}
-\frac{4\pi \hbar^2}{M} f({\bf k', k}) = \tilde V_{\bf k'-k}
-4\pi\int \frac{d^3 p}{(2\pi)^3}\,\frac{\tilde V_{\bf k'-p}\,f({\bf p,
k})}{k^2-p^2+i\eta},\label{eq:scattamp}
\end{equation}
which requires knowledge of the scattering amplitude for $p^2 \neq
k^2$ (known as ``off-the-energy-shell'').  Note that the leading term
of eq.\ (\ref{eq:scattamp}) is simply the Born approximation $f_B$,
and an iteration gives the first correction

\begin{equation}
-\frac{4\pi \hbar^2}{M} f({\bf k', k}) \approx \tilde V_{\bf k'-k}
+\frac{M}{\hbar^2} \int \frac{d^3 p}{(2\pi)^3}\,\frac{\tilde V_{\bf
k'-p}\tilde V_{\bf p-k}}{k^2-p^2+i\eta}+\cdots.
\end{equation}
In the limit $k'^2 = k^2 \to 0$, this equation becomes a perturbation
expansion for the scattering length

\begin{equation}
\frac{4\pi \hbar^2 a}{M} \approx \tilde V_0 -\frac{M}{\hbar^2} \int
\frac{d^3 p}{(2\pi)^3}\,\frac{|\tilde V_{\bf
p}|^2}{p^2}+\cdots,\label{eq:scattamp1}
\end{equation}
and the first correction is finite if $\tilde V_{\bf p} $ vanishes
sufficiently rapidly as $|{\bf p}|\to \infty$.

The situation is different in the singular case of the
pseudopotential $\tilde V_{\bf p} = g$, for eq.\ (\ref{eq:scattamp1})
reduces to

\begin{equation}
\frac{4\pi \hbar^2 a}{M} \approx g -\frac{Mg^2}{\hbar^2} \int
\frac{d^3 p}{(2\pi)^3}\,\frac{1}{p^2}+\cdots,\label{eq:scattamp2}
\end{equation}
To leading order, this equation relates the  pseudopotential $g$ to
the measurable scattering amplitude

\begin{equation}
g\approx \frac{4\pi \hbar^2 a}{M}.\label{eq:pseudo}
\end{equation}
It typifies the general prescription that a first-order perturbation
calculation
can usually  be extended to treat a dilute hard-core gas by substituting
the true
$s$-wave scattering length $a$
 for the first Born approximation $a_B = M\tilde V_0/4\pi \hbar^2$;  in
the present context of a dilute Bose gas, this procedure was
suggested by Landau (see last footnote of ref.
\cite{Bog}). In contrast,  the second-order correction in
eq.~(\ref{eq:scattamp2}) diverges for large values of
$|\bf p|$.  As seen below, the Bogoliubov theory yields a similar
divergence for the ground-state energy that disappears when
re-expressed in terms of the physical quantity $a$ (for an alternative
treatment of the divergence
 arising from the use of the pseudopotential, see
ref.~\cite{LHY}).

There is one remaining subtlety, for the $s$-wave scattering length
has been defined as if the particles were distinguishable, whereas the
true overall wave function is symmetric.  Consequently, the actual
differential cross section is obtained from a symmetrized scattering
amplitude

\begin{equation}
\frac{d\sigma}{d\Omega} = |f(k,\theta) + f(k,\pi - \theta)|^2,
\end{equation}
 which reduces to $|2a|^2 = 4a^2$ in the low-energy limit, four times
that for distinguishable particles.  The corresponding total cross
section
$\sigma_T= 8\pi a^2$ is obtained by integrating over {\it one\/}
hemisphere, because the particles are indistinguishable.

\subsection{Bogoliubov quasiparticles}
Given the approximate relation in eq.~(\ref{eq:pseudo}) between the
pseudopotential
$g$ and the
$s$-wave scattering length $a$, it is now feasible to proceed with Bogoliubov's
\cite{Bog} analysis of the model hamiltonian in eq.~(\ref{eq:ham2}),
following the treatment of ref.~\cite{FW}, secs.\ 35 and 55.  The
basic observation is that an imperfect Bose gas at low
temperature should resemble an ideal Bose gas to the extent that a
macroscopic number $N_0$ of particles  occupies one
single-particle state. In the simplest case of a stationary uniform
system, this condensation occurs in the mode with $\bf k = 0$, but
many other possibilities also can occur (for example, a condensate moving with
velocity $\bf v$).  The effect of the repulsive interactions is to scatter
particles from the ideal-gas ground state
$\bf k = 0$ to pairs of higher-momentum states with $\pm \bf k \neq
0$. This  depletion of  the zero-momentum condensate implies  that $N_0
$ is definitely less than
$N$, even at $T=0$, but the ``condensate fraction''
$N_0/N$ is assumed to remain finite in the thermodynamic limit ($N\to
\infty$, $V\to \infty$, with $N/V$ fixed);  this assumption
must naturally  be verified at the end of the calculation.

Recall that the ground state of an ideal uniform   Bose gas
$|\Phi_0(N)\rangle = |N, 0, 0, \cdots\rangle \equiv
(a_0^\dagger)^N(N!)^{-1/2}|0, 0,
0,
\cdots\rangle$ has all
$N$ particles in the single-particle mode with $\bf k = 0$.  The associated
annihilation and creation   operators then yield macroscopic
coefficients when applied to this ground state

\begin{equation}
a_0|\Phi_0(N)\rangle = \sqrt
N\,|\Phi_0(N-1)\rangle,\quad a_0^\dagger|\Phi_0(N)\rangle =
\sqrt{N+1}\,|\Phi_0(N+1)\rangle,
\end{equation}
whereas their commutator
\begin{equation}
[a_0,a_0^\dagger]\,|\Phi_0(N)\rangle = |\Phi_0(N)\rangle
\end{equation}
gives a coefficient unity.  To the extent that the
thermodynamic properties of an ideal Bose gas are independent of the
addition or subtraction of one particle, it is natural to treat the
condensed mode differently from all the others, replacing the
operators $a_0$ and $a_0^\dagger$ by pure numbers, because their
commutator is of order $N^{-1/2}$ relative to their individual
matrix elements.

 The same situation is assumed to describe the
ground state $|\Phi\rangle$ of an imperfect uniform Bose gas, as long as the
condensate density
\begin{equation}\frac{\langle\Phi|a_0^\dagger a_0|\Phi\rangle}{V} =
\frac{N_0}{V} \equiv n_0
\end{equation}
remains finite in the thermodynamic limit.  Consequently, Bogoliubov
proposed  the  now famous prescription
\begin{equation}
a_0, a_0^\dagger \to \sqrt{N_0},\label{eq:Bogpres}
\end{equation}
with the operators $a_0$ and $a_0^\dagger$ reinterpreted as pure
numbers that commute. For simplicity, he also imposed the additional stronger
condition of small total depletion
$N-N_0\ll N$, which limits the strength of the
interparticle potential and requires that the temperature $T$ be small
compared to the critical temperature $T_c$ for the onset of Bose
condensation.  It is important to realize that this second assumption
is distinct from the Bogoliubov  prescription in
eq.~(\ref{eq:Bogpres}), which merely requires the existence of a finite
condensate density $n_0$ in the thermodynamic limit (the more general
  case at finite temperature is described,  for example, in
refs.~\cite{AG1,AG2} as well as in Griffin's  lectures  in this
volume).

Substitute the Bogoliubov prescription (\ref{eq:Bogpres}) into
the model hamiltonian (\ref{eq:ham2}) and keep only the leading terms
of order $N_0^2$ and $N_0$ (for a uniform gas, the
terms of order
$N_0^{3/2}$ vanish identically because of momentum conservation, but a
similar situation holds in general).  The resulting interaction part of the
hamiltonian becomes

\begin{equation}
\hat H_{\rm int} \approx \frac{g}{2V} \left[N_0^2 + 2N_0\sum_{\bf k
\neq 0} \left(a_{\bf k}^\dagger a_{\bf k} + a_{-\bf k}^\dagger a_{-\bf
k}\right) + N_0\sum_{\bf k\neq 0} \left(a_{\bf k}^\dagger a_{-\bf
k}^\dagger + a_{\bf k}a_{-\bf k}\right)\right].\label{eq:hint1}
\end{equation}
In the present context, it is convenient to consider only states with
a fixed number $N$ of particles (see sec.~\ref{sec:trap} for a more general
approach using the grand canonical ensemble).   The number operator

\begin{equation}\hat N =
\sum_{\bf k} a_{\bf k}^\dagger a_{\bf k} \approx N_0 +\case{1}{2}
\sum_{\bf k
\neq 0}(a_{\bf k}^\dagger a_{\bf k}+ a_{-\bf k}^\dagger a_{-\bf
k})\label{eq:num}
\end{equation}
can then be replaced by its eigenvalue $N$, and eq.~(\ref{eq:num})
thus
serves to eliminate
$N_0$ in favor of $N$ through terms of order
$N^2$ and $N$. As a result, the model hamiltonian (\ref{eq:ham2})
becomes

\begin{equation}
\hat H \approx \frac{gN^2}{2V} + \case{1}{2} \sum_{\bf k \neq 0}
\left[\left(\epsilon_k^0+ng\right)\left(a_{\bf k}^\dagger a_{\bf k} +
a_{-\bf k}^\dagger a_{-\bf k}\right)+ ng\left(a_{\bf k}^\dagger a_{-\bf
k}^\dagger + a_{\bf k} a_{-\bf k}\right)\right]. \label{eq:ham3}
\end{equation}
 This approximation
evidently neglects terms like $(\sum_{\bf k \neq 0} a_{\bf k}^\dagger
a_{\bf k})^2$, which are small for $N~-~N_0~\ll~N$.  The Bogoliubov
hamiltonian  (\ref{eq:ham3})  is a
quadratic form in the creation and annihilation operators and
 can be diagonalized with a canonical transformation.

		To demonstrate this remarkable feature, Bogoliubov
introduced a new
set of ``quasiparticle'' operators $\alpha_{\bf k}$ and $\alpha_{\bf
k}^\dagger$, defined by the linear transformation

\begin{mathletters}
\label{eq:canon}
\begin{equation}
a_{\bf k}=u_k\alpha_{\bf k} -v_k\alpha_{-\bf
k}^\dagger,\label{eq:canona}
\end{equation}
\begin{equation}
a_{-\bf k}^\dagger=u_k\alpha_{-\bf k}^\dagger -v_k\alpha_{\bf
k},\label{eq:canonb}
\end{equation}
\end{mathletters}
with real isotropic coefficients $u_k$ and $v_k$.  Note that both
operators $a_{\bf k}$ and $a_{-\bf k}^\dagger$ reduce the total
momentum by $\hbar \bf k$.  This  transformation is canonical if
the new operators also obey bosonic commutation relations, with
$[\alpha_{\bf k},\alpha_{\bf k'}^\dagger] = \delta_{\bf k, k'}$ and
$[\alpha_{\bf k},\alpha_{\bf k'}]= [\alpha_{\bf k}^\dagger,\alpha_{\bf
k'}^\dagger]=0$.  It is easy to verify that this condition  requires

\begin{equation}
u_k^2 - v_k^2 = 1\quad\hbox{for all $\bf k \neq 0$}.\label{eq:boseqp}
\end{equation}

Direct substitution of eq.\ (\ref{eq:canon}) into the Bogoliubov
hamiltonian (\ref{eq:ham3}) yields

\begin{eqnarray}
\hat H&=&\case{1}{2} gn^2V + \sum_{\bf k\neq
0}\left[\left(\epsilon_k^0+ng\right)v_k^2 -ng\,u_kv_k\right]\nonumber\\
& &+\case{1}{2}\sum_{\bf k\neq
0}\left[\left(\epsilon_k^0+ng\right)\left(u_k^2+v_k^2\right)
-2ng\,u_kv_k\right]\left(\alpha_{\bf k}^\dagger\alpha_{\bf
k}+\alpha_{-\bf k}^\dagger\alpha_{-\bf k}\right)\nonumber\\
& &+\case{1}{2}\sum_{\bf k\neq
0}\left[ng\left(u_k^2+v_k^2\right)
-2\left(\epsilon_k^0+ng\right)\,u_kv_k\right]\left(\alpha_{\bf
k}^\dagger\alpha_{-\bf k}^\dagger+\alpha_{\bf k}\alpha_{-\bf
k}\right).\label{eq:diag}
\end{eqnarray}
Here, the first line is a pure number with no operator character, and  the
second line is diagonal in the quasiparticle number operator
$\alpha_{\bf k}^\dagger\alpha_{\bf k}$.  In contrast,  the third  is
{\it off-diagonal\/} in the quasiparticle number,  but it can be eliminated
entirely by choosing  the coefficients
$u_k$ and
$v_k$ to satisfy

\begin{equation}
ng\left(u_k^2+v_k^2\right)
=2\left(\epsilon_k^0+ng\right)\,u_kv_k.\label{eq:elim}
\end{equation}
The subsidiary condition $u_k^2-v_k^2=1$ in eq.~(\ref{eq:boseqp}) can
be incorporated automatically by writing

\begin{equation}
u_k = \cosh\theta_k \quad\hbox{and}\quad v_k = \sinh\theta_k,
\end{equation}
and a combination with eq.\ (\ref{eq:elim}) readily yields

\begin{equation}
\tanh\,2\theta_k = \frac{ng}{\epsilon_k^0+ ng}.
\end{equation}
If $g$
is negative, the right-hand side of this equation will diverge for
some $k$, implying  unphysical values for a range of the $\theta_k$'s.
The present case of a uniform dilute Bose condensate
therefore  requires repulsive interactions, with both
$g$ and
$a$   positive.

	The resulting equations for $u_k^2$ and $v_k^2$ are easily solved to
yield

\begin{equation}
v_k^2 = u_k^2 -1 = \case{1}{2}
\left(\frac{\epsilon_k^0+ng}{E_k}-1\right),\label{eq:uv}
\end{equation}
where
\begin{equation}
E_k = \sqrt{\left(\epsilon_k^0+ng\right)^2 -\left(ng\right)^2} =
\sqrt{\left(\epsilon_k^0\right)^2 + 2ng\epsilon_k^0}\label{eq:Ek}
\end{equation}
is defined as the positive square root. Substitution into eq.\ (\ref{eq:diag})
gives the final and simple quasiparticle hamiltonian

\begin{equation}
\hat H = \case{1}{2} gn^2V - \case{1}{2} \sum_{\bf k \neq
0}\left(\epsilon_k^0 + ng -E_k\right) +\case{1}{2}\sum_{\bf k \neq 0}
E_k\left(\alpha_{\bf k}^\dagger\alpha_{\bf k} + \alpha_{-\bf
k}^\dagger\alpha_{-\bf k}\right),\label{eq:hqp}
\end{equation}
which involves only the quasiparticle number operator for each
normal mode $\bf k$.  Since $\alpha_{\bf k}^\dagger\alpha_{\bf k}$
has the eigenvalues $0, 1, 2, \cdots$,  it follows immediately  that
the ground state
$|\Phi\rangle$ of this model hamiltonian is simply the quasiparticle
vacuum, defined by

\begin{equation}
\alpha_{\bf k} |\Phi\rangle = 0 \quad\hbox{for all $\bf k \neq 0$};
\end{equation}
evidently, $|\Phi\rangle$  is a very complicated combination of
unperturbed particle  states, for neither $a_{\bf k}$ nor $a_{\bf
k}^\dagger$ annihilates it.  In addition, the ground-state energy is
given by

\begin{equation}
E_g = \langle \Phi|\hat H|\Phi\rangle = \case{1}{2} gn^2V -\case{1}{2}
 \sum_{\bf k \neq
0}\left(\epsilon_k^0 + ng -E_k\right)=  \case{1}{2} gn^2V -
 \sum_{\bf k \neq 0}E_k\,v_k^2.\label{eq:grndst}
\end{equation}

The physics of the quasiparticle hamiltonian is very transparent, for
all the  excited states correspond to various numbers of non-interacting
bosonic quasiparticles, each with  a wave vector $\bf k$ and an excitation
energy

\begin{eqnarray}
E_k & = & \sqrt{\frac{ng\hbar^2k^2}{M} +
\left(\frac{\hbar^2k^2}{2M}\right)^{\!\!2}}\label{eq:Esubk}\\
\noalign{\vspace{.2cm}} &\approx&\cases{{\displaystyle{\sqrt{\frac{ng}{M}}\
\hbar k=
\sqrt{\frac{4\pi \hbar^2an}{M^2}}\,\hbar k}}, &for
$k\to 0$,\cr\noalign{\vspace{.2cm}} {\displaystyle{\epsilon_k^0 +
\frac{4\pi
\hbar^2 a n}{M}}},& for
$k\to
\infty$.\cr}
\end{eqnarray}
In the long-wavelength limit ($k\to 0$), the elementary excitations
represent sound
waves with the propagation speed

\begin{equation}
s = \sqrt{\frac{ng}{M}} = \frac{\hbar}{M}\sqrt{4\pi
an},\label{eq:speed}
\end{equation}
which again shows that $g$ and $a$ must both be positive.  For short
wavelengths, in contrast, the spectrum is that of a free particle, but
shifted upward by a constant  ``optical potential'' $gn= 4\pi \hbar^2
an/M$ arising from the Hartree interaction with the remaining particles
(equivalently, this shift reflects  the forward scattering from the
background medium, producing an  effective index of refraction).

The cross-over between the linear  and quadratic regimes occurs when
$\epsilon_k^0/ng\approx 1$.  This ratio  can be rewritten as $k^2/8\pi
na\approx 1$, suggesting the introduction of a characteristic length

\begin{equation}
\xi = \frac{1}{\sqrt{8\pi na}}\label{eq:healing}
\end{equation}
that is frequently known as the ``correlation length'' \cite{LHY}
or the ``coherence length;''  as shown in sec.~\ref{sec:trap}, however, the
term
``healing length'' is usually preferable  in the context of a
non-uniform dilute trapped Bose gas. Note that $\xi\to \infty$ for a
non-interacting gas ($a\to 0$);  in this limit,  $E_k$ reduces to the
quadratic free-particle spectrum for all $k$, and the linear
phonon-like region then vanishes entirely.  The energy spectrum in
eq.~(\ref{eq:Esubk}) can be rewritten as $E_k/ng ~=~E_kM/(4\pi
\hbar^2an)~=~k\xi\,\sqrt{2 + k^2\xi^2}$,  shown in  fig.~1
 as a function of the dimensionless combination
$k\xi$.

The Bogoliubov transformation (\ref{eq:canon}) can be inverted to
express the quasiparticle operators in terms of the original particle
operators
\begin{mathletters}\label{eq:canoninv}
\begin{equation}
\alpha_{\bf k} = u_ka_{\bf k} +v_k a_{-\bf
k}^\dagger,\label{eq:canoninva}
\end{equation}
\begin{equation}
\alpha_{-\bf k}^\dagger = u_ka_{-\bf k}^\dagger +v_k a_{\bf
k}.\label{eq:canoninvb}
\end{equation}
\end{mathletters}
The quasiparticle operators are coherent linear
superpositions of particle and hole operators, with $u_k$ and $v_k$ as
the weight factors.  In the phonon regime ($k\xi\ll1$), these
coefficients are both large, with $u_k^2\approx v_k^2 \approx (\sqrt
8 k\xi)^{-1} \gg 1$, and the
long-wavelength quasiparticle operators  therefore
represent a coherent nearly  equal admixture of a particle and a
hole.   In the free-particle regime ($k\xi \gg 1$), however, the
behavior changes to $u_k^2\approx 1$ and $v_k^2 \approx
(4k^4\xi^4)^{-1}\ll 1$, so that the short-wavelength quasiparticle
creation operator is effectively a pure particle creation operator.
Figure 2 shows the quantities $u_k^2$ and $v_k^2$ as a function of
the dimensionless wavenumber $k\xi$, illustrating clearly
the qualitative effect of the repulsive interactions on the long-wavelength
properties of a uniform dilute Bose gas.

\subsection{Macroscopic properties}

Within the Bogoliubov approximation, the low-temperature behavior can
be described with   the  quasiparticle hamiltonian $\hat H$
in eq.~(\ref{eq:hqp}) and an associated Gibbs ensemble density operator

\begin{equation}
\hat \rho = \frac{\exp(-\beta\hat H)}{{\rm Tr}\left[\,\exp(-\beta\hat
H)\,\right]},\label{eq:densop}
\end{equation}
where $\beta^{-1} = k_BT$ and the denominator is simply the partition
function (the notation $\rm Tr$ denotes a  sum over all states).
Manipulations with this density operator are straightforward,
for $\hat H$ is simply a sum of non-interacting bosonic contributions,
yielding
 a thermal average over familiar harmonic-oscillator states.   For
example, consider the total particle number at low temperature

\begin{equation}
N =N_0 + \case{1}{2}\sum_{\bf k \neq 0}\,{\rm Tr}\left[\hat
\rho\left(a_{\bf k}^\dagger a_{\bf k} + a_{-\bf k}^\dagger a_{-\bf
k}\right)\right].\label{eq:number}
\end{equation}
The trace is invariant under a canonical transformation, and it is
most readily  evaluated in the quasiparticle basis where $\hat H$ is
diagonal.  Use of the Bogoliubov transformations from
eq.~(\ref{eq:canon}) leads to various  products of two
quasiparticle  and quasihole operators.  Those involving off-diagonal
quantities
like
${\rm Tr}\left(\hat\rho\,\alpha_{\bf k}^\dagger \alpha_{-\bf k}^\dagger
\right)$ and ${\rm Tr} \left(\hat\rho\,\alpha_{\bf k} \alpha_{-\bf k}
\right)$ vanish identically, and  the diagonal quantities are just the
finite-temperature  harmonic-oscillator occupation numbers

\begin{equation}
{\rm Tr}\left(\hat\rho\,\alpha_{\bf k}^\dagger \alpha_{\bf k}\right) =
f(E_k) \equiv \frac{1}{\exp(\beta E_k)-1}\quad \hbox{and}\quad
 {\rm Tr}\left(\hat\rho\,\alpha_{\bf k} \alpha_{\bf k}^\dagger\right) =
1+ f(E_k),
\end{equation}
involving the familiar Planck (or Bose-Einstein) distribution
function $f(E_k)$  for a quasiparticle with energy $E_k$ in thermal equilibrium
at temperature $T$ [note that
  $f(E_k)$ has  no chemical
potential because quasiparticles are not conserved].

	A combination of these results with eq.~(\ref{eq:number}) immediately
yields the relation

\begin{equation}
N = N_0 + \sum_{\bf k \neq 0} \left[v_k^2 +\frac{u_k^2
+v_k^2}{\exp(\beta E_k)-1}\right]= N_0 + \sum_{\bf k \neq 0}
\left[v_k^2 +\left(u_k^2
+v_k^2\right)f(E_k)\right].\label{eq:number1}
\end{equation}
Since $N$ is fixed, this equation determines  implicitly the
temperature-dependent condensate number $N_0(T)$. In a similar way, the
low-temperature internal energy $E(T)$ is simply the finite-temperature
average of the Bogoliubov quasiparticle hamiltonian in
eq.~(\ref{eq:hqp})

\begin{equation}
E(T) = {\rm Tr}\left(\hat\rho\hat H\right) = E_g + \sum_{\bf k\neq
0}\frac{E_k}{\exp(\beta E_k)-1}=  E_g + \sum_{\bf k\neq
0}E_k\, f(E_k);\label{eq:inten}
\end{equation}
expressed as the  zero-temperature ground-state energy $E_g$ plus the sum
over all thermally weighted excited states.

For simplicity, consider the  total number of non-condensate
particles $N'\equiv N-N_0$ (the depletion of the condensate arising
from the repulsive interactions).  At zero temperature,  the
thermal factors vanish, yielding

\begin{equation}
N'(T=0) = N-N_0(T=0) = \sum_{\bf k\neq 0} v_k^2.\label{eq:number2}
\end{equation}
This equation shows that $v_k^2$ can be interpreted as the number
$N_{\bf k}'$ of non-condensate particles  with wave number
$\bf k$ in the Bogoliubov ground state $|\Phi\rangle$;  since $v_k^2\propto
(k\xi)^{-1} $ is large for $k\xi\ll 1$, the low-lying phonon-like modes
have large occupation numbers (this behavior reflects the Bose
condensation in the mode with $\bf k = 0$).  Nevertheless, the total
noncondensate number remains finite because $v_k^2$ vanishes rapidly
for $k\xi \gg 1$.

The sum over the excited plane-wave states $\bf k$ in
eq.~(\ref{eq:number2}) can be approximated as an integral $\sum_{\bf
k\neq 0}\,\cdots  \approx V(2\pi)^{-3}\int d^3 k\,\cdots$, and the
total zero-temperature fractional depletion becomes \cite{Bog,FW,LHY}

\begin{eqnarray}
\frac{N'(T=0)}{N} &= &\frac{N-N_0(T=0)}{N}\approx
\frac{1}{n}\int
\frac{d^3k}{(2\pi)^3}\,v_k^2\nonumber\\
&=& 4\left(\frac{2na^3}{\pi}\right)^{\!\!1/2}\int_0^\infty
y^2\,dy\bigg[\frac{y^2 + 1}{\left(y^4 +
2y^2\right)^{1/2}}-1\bigg]\nonumber\\
&=&\frac{8}{3}\left(\frac{na^3}{\pi}\right)^{\!\!1/2}.\label{eq:depl1}
\end{eqnarray}
This result exhibits $\sqrt{na^3}$ as the relevant expansion
parameter, and the assumption of small total depletion requires
that $\sqrt{na^3}\ll 1$.  Note that the fractional depletion is non-analytic in
the interaction strength
$a$ (or $g$), for the $\frac{3}{2}$ power requires a branch cut in the
complex $a$
plane.  As a result, this and other corrections to the physical properties
of an
ideal Bose gas are
 non-perturbative;  they cannot be obtained with conventional perturbation
theory.  Such a conclusion here presents no conceptual difficulty, however,
for it
relies  a (non-perturbative) canonical transformation.   It is not difficult to
obtain the first low-temperature contribution to the depletion

\begin{equation}
\frac{N'(T)}{N} = \frac{N-N_0(T)}{N}
\approx\frac{8}{3}\left(\frac{na^3}{\pi}\right)^{\!\!1/2}
+\frac{M(k_BT)^2}{12\hbar^3sn}.\label{eq:depl2}
\end{equation}

It is also interesting to evaluate the ground-state energy from
eq.~(\ref{eq:grndst}), which has the form

\begin{equation}
E_g =  \case{1}{2}gn^2V -\case{1}{2}
 \sum_{\bf k \neq
0}\left(\epsilon_k^0 + ng -E_k\right).\label{eq:energy1}
\end{equation}
An expansion of $E_k$ for large $k\xi$ shows that the sum diverges
like $g^2\sum_{\bf k\neq 0} k^{-2}$, which reflects a failure of
second-order perturbation theory for a dilute Bose gas.  Here, it
arises from the assumption that the Fourier transform $g$ of the
interparticle potential remains constant for large wave numbers, and
the same divergence appeared in eq.~(\ref{eq:scattamp2}) for the
scattering length to order $g^2$.  Equation (\ref{eq:energy1}) can be
rewritten by adding and subtracting the divergent second-order
contribution

\begin{equation}
E_g = \case{1}{2} gn^2 V- \case{1}{2}g^2n^2\sum_{\bf k \neq
0}\frac{M}{\hbar^2k^2} +\case{1}{2}
 \sum_{\bf k \neq
0}\left(E_k-\epsilon_k^0 - ng +\frac{Mg^2n^2}{\hbar^2k^2}\right),
\end{equation}
and it is easy to verify that the last sum  converges.  In
addition, the first two terms on the right-hand side are precisely
$\frac{1}{2}n^2V$ times those in eq.~(\ref{eq:scattamp2}) for
$4\pi \hbar^2a/M$, where $a$ is the physical
scattering length.  A detailed
evaluation yields the ground-state energy per particle~\cite{FW,LY}

\begin{eqnarray}
\frac{E_g}{N} &= &\case{1}{2}n\left(g -
\frac{Mg^2}{\hbar^2}\int\frac{d^3 k}{(2\pi)^3}\,\frac{1}{k^2}\right)
+\case{1}{2}g\int
\frac{d^3k}{(2\pi)^2}\left(\frac{E_k}{ng}-\frac{\epsilon_k^0}{ng} -1
+\frac{ng}{2\epsilon_k^0}\right) \nonumber  \\
&=&\frac{2\pi
\hbar^2 an}{M}\left\{1+
8\left(\frac{2na^3}{\pi}\right)^{\!\!1/2}\int_0^\infty
y^2\,dy\bigg[\left(y^4 +
2y^2\right)^{1/2}-y^2-1+\frac{1}{2y^2}\bigg]\right\}\nonumber\\
 &=
&\frac{2\pi
\hbar^2 an}{M}\left[1+
\frac{128}{15}\left(\frac{na^3}{\pi}\right)^{\!\!1/2}\right],
\label{eq:energy2}
\end{eqnarray}
which again exhibits the same non-analytic dependence on $\sqrt{na^3}$.

Equation (\ref{eq:energy2}) expresses the ground-state energy as an
explicit function of the number
$N$ and volume $V$, and standard thermodynamics provides many other
important  ground-state properties.  For example, the
chemical potential $\mu$  and pressure $p$ follow from

\begin{equation}
\mu = \left(\frac{\partial E_g}{\partial N}\right)_{\!\!V}=
\frac{4\pi\hbar^2an}{M}\left[ 1 +
\frac{32}{3}\left(\frac{na^3}{\pi}\right)^{\!\!1/2}\right],
\end{equation}
and

\begin{equation}
p = -\left(\frac{\partial E_g}{\partial V}\right)_{\!\!N}=
\frac{2\pi\hbar^2an^2}{M}\left[ 1 +
\frac{64}{5}\left(\frac{na^3}{\pi}\right)^{\!\!1/2}\right].
\end{equation}
In addition, the speed of sound $s$ follows from the compressibility
through the general formula $s^2= M^{-1}\left(\partial p/\partial
n\right)$,  giving

\begin{equation}
s =
\frac{\hbar}{M} \sqrt{4\pi an}\left[ 1 +
8\left(\frac{na^3}{\pi}\right)^{\!\!1/2}\right].
\end{equation}
It is striking that the speed of sound obtained here with macroscopic
thermodynamics agrees to leading order with the slope of the
quasiparticle spectrum given in eq.~(\ref{eq:speed});  although this
zero-temperature relation is not entirely obvious, it has, in fact, been
verified to all orders in perturbation theory \cite{GN,HM} that the
slope of the quasiparticle spectrum agrees precisely with the
thermodynamic speed of sound.

\subsection{Moving condensate}

It is not difficult to generalize
this analysis to the case of Bose condensation in a mode with
non-zero momentum $\hbar\bf q$, which implies that the condensate
moves with velocity ${\bf v} = \hbar{\bf q}/M$.   The
quasiparticle energy spectrum $E_{\bf k}$ for the state with
the proper positive normalization $|u|^2-|v|^2 = 1$ is simply related
to the energy
$E_k^0$  for a stationary condensate:

\begin{equation}
E_{\bf k} = \hbar{\bf k\cdot v} + E_k^0.\label{eq:doppler}
\end{equation}
This expression can be considered a Doppler shift in the frequency;
its form  agrees precisely with Landau's general result  in his celebrated 1941
paper on superfluid
${}^4$He
\cite{LDL,LP1}. In the long-wavelength limit, the excitation energy
reduces to
$E_{\bf k} \approx \hbar k(v\cos\theta+s)$, where $s$ is the
Bogoliubov speed of sound and $\theta$ is the angle between $\bf k$ and
the flow velocity
$\bf v$.  For $v\le s$, the quasiparticle  energy is positive for all
physical angles $\theta$;   if $v$ exceeds $s$, however, the
quasiparticle energy becomes negative in a cone around the backward
direction where $\cos\theta\approx -1$.  Consequently, the
quasiparticle hamiltonian in eq.\ (\ref{eq:hqp}) becomes unstable, and
the system can lower its energy arbitrarily by creating
quasiparticles with negative energy.  This behavior corresponds to the
well-known Landau critical velocity $v_c$ for the onset of dissipation in a
superfluid
\cite{LDL,LP1}. In the Bogoliubov model, $v_c$  is equal to the speed of sound
$s$.  Physically, the finite $s$  reflects the presence of repulsive
interactions
[see eq. (\ref{eq:speed})] and therefore $s$ vanishes for an ideal Bose
gas.  As a
corollary, an ideal Bose gas also has $v_c=0$ and hence
 cannot sustain  superfluid flow.

At $T=0$, it is straightforward to prove that the {\it total\/} momentum
$\bf P$ carried by both the condensed and non-condensed particles is
 ${\bf P } = NM{\bf v}$, where $N$ is the {\it total\/}
number of particles,  {\it not\/} the condensate number $N_0$.  This result
implies that the  Bose condensation induces a coherence or rigidity,
ensuring that, in effect,  {\it all\/} the particles participate in the
motion.  The same conclusion  also follows with a Galilean
transformation from the rest frame of the condensate to a  frame
moving with velocity
$-\bf v$.

At finite temperature $T\neq 0$, the corresponding
total momentum   decreases, according to
the expression ${\bf P} = (\rho_s/\rho)\,NM{\bf v}$, where the
reduction factor
$\rho_s(T)/\rho\le 1$ is the temperature-dependent superfluid fraction
 first calculated by Landau
\cite{LDL,LP1}.  In his  model for superfluidity, the total
density $\rho$ consists of two separate parts, the superfluid density
$\rho_s$ and the normal fluid density $\rho_n$, with $\rho_s +
\rho_n =
\rho$.  The normal fluid density vanishes at $T=0$ and   depends
solely on the excitation spectrum in the rest frame of the
condensate.  For the Bogoliubov energy
$E_k$, the normal fluid fraction $\rho_n/\rho$ is proportional to  $T^4$ at
low temperatures, similar to the phonon contribution in superfluid
${}^4$He  \cite{Wilks}.

\section{Dilute Bose gas in a harmonic trap}\label{sec:trap}

The basic physics of a uniform dilute Bose gas has  been well understood for
over 30 years \cite{AG1,HM}.  The  recent intense renewed interest in
such systems arises from the remarkable experimental demonstration of
Bose-Einstein condensation of dilute ultra-cold alkali atoms in harmonic
traps \cite{And,Dav,Brad}.  The presence of a trap alters the physics in
several
important ways, and it is essential to reformulate the theory to
incorporate the
many  qualitatively new features \cite{RMP}.

	\subsection{Ideal Bose gas in a harmonic trap}

In the usual case,  the trap is an  axisymmetric harmonic-oscillator
potential

\begin{equation}
V_{\rm tr}({\bf r}) = V_{\rm tr}(r_\perp, z)=
\case{1}{2}M(\omega_\perp^2r_\perp^2+\omega_z^2z^2),\label{eq:trap}
\end{equation}
where the coordinate vector $\bf r$ is expressed in cylindrical polar
coordinates ($r_\perp, \phi, z$).  Here,  $\omega_\perp$ and $\omega_z$ are the
radial and axial angular frequencies, and it is common to introduce the
``anisotropy parameter''

\begin{equation}
\lambda\equiv\frac{\omega_z}{\omega_\perp}. \label{eq:aniso}
\end{equation}
A particle with mass $M$ in such a  potential has
a Gaussian ground-state wave function

\begin{equation}
\psi_g(r_\perp,z) \propto
\exp\bigg[-\frac{1}{2}\left(\frac{r_\perp^2}{d_\perp^2}
+\frac{z^2}{d_z^2}\right)\bigg],\label{eq:psig0}
\end{equation}
with radial and axial dimensions

\begin{equation}
d_\perp=\sqrt{\frac{\hbar}{M\omega_\perp}}\quad\hbox{and}\quad
d_z=\sqrt{\frac{\hbar}{M\omega_z}}=
\frac{d_\perp}{\lambda^{1/2}},\label{eq:osc}
\end{equation}
which are the relevant oscillator lengths (the mere presence of the trap
introduces
a new characteristic length scale that will play a significant role in the
physics of these systems).  The  anisotropy in the ideal-gas density profile is
simply

\begin{equation}
\frac{d_z^2}{d_\perp^2} = \frac{1}{\lambda}\quad\hbox{for a non-interacting
gas},\label{eq:aniso-id}
\end{equation}
where  $\lambda\gg 1 $ represents  a ``pancake''  and
$\lambda\ll 1$ represents a ``cigar.''

What is  the transition temperature $T_c$ for the onset of Bose-Einstein
condensation in an ideal gas in such a trap?   For many qualitative
discussions, it is  convenient to define a mean frequency $\omega_0=
(\omega_\perp^2\omega_z)^{1/3}$ and a mean oscillator length
$d_0=(d_\perp^2d_z)^{1/3}$. Consider the classical limit, with
$k_BT \gg \hbar\omega_0$.  Instead of the ground-state density $|\psi_g({\bf
r})|^2$,
 the  classical density profile  has the
Maxwell-Boltzmann form

\begin{equation}
n({\bf r})\propto \exp\left[-\beta V_{\rm tr}({\bf
r})\right]= \exp\bigg[-\frac{V_{\rm tr}({\bf r})}{k_BT}\bigg],
\end{equation}
where $\beta^{-1} = k_BT$.  For simplicity, consider a spherical trap
($\lambda = 1$), when the density

\begin{equation}
n(r) \propto \exp\left(-\frac{r^2}{R_T^2}\right),
\end{equation}
 is isotropic with a mean thermal radius $R_T$ given by

\begin{equation}
R_T^2 = \frac{2k_BT}{M\omega_0^2}=
d_0^2\,\frac{2k_BT}{\hbar\omega_0} \gg d_0^2.\end{equation}
In this classical limit, the density is again Gaussian, but  its mean
dimension
 $R_T$   necessarily exceeds the mean ground-state
radius $d_0$.  Correspondingly, the mean number density is of order $n \sim
N/R_T^3$.

Recall the estimate $k_BT_c \sim \hbar^2n^{2/3}/M$ in eq.~(\ref{eq:Tc}) for the
onset of quantum degeneracy in an ideal Bose gas.  A combination of these
expressions readily shows that

\begin{equation}
k_BT_c\sim N^{1/3}\hbar\omega_0,\label{eq:Tc2}
\end{equation}
and a detailed calculation gives the more accurate value $k_BT_c=
[\zeta(3)]^{-1/3} N^{1/3}\hbar\omega_0\approx  0.941 N^{1/3}\hbar\omega_0$.
Since $N$ typically is large in most cases of interest, the
Bose-Einstein transition  indeed occurs in the classical regime
($k_BT_c\gg
\hbar\omega_0$). For $T<T_c$, a Bose condensate forms in the lowest
single-particle state of the harmonic oscillator potential, whose  width
is the
oscillator length $d_0$,  much smaller than the thermal width $R_T$ of
the non-condensate for $T\lesssim T_c$.  The appearance of this sharp
narrow spike
in the particle density served as a dramatic  signal of the first  Bose
condensation in a dilute alkali gas \cite{And}.

In a typical situation (see, for example, ref.~\cite{Mewes2}), the trap has a
frequency
$\omega_0/2\pi
\sim 100\ \rm Hz$ and an oscillator length $d_0\sim 2\ \rm \mu m$, containing
$N\sim10^6$ particles. The ground-state energy corresponds to a temperature
$\hbar\omega_0 /k_B\sim 5\times 10^{-9}\ \rm K$.  In this case,  $N^{1/3}\sim
100$, so that $T_c\sim 0.5\times 10^{-6}\ \rm K$;  at this temperature,  the
ratio
$R_T/d_0\sim N^{1/6}$ is of order 10.

\subsection{Effect of interactions on a trapped Bose condensate}

As in the case of a uniform condensate, the Bose gas is assumed to be
dilute, in
the sense that the interparticle spacing $l\sim n^{-1/3}$ is large compared to
the $s$-wave scattering length $a$ and the range of the interactions.  Thus the
interparticle potential is again approximated as a pseudopotential $V({\bf r} )
\approx g\delta^{(3)}({\bf r})$, with $g\approx 4\pi \hbar^2 a/M$, as in
eq.~(\ref{eq:pseudo}).  Typically this low-density condition is  well
satisfied, with $a\sim$ a few nm and mean number density $n\sim 10^{20}\
\rm m^{-3}$, implying $l\sim$  a  few $\times 10^{-7} $ m and $a\ll
l$.  In addition, the condition $a\ll d_0$ holds whenever $na^3\ll 1$,
since the existence of a macroscopic Bose condensate implies  that $l\ll
d_0$.

The first question is how to generalize the Bogoliubov theory to describe a
non-uniform Bose gas in a trap.  Although  various approaches lead to
essentially the same final results, the treatment here \cite{AP} emphasizes the
field-theory aspects of the many-body hamiltonian

\begin{equation}
\hat H = \int d^3r\left[\hat\psi^\dagger\left(T + V_{\rm tr}\right) \hat\psi +
\case{1}{2}g\,\hat\psi^\dagger\hat\psi^\dagger\hat\psi\hat\psi\right],
\label{eq:hatH}
\end{equation}
where

\begin{equation}
T= -\frac{\hbar^2\nabla^2}{2M}
\end{equation}
is the kinetic-energy operator,  $V_{\rm tr}$ is the harmonic potential given
in eq.~(\ref{eq:trap}), and the two-body potential has been replaced by
a short-range pseudopotential $V({\bf r} )
\approx g\delta^{(3)}({\bf r})$.  Here, $\hat\psi$ and $\hat\psi^\dagger$ are
field operators that obey bosonic commutation relations

	\begin{equation}
\left[\hat\psi({\bf r}), \hat\psi^\dagger({\bf r'})\right] = \delta^{(3)}({\bf
r-r'}), \quad \left[\hat\psi({\bf r}), \hat\psi({\bf r'})\right]=
\left[\hat\psi^\dagger({\bf r}), \hat\psi^\dagger({\bf r'})\right]= 0.
\end{equation}

As an aside, the field operator for the uniform system has the simple form

\begin{equation}
\hat\psi({\bf r} ) = \sum_{\bf k} \frac{1}{\sqrt V}\,e^{i{\bf k\cdot
r}}\,a_{\bf
k},
\end{equation}
using the normalized plane-wave single-particle wave functions. More generally,
given any complete set of normalized single-particle wave functions
$\{\chi_j({\bf r})\}$, the corresponding field operator is  a sum over all
normal modes

\begin{equation}
\hat\psi({\bf r}) = \sum_j\chi_j({\bf r}) a_j,
\end{equation}
where $a_j$ is a corresponding bosonic annihilation operator for the
single-particle mode
$j$.

In the Bogoliubov approximation for a non-uniform system, one single mode is
assumed to be   macroscopically occupied.   The exact field operator is
therefore separated into two terms,

\begin{equation}
\hat\psi({\bf r} ) \approx \Psi({\bf r}) +
\hat\phi({\bf r}),\label{eq:fluct}
\end{equation}
where $\Psi$ is the (large) condensate wave function [analogous to the
Bogoliubov
prescription in eq.~(\ref{eq:Bogpres})] and
$\hat\phi$ is the (small) operator characterizing the remaining non-condensate.
Formally, the condensate wave function $\Psi =\langle\hat\psi\rangle$ is an
ensemble
average of the full field operator  in a particular ensemble that breaks the
symmetry with respect to the phase \cite{HM}, and  the remaining fluctuation
part $\hat\phi$ can be defined by the condition $\langle\hat\phi\rangle =
0$. In
particular, the condensate density
$n_0({\bf r})$ is given by

\begin{equation}
n_0({\bf r}) = |\Psi({\bf r})|^2,
\end{equation}
and the total number of condensate particles is

\begin{equation}
N_0 = \int d^3 r \,|\Psi|^2.
\end{equation}

It is most convenient to introduce the grand canonical ensemble at temperature
$T$ and chemical potential $\mu$,  studying the modified hamiltonian

\begin{equation}
\hat K = \hat H -\mu \hat N,\label{eq:hatK}
\end{equation}
where $\hat N = \int d^3r\,\hat\psi^\dagger\hat\psi$ is the number operator.
 For a uniform system, the condensate number density $n_0$
is simply a spatial constant.  For a non-uniform system, however,  the
determination of
$\Psi$ and
$n_0$   is in general quite complicated.  In analogy
with the leading  term of  the Bogoliubov hamiltonian in eq.~(\ref{eq:ham3}),
where all the creation and annihilation operators were replaced by the
condensate
contributions
$\sqrt{N_0}$, the condensate part of the modified  hamiltonian

\begin{equation}
K_0 = H_0-\mu N_0=\int d^3 r\left[\Psi^*\left(T+V_{\rm
tr}-\mu\right)\Psi+\case{1}{2}g\,\Psi^*\Psi^*\Psi\Psi\right]\label{eq:K0}
\end{equation}
    is assumed to  dominate the physics at low temperature.

In the grand canonical ensemble at zero temperature,
equilibrium corresponds to the minimum value of  the thermodynamic potential,
which is
 the ensemble average of the modified hamiltonian $\langle \hat K\rangle$.  For
the present situation of a macroscopic non-uniform condensate, this quantity is
approximately  $K_0$ in eq.~(\ref{eq:K0}), and the only available
parameter to vary is the condensate wave function itself.  Thus, the value of
the  integral $K_0$ must be stationary  under the variation
$\Psi^*\to\Psi^* + \delta\Psi^*$ \cite{AP}.  The corresponding
Euler-Lagrange equation

\begin{equation}
\left(T+ V_{\rm tr}-\mu\right) \Psi + g|\Psi|^2\Psi = 0\label{eq:GP}
\end{equation}
is known as the Gross-Pitaevskii (GP) equation  \cite{EPG,LPP}. This non-linear
Schr\"odinger equation for the condensate wave function is expected to hold at
low temperature;  it is  similar in form (but not in spirit) to the non-linear
Ginzburg-Landau equation for the superconducting order parameter that
applies  near
the superconducting transition temperature (see, for example, ref.\ \cite{LP1},
sec.\ 45).

It is
convenient to  introduce the ``Hartree'' potential energy
$V_H({\bf r})$, which is the  interaction potential of a particle at
$\bf r$ with all the other condensed particles

\begin{equation}
V_H({\bf r}) \approx \int d^3 r'\, V({\bf r-r'})\,n_0({\bf r'}) \approx
gn_0({\bf
r})= g|\Psi({\bf r})|^2 ,\label{eq:VH}
\end{equation}
so that the GP equation assumes the equivalent form

\begin{equation}
\left(T + V_{\rm tr} +V_H-\mu\right)\Psi = 0.
\end{equation}
 The {\it  interaction  energy\/}  of the condensate is the last term of eq.\
(\ref{eq:K0})
\begin{equation}
E_{\rm int}=\langle V_{\rm
int} \rangle_0=
 \int d^3 r\,\case{1}{2}g|\Psi|^4 = \case{1}{2} \int d^3 r\,V_H({\bf
r})\,|\Psi({\bf r})|^2= \case{1}{2} \langle V_H\rangle_0,
\end{equation}
where the angular bracket here denotes an integral $\langle \cdots\rangle_0 =
\int d^3 r\,\Psi^*\cdots\Psi$ over the ground-state condensate wave
function; note
that
$E_{\rm int}$ is half the mean Hartree energy of the condensate.

 In time-dependent form, the GP
equation becomes

\begin{equation}
i\hbar\frac{\partial \Psi}{\partial t} = \left(T + V_{\rm tr} +
V_H\right)\Psi ,
\label{eq:TDGP}\end{equation}
where the condensate wave function is interpreted as an off-diagonal matrix
element of the destruction operator  $\Psi({\bf r},t)
=
\langle N-1|\hat\psi({\bf r},t)|N\rangle$.   The time-dependence of the
operator is
given by the usual Heisenberg
picture $\hat\psi({\bf r},t)=\exp(i\hat H t/\hbar)\, \hat\psi({\bf
r})\exp(-i\hat H t/\hbar)$.  The time-dependent  phase factor in the  matrix
element thus
  involves the change  in the ground-state energy $E_g(N-1)-E_g(N)$ when a
particle is removed from the system, which immediately reproduces the
original GP eq.\ (\ref{eq:GP}) because $\mu\approx
\partial E_g/\partial N$ at $T=0$.

The ground-state  solution of the GP equation satisfies several important
relations.  The first is the ground-state energy $E_g$, which is the
expectation
value of the condensate hamiltonian in eq. (\ref{eq:K0})

\begin{equation}
E_g = \langle H_0\rangle_0 = \langle T\rangle_0 + \langle V_{\rm tr}\rangle_0
+\langle V_{\rm int}\rangle_0=\langle T\rangle_0 + \langle V_{\rm tr}\rangle_0
+\case{1}{2}\langle V_H\rangle_0,\label{eq:Eg}
\end{equation}
 Second, the
expectation value of the GP equation itself yields an expression for the
chemical potential

\begin{equation}
N_0\mu = \langle T\rangle_0 + \langle V_{\rm tr}\rangle_0
+\langle V_H\rangle_0= \langle T\rangle_0 + \langle V_{\rm tr}\rangle_0
+2\langle V_{\rm int}\rangle_0.\label{eq:mu}
\end{equation}
Finally, consider a dimensionless scale transformation on the coordinates ${\bf
r} \to \zeta{\bf r}$.  The scaling of the various terms in $H_0$ follows by
inspection to give

\begin{equation}
E_g\to  \zeta^{-2}\langle T\rangle_0 +\zeta^2 \langle V_{\rm tr}\rangle_0
+\zeta^{-3}\langle V_{\rm int}\rangle_0;
\end{equation}
since this scale transformation cannot change the  ground-state energy, it
follows that $\left(\partial E_g/\partial \zeta\right)|_{\zeta = 1} = 0$, which
readily yields the virial theorem \cite{DS}

\begin{equation}
  2 \langle V_{\rm tr}\rangle_0= 2\langle T\rangle_0
+3\langle V_{\rm int}\rangle_0=2\langle T\rangle_0
+\case{3}{2}\langle V_H\rangle_0.\label{eq:vir}
\end{equation}
A combination of this identity with Eqs.\ (\ref{eq:Eg}) and (\ref{eq:mu})
provides the simpler expressions

\begin{mathletters}
\label{eq:virtot}
\begin{equation}
E_g = \case{1}{3}\langle  T\rangle_0 + \case{5}{3}\langle V_{\rm
tr}\rangle_0,\label{eq:virEg}
\end{equation}
\begin{equation}
N_0\mu  = -\case{1}{3}\langle  T\rangle_0 + \case{7}{3}\langle V_{\rm
tr}\rangle_0\label{eq:virmu}
\end{equation}\end{mathletters}
that do not involve the Hartree or interaction energy.

\subsection{Basic physics of the Gross-Pitaevskii equation}

The presence of the trap introduces a new characteristic energy $\hbar\omega_0$
and a new length scale $d_0=\sqrt{\hbar/M\omega_0}$.  The repulsive
interparticle interactions tend to expand the condensate, so that the true mean
size $R_0$ generally exceeds the size  $d_0$ of the ideal Bose gas in
the Gaussian ground state. To estimate the dimensionless  expansion ratio
$R\equiv R_0/d_0$, consider the order-of-magnitude of the various terms in the
ground-state energy.  Since the ground-state wave function is nodeless, the
kinetic energy per particle has the order-of-magnitude

\begin{mathletters}
\begin{equation}
\frac{\langle T\rangle_0 }{N_0}\sim \frac{\hbar^2}{MR_0^2}
\sim\frac{\hbar\omega_0}{R^2}.
\end{equation}
Similarly,
\begin{equation}
\frac{\langle V_{\rm tr}\rangle_0 }{N_0}\sim \hbar\omega_0\,R^2,
\end{equation}
and
\begin{equation}
\frac{\langle V_H\rangle_0 }{N_0}\sim gn\sim
\frac{\hbar^2an}{M}\sim\frac{\hbar^2aN}{MR_0^3}
\sim\frac{\hbar\omega_0}{R^3}\,\frac{Na}{d_0},
\end{equation}
\end{mathletters}
with  the  {\it new\/} dimensionless parameter $Na/d_0$
that specifically reflects the presence of the trap. Although the ratio
$a/d_0$ is generally small, of order  $\sim10^{-3}$, $N$ is often of order
$10^6$-$10^7$, so that the product $Na/d_0$ can itself be large, of order
$10^3$-$10^4$.

The total energy per particle is the sum of these three contributions

\begin{equation}
\frac{E_g}{N_0} \sim\hbar\omega_0 \left(\frac{1}{R^2} + R^2+
\frac{Na}{d_0}\,\frac{1}{R^3}\right),\label{eq:Evar}
\end{equation}
with the dimensionless expansion ratio $R$ as  a variational parameter.
Two limits are easy to analyze.

1. If $Na/d_0\ll 1$ (a nearly ideal Bose gas), the last term of eq.\
(\ref{eq:Evar}) can be omitted, and the minimum ground-state energy occurs at
$R=1$, which is just the familiar variational solution for the simple harmonic
oscillator with $\langle T\rangle_0= \langle V_{\rm tr}\rangle_0$.

2.  In the opposite limit that $Na/d_0\gg 1$, the repulsive interactions are
crucial (even though the gas is assumed to remain dilute with $na^3\ll 1$).  In
this case,  the kinetic energy ($\propto R^{-2}$) can be neglected because the
condensate is significantly larger than its ideal-gas dimension, and the
minimum ground-state energy occurs when the last two terms in eq.\
(\ref{eq:Evar})
are comparable, namely when
\begin{equation}R^5= R_0^5/d_0^5 \sim Na/d_0\gg 1.\label{eq:TFexpan}
\end{equation}

Before proceeding with an analysis of this  very important scaling
relation that holds in what has become known as the Thomas-Fermi (TF) limit, it
is helpful to recall the healing (or coherence) length
$\xi  = (8\pi
na)^{-1/2}$ defined in eq.~(\ref{eq:healing}) for a uniform unbounded
condensate.
Physically, $\xi$  arises from the balance between the kinetic energy
$\hbar^2/2M\xi^2 $ and the Hartree energy $gn=4\pi \hbar^2a/M$. When the bulk
condensate wave function $\Psi$ is perturbed locally
(for example,  an otherwise uniform Bose gas in a half space bounded by an
external
plane), the length
$\xi$ characterizes the distance required for $\Psi$ to heal back to the
previous
background value, justifying the  name ``healing length.''

The presence of a closed boundary  (with an  additional associated length
scale) complicates the situation, even in the absence of an external confining
potential like $V_{\rm tr}$.  One simple and exactly soluble example is a
three-dimensional   dilute Bose gas  with uniform areal density $N/A$ in the
$yz$ plane and   confined to  a slab
$|x|\le
\frac{1}{2} L$ by  rigid transverse walls~\cite{Wu}.  Here the real
condensate wave
function
$\Psi(x)$ obeys a one-dimensional GP equation $-(\hbar^2/2M)\Psi^{\prime\prime}
-\mu\Psi + g\Psi^3 = 0$ and vanishes at
$x=\pm
\frac{1}{2} L$.  The nearly ideal limit can be characterized by the condition
$\xi^2\sim AL/Na \gg L^2$, when  the   condensate
density $n_0(x) = |\Psi(x)|^2$ is everywhere small and proportional to
$\cos^2(\pi
x/L)$.  In this case,  the length scale $L$ characterizes the spatial
variation of $n_0$, so that $L$ itself  plays the role of  the healing
length in
the nearly ideal limit.   As
$N$ increases (and $\xi$ decreases), however,  the  repulsive interactions
act to
flatten and spread  the central density maximum. In the opposite limit
$\xi^2\ll
L^2$,  the detailed solution shows that the condensate density  $n_0$ forms
a
plateau with the density falling to zero at the walls in a skin thickness of
order $\xi$.  This simple model shows that  the physical  healing length in a
confined Bose condensate  should be taken as the smaller of the two lengths
$L$ and
$\xi$

A similar situation holds for a dilute three-dimensional  Bose gas in a
harmonic
trap, apart from the additional complication that the condensate's volume
$\sim R_0^3$ can exceed the ideal-gas volume $\sim d_0^3$ when the interactions
are sufficiently strong.  The squared coherence length
$\xi^2\sim 1/na\sim R_0^3/Na$   provides an
instructive way to think about the various limits discussed
above in connection with the variational energy in eq.~(\ref{eq:Evar}).
In the
nearly ideal limit  ($Na/d_0\ll 1$ and
$R_0
\approx  d_0$), it is easy to see that
$\xi^2 \gg d_0^2$;   the ground-state Gaussian condensate density
varies on the length scale $d_0$, which thus acts as the healing length.
 As  $N$ increases, however, the coherence length  $\xi$ shrinks,  becoming
comparable with the ideal-gas  condensate dimension
$d_0$   when $Na/d_0$ is of order 1.

 In the opposite (TF) limit   $Na/d_0 \gg1$,
the coherence  length $\xi$ falls significantly  below
$d_0$, and $\xi$ itself now plays the role of the healing length (for
example, near
a hole produced by   a localized blue-detuned laser beam).
In addition, the characteristic condensate dimension
$R_0$  expands beyond
$d_0$.  To analyze this last result in detail, note that the definition of the
coherence
 length implies that
$\xi^2R_0^2 \sim R_0^5/Na$, and  use of  the preceding qualitative
 TF relation
$R_0^5/d_0^5 \sim Na/d_0$ from eq.~(\ref{eq:TFexpan}) yields

\begin{equation}
\xi^2R_0^2 \sim d_0^4\,\frac{R_0^5}{d_0^5}\,\frac{d_0}{Na} \sim
d_0^4,\quad\hbox{or, equivalently,}\quad
\frac{\xi}{d_0}\sim\frac{d_0}{R_0}\ll1. \label{eq:healingTF0}
\end{equation}
Consequently, in the TF limit,  the oscillator length $d_0$ is
approximately the
geometric mean of the small healing length $\xi$ and the large condensate
radius
$R_0$.

If the gas remains dilute with
$na^3\ll1$, the healing length is considerably larger than the interparticle
spacing.  The situation changes in  the extreme large-$N$  (high-density) limit
with $na^3\sim 1$,  for  the healing length
$\xi$ then  becomes comparable with the interparticle spacing $l\sim
n^{-1/3}$. Although this dense regime is probably unattainable for the trapped
alkali gases because of three-body recombination effects, it is worth
mentioning
that  bulk superfluid
${}^4$He does have $na^3\sim 1$, for its healing length $\xi$ is
comparable with an atomic dimension (and thus with the interparticle spacing).

The original Bogoliubov theory of a dilute uniform
 Bose gas with $na^3\ll1$ has no independent length scale
for the dimension of the condensate. In contrast, the new dimension
$d_0$ for the trapped Bose gas correspondingly yields two  distinct physical
regimes (although both are dilute with $na^3\ll1$):

 1.  the nearly ideal regime  with $\xi \gg d_0$ and $Na/d_0\ll 1$;

2. the interacting regime with $\xi \ll d_0$ and $Na/d_0\gg 1$.

\noindent This comparison of the uniform and trapped dilute Bose gases
emphasizes the qualitatively new physics associated with the presence of the
trap.

\subsection{The behavior of the condensate for large $N$}

In the limit $Na/d_0\gg 1$, the GP equation (\ref{eq:GP}) can be simplified by
omitting the kinetic energy \cite{HS,BP},  yielding a simple algebraic
equation

\begin{equation}
(V_{\rm tr} + V_H -\mu)\Psi = 0.
\end{equation}
Either $\Psi = 0$ or  the Hartree potential and condensate density are   given
by $V_H = g|\Psi|^2 = \mu - V_{\rm tr}$.  In the present case of a harmonic
potential, the approximate condensate density  has a parabolic profile

\begin{equation}
|\Psi({\bf r})|^2 = n_0({\bf r}) = n(0)\left(1-\frac{r_\perp^2}{R_\perp^2} -
\frac{z^2}{R_z^2}\right)\,\Theta\!\left(1-\frac{r_\perp^2}{R_\perp^2} -
\frac{z^2}{R_z^2}\right),\label{eq:TFdens}
\end{equation}
where $\Theta$ denotes the unit positive step function.  Here,
\begin{equation}
n(0) = \frac{\mu}{g}= \frac{\mu M}{4\pi \hbar^2 a}\label{eq:n(0)}
\end{equation}
is the central density in the trap and

\begin{equation}R_\perp^2 = \frac{2\mu}{M\omega_\perp^2}\quad\hbox{and}\quad
R_z^2 = \frac{2\mu}{M\omega_z^2}\label{eq:TFsize}
\end{equation}
characterize the radial and axial semiaxes of the ellipsoidal condensate
density.   In this large-$N$ limit, the slowly varying trap potential
determines the local density [often called the Thomas-Fermi (TF) limit in
analogy
with the corresponding treatment of the electron density in large atoms].  Note
that the anisotropy of the condensate in this TF limit is now given
by
\begin{equation}
\frac{R_z^2}{R_\perp^2} = \frac{\omega_\perp^2}{\omega_z^2} =
\frac{1}{\lambda^2}\quad\hbox{in TF limit},\label{eq:aniso-TF}
\end{equation}
differing from the anisotropy $1/\lambda$ in eq.\ (\ref{eq:aniso-id}) for the
near-ideal condensate.

Typically, the fractional depletion $N'/N=1-N_0/N$ is small, and it is
sufficient to set $N_0\approx N$ and hence $n_0\approx n$.  In this case, the
normalization of the condensate wave function requires

\begin{equation}\int d^3r\,n({\bf r}) \approx \int d^3r\,|\Psi({\bf r})|^2
\approx N.
\end{equation}
A straightforward calculation \cite{BP} with the TF density from eq.\
(\ref{eq:TFdens}) yields

\begin{equation}
N = \frac{1}{15}\,\frac{R_0^5}{ad_0^4},
\end{equation}
where $R_0^3 = R_\perp^2R_z$.  This result can be rewritten in
dimensionless form
as

\begin{equation}
R^5\equiv \frac{R_0^5}{d_0^5} = 15 \frac{Na}{d_0},\label{eq:TFlimit}
\end{equation}
which has many important implications for the behavior of the condensate in the
large-$N$ (or TF) limit.

	1.  The dimensionless expansion factor $R$ grows relatively slowly,
like
$N^{1/5}$, which thus characterizes the expansion of the mean condensate radius
$R_0$.  Equivalently, for a given radius $R_0$, the total number increases like
$N\propto R_0^5$.

2.  The central density $n(0)$ is of order $N/R_0^3$, and the preceding
dependence
implies that $n(0)\propto R_0^2$.

3.  The chemical potential is $\mu \approx \frac{1}{2} \hbar\omega_0 R^2$, with
$R\gg 1$, so that $\mu $ far exceeds the ground-state energy of a single
particle in the trap.

4.  The ground-state condensate energy $E_g \approx \frac{5}{3}\langle V_{\rm
tr}\rangle_0 = \frac{5}{6}M \omega_0^2\langle r^2\rangle_0=
\frac{5}{14}\hbar\omega_0R^2  N\propto N^{7/5}$ follows from the virial theorem
in eq.\ (\ref{eq:virEg}) and direct evaluation of
$\langle r^2\rangle_0$,
since
$\langle T\rangle_0$ is negligible in the TF limit.

5.  The TF chemical potential given above in item 3 is consistent with the
general thermodynamic identity $\mu = (\partial E_g/\partial N)
\approx
\frac{7}{5} E_g/N \propto N^{2/5}$,  as well as the virial theorem in eq.\
(\ref{eq:virmu}).

6.  It is natural to define the healing length for the trapped condensate
 in terms of the central density, with $\xi = [8\pi n(0)a]^{-1/2}$.  In
the TF limit, this choice implies that
\begin{equation}
\xi R_0 = d_0^2,\quad\hbox{or, equivalently,}\quad \frac{\xi}{d_0}=
\frac{d_0}{R_0} \ll 1,\label{eq:healingTF}
\end{equation}
quantifying the previous qualitative TF conclusion from
eq.~(\ref{eq:healingTF0})
that the bare trap size
$d_0$ is the geometric mean of the  healing length
$\xi$ and the  mean condensate  radius $R_0$.

7.  This whole TF analysis is easily generalized to an arbitrary anisotropic
(triaxial) trap with
$\omega_x\neq\omega_y\neq\omega_z$  by reinterpreting the mean frequency as
$\omega_0^3 = \omega_x\omega_y\omega_z$, with similar expressions for $d_0^3 =
d_xd_yd_z$ and
$R_0^3 = R_xR_yR_z$.

Recall  that a uniform Bose gas has the speed of sound $s = \sqrt{gn/M}$
from eq.\ (\ref{eq:speed}).  In a trap, the central density $n(0)$ provides the
corresponding estimate $s\approx \sqrt{gn(0)/M}$, and item~2 above shows that
the central density $n(0)\propto R_0^2 $ scales with the square of the
condensate
radius $R_0$ in the TF limit.  Thus, the TF speed of sound in a trapped
condensate  increases linearly with $R_0$. With item 6 above,  it is easy
to find the quantitative expression

\begin{equation}
s= \frac{\hbar}{\sqrt 2 M}\,\frac{R_0}{d_0^2}= \frac{\hbar}{\sqrt 2
M\xi}.\label{eq:speedTF}
\end{equation}
This linear dependence on the condensate radius has the following  remarkable
implication for the lowest-lying compressional modes of the condensate. The
fundamental mode has a wavelength of order $R_0$, and the lowest frequency will
therefore be of order $\omega\sim s/R_0\sim \hbar/Md_0^2\sim \omega_0$, namely
the mean trap frequency $\omega_0$. Consequently, the lowest normal-mode
frequencies of the condensate should be independent of the condensate radius
$R_0$ (and hence the condensate number $N$) in the TF limit of a large
condensate.  Several experimental studies \cite{Jin,MOM} have verified this
prediction (sec.~\ref{sec:III} below contains an account of the theoretical
analysis\cite{SS}).

It is also instructive to consider the Landau critical velocity $v_c$ for
the onset
of dissipation in the condensate, which is simply  the speed of sound in
the Bogoliubov description of a  uniform dilute Bose gas.    In the present
case
of a trapped condensate, it is not obvious how to send an   impurity rapidly
through the sample, but a rigid rotation with angular velocity $\Omega$ can
accomplish much the same effect.  The equatorial speed is $\sim \Omega
R_0$, and
the onset of dissipation is expected when this speed equals $v_c\approx s$.
Thus the critical rotation speed should be of order $\Omega_c \sim v_c/R_0
\approx s/R_0\sim \omega_0$ as shown above.  This result means that the
condensate should become unstable if the rotation speed exceeds any of the trap
frequencies  (a similar instability occurs for a particle in a harmonic
potential).

\subsection{Effect of an attractive interaction}

Bose-Einstein condensation has been observed \cite{Brad} in  ${}^7$Li, which is
known to have  a negative scattering length
$a$,  corresponding to an attractive interaction.  For a uniform dilute Bose
gas, the corresponding energy spectrum in eq.\ (\ref{eq:Esubk}) becomes
imaginary at long wavelengths (see fig.\ 3)

\begin{equation}
E_k^2 = \left(\frac{\hbar^2}{2M}\right)^{\!\!2}k^2\left(k^2-16\pi
n|a|\right),\label{eq:unstable}
\end{equation}
 implying an instability for those modes with $k^2\le 16\pi n|a|$.
Equivalently,
the speed of sound becomes imaginary.  For a trap, however, the wavenumber
cannot be arbitrarily small, and the minimum value is of order $k_{\rm min}
\approx \pi/R_0$.  Hence the system can remain stable if
$\pi^2/R_0^2\gtrsim 16\pi
n|a|$.  Since the density is of order $n\sim N/R_0^3$, this analysis predicts
the critical number

\begin{equation}
N_c\approx \frac{\pi}{16}\,\frac{R_0}{|a|}
\approx\frac{\pi}{16}\,\frac{d_0}{|a|},\label{eq:Nc1}
\end{equation}
because $R_0\approx d_0$ for such small values of the ratio $N_c|a|/d_0\approx
\pi/16$.  For
$N\lesssim N_c$, the trapped condensate with attractive interactions should
remain stable because of the positive kinetic energy arising from the
confinement;  for $N\gtrsim N_c$, however, the attractive interaction energy
predominates, and the condensate should collapse.

It is not difficult to make a variational estimate of the critical condensate
number $N_c$ for a spherical trap with attractive interactions.  As a simple
model, take a   Gaussian trial function [a similar
variational estimate was used by Baym and Pethick \cite{BP} in their initial
study of the GP equation for repulsive interactions; compare also the
qualitative estimate in eq. (\ref{eq:Evar})],

\begin{equation}
\Psi(r)\propto \exp\left(-\frac{r^2}{2\beta^2d_0^2}\right)
\end{equation}
with $\beta$ as the variational parameter;  the
condensate radius is $\approx \beta d_0$ with  $\beta<1$  for attractive
interactions. The ground-state energy is readily evaluated to be
\cite{ALF1,Stoof,UL}
\begin{equation}
E_g(\beta)  =
\frac{1}{2}N\hbar\omega_0\left[\frac{3}{2}\left(\frac{1}{\beta^2}+\beta^2\right)
-\sqrt{\frac{2}{\pi}}\,\frac{N|a|}{d_0}\,\frac{1}{\beta^3}\right],
\label{eq:unstable1}
\end{equation}
and fig.\ 4 shows this function of $\beta$ for several values of the
interaction
parameter $N|a|/d_0$.  Evidently, $E_g(\beta)$ has no global minimum, for it
becomes arbitrarily negative as $\beta\to 0$.  Nevertheless, the energy
does have
a {\it local\/} minimum if $N< N_c$.  A straightforward analysis shows that the
minimum disappears at the critical value $N_c|a|/d_0\approx 0.671$, and
that the
critical condensate radius is reduced by a factor $\beta_c = 5^{-1/4} \approx
0.669$. For comparison, a numerical study of the GP equation \cite{Rup} yields
the value $N_c|a|/d_0\approx 0.575$, which differs from the variational
estimate
by $\approx 17\%$.

\subsection{Surface region of Thomas-Fermi condensate}

For a spherical trap, the slope of the TF condensate wave function

\begin{equation}
\Psi_{\rm TF}(r)\propto
\sqrt{1-\frac{r^2}{R_0^2}}\>\Theta\!\left(R_0-r\right)
\end{equation}
 becomes infinite  at the surface ($r\to R_0$), leading to logarithmic
singularity in the kinetic energy.  Consequently, it is essential to
introduce a
boundary layer near the surface, where the true wave function is modified from
its  approximate TF form
\cite{DPS}.  To study this surface region, it is convenient to focus on the
region $|r-R_0|\ll R_0$, writing
$r = R_0(1+X\delta)$, where $X$ is a new scaled dimensionless variable
measured
from
the TF surface,  and $\delta\ll 1$ is the
dimensionless thickness of the boundary  layer, measured in units of $R_0$.
Equivalently,  $X=( r-R_0)/(R_0\delta)$. With this change of variable,
the GP equation for a spherical trap

\begin{equation}
-\frac{\hbar^2}{2M}\,\frac{d^2}{dr^2}\Psi
-\frac{\hbar^2}{Mr}\,\frac{d}{dr}\Psi
+\left(\case{1}{2}M\omega_0^2r^2-\mu\right)\Psi+\frac{4\pi \hbar^2
a}{M}|\Psi|^2\Psi = 0
\end{equation}
takes the approximate form

\begin{equation}
-\frac{\hbar^2}{2MR_0^2}\,\frac{1}{\delta^2}\frac{d^2}{dX^2}\Psi
+M\omega_0^2R_0^2\delta\,X\Psi
+\frac{4\pi \hbar^2a}{M}|\Psi|^2\Psi=0.
\end{equation}

The balance between the first two terms gives the relation

\begin{equation}
2\delta^3 =\frac{\hbar^2}{M^2\omega_0^2R_0^4}= \frac{d_0^4}{R_0^4}=R^{-4}
\propto
N^{-4/5},
\end{equation}
verifying that the boundary layer is indeed thin and showing  that $\delta
\propto R^{-4/3}$ scales with the condensate number like $N^{-4/15}$.  In
contrast, the balance between the last two terms shows that the condensate
density
$|\Psi|^2$ near the surface is small (of order
$\delta$).  A detailed study \cite{DPS,LPS,FF} of the condensate wave
function in
the surface region involves a non-linear differential equation.  Its solution
 matches the TF behavior in the interior of the condensate  for $X\ll -1$ and
vanishes exponentially for $X\gg 1$, thus providing a uniform  extension
across
the boundary layer. The resulting  kinetic energy per particle

\begin{equation}
\frac{\langle T\rangle}{N} \approx
\frac{5}{2}\frac{\hbar^2}{MR_0^2}\,\ln\left(\frac{R_0}{1.3d_0}\right)
\end{equation}
exhibits the role of the boundary layer in cutting off the logarithmic
singularity. This contribution to the total energy is much smaller than the
other dominant  contributions $\langle V_{\rm tr}\rangle_0 /N \approx
\frac{3}{14} M\omega_0^2R_0^2$ and $ \langle V_{\rm int}\rangle
/N=\frac{1}{2} \langle V_H\rangle /N \approx \frac{1}{7} M\omega_0^2R_0^2$,
whose sum gives the previous expression $E_g/N\approx \frac{5}{14}
M\omega_0^2R_0^2$.

\subsection{Excited states of a trapped Bose condensate}

In the case of the uniform stationary  dilute Bose gas, the condensate wave
function is a spatial constant.  This simple structure  allows a complete
 explicit solution for the elementary excitations, which are the Bogoliubov
quasiparticles.  For a non-uniform trapped Bose condensate, however, the
situation
is more complicated because the determination of the condensate wave function
$\Psi({\bf r})$  already constitutes a significant problem.  Nevertheless,
given
$\Psi$,  a well-defined  procedure leads to the corresponding
equations for the small-amplitude normal modes of the condensate.  These
equations were first studied by Pitaevskii \cite{LPP} in the context of a
long straight vortex line in an unbounded condensate, and they have
subsequently
been generalized as operator equations in ref.\ \cite{AP}.

Recall that eq.\ (\ref{eq:fluct})  separated the field operator into a
(large) condensate part and a (small) fluctuation operator

\begin{equation}
\hat\psi = \Psi + \hat\phi, \label{eq:fluct1}
\end{equation}
where $\Psi=\langle\hat\psi\rangle $ and $\langle\hat\phi\rangle= 0$,  and
the angular brackets
denote an ensemble  average  in a restricted  grand-canonical ensemble that
breaks the symmetry with respect to the phase \cite{HM}.
The derivation of the GP equation for the
condensate wave function  retained only the dominant terms involving
$\Psi$, giving   the leading condensate approximation $K_0$ to the
thermodynamic
potential  in eq.\ (\ref{eq:K0}), thereby  ignoring the contribution of the
fluctuation operators.  To include the first corrections, it is necessary to
retain the fluctuation contributions $\hat\phi$ and $\hat\phi^\dagger$  to the
original field operators;  they   obey the approximate bosonic
commutation relations \cite{AP}

\begin{equation}
\left[\hat\phi({\bf r}),\hat\phi^\dagger({\bf r'})\right] \approx
\delta^{(3)}({\bf r-r'}), \quad\left[\hat\phi({\bf r}),\hat\phi({\bf
r'})\right]=\left[\hat\phi^\dagger({\bf r}),\hat\phi^\dagger({\bf r'})\right]=
0.\label{eq:phicomm}
\end{equation}
 Substitution into  the full modified hamiltonian $\hat K
=\hat H -\mu\hat N$ from eqs.\ (\ref{eq:hatH}) and (\ref{eq:hatK})  gives the
previous
$K_0$ plus a quadratic correction

\begin{equation}
\hat K' = \int d^3r \hat\phi^\dagger\left( T+ V_{\rm tr}-\mu
\right)\hat\phi+\frac{2\pi \hbar^2 a}{M}\int
d^3r\left[4|\Psi|^2\hat\phi^\dagger\hat\phi +
\Psi^2\hat\phi^\dagger\hat\phi^\dagger+
\left(\Psi^*\right)^2\hat\phi\hat\phi\right].\label{eq:K'}
\end{equation}
 The linear contribution to $\hat K$ vanishes identically because
 $\hat K$ is
stationary with respect to small changes around the GP solution $\Psi$;
this condition merely  requires that  $\Psi$  obey the appropriate
Euler-Lagrange
equation.

The mean number density in the ground state is given by $n({\bf r}) = \langle
\hat\psi^\dagger({\bf r})\hat\psi({\bf r})\rangle$.  The expansion in eq.\
(\ref{eq:fluct1}) shows that $n({\bf r})= |\Psi({\bf r})|^2 + \langle
\hat\phi^\dagger({\bf r})\hat\phi({\bf r})\rangle$, because, by assumption,
the
linear contributions have zero ground-state ensemble average. Consequently,
the
mean  total number of particles  is a sum of condensate and noncondensate
contributions
$N = N_0+N'$, where  $N_0 = \int d^3 r\,n_0({\bf r})$ and

	\begin{equation}
N'=\int d^3r\,\langle \hat\phi^\dagger({\bf
r})\hat\phi({\bf r})\rangle.\label{eq:nonconddens}
\end{equation}

The dynamical equations for the fluctuation operators follow immediately from
the commutators with $\hat K'$:

\begin{equation}
i\hbar\,\frac{\partial\hat\phi}{\partial t} = \left[\hat\phi,\hat
K'\right]\quad\hbox{and}\quad
i\hbar\,\frac{\partial\hat\phi^\dagger}{\partial t}
=
\left[\hat\phi^\dagger,\hat K'\right].
\end{equation}
A straightforward calculation yields the coupled operator equations
\begin{mathletters}
\label{eq:fieldeq}
\begin{equation}
i\hbar\,\frac{\partial\hat\phi}{\partial t} = \left(T+V_{\rm tr}
-\mu+\frac{8\pi\hbar^2 a}{M}\,|\Psi|^2\right)\hat\phi +
\frac{4\pi\hbar^2a}{M}\,\Psi^2\hat\phi^\dagger,
\end{equation}
\begin{equation}
-i\hbar\,\frac{\partial\hat\phi^\dagger}{\partial t} = \left(T+V_{\rm tr}
-\mu+\frac{8\pi\hbar^2 a}{M}\,|\Psi|^2\right)\hat\phi^\dagger +
\frac{4\pi\hbar^2a}{M}\left(\Psi^*\right)^2\hat\phi.
\end{equation}
\end{mathletters}
In analogy with Bogoliubov's treatment of a uniform dilute Bose gas, assume
that
the fluctuation field operator has an expansion in a set of self-consistent
normal modes

\begin{equation}
\hat\phi({\bf r},t) = {\sum_j}'\left[u_j({\bf r})\alpha_j(t) -v_j^*({\bf
r})\alpha_j^\dagger(t)\right],\label{eq:Bogexpan}
\end{equation}
where  the primed sum runs over all the {\it excited\/} states.  Here, the
quasiparticle operators have harmonic time dependence $\alpha_j(t) =
\alpha_j\exp(-iE_j/\hbar)$ and $\alpha_j^\dagger (t) =
\alpha_j^\dagger\exp(iE_j/\hbar)$, obeying bosonic commutation relations at
equal times.  Substitution into eqs. (\ref{eq:fieldeq}) leads to the coupled
``Bogoliubov equations'' \cite{deG}

\begin{mathletters}\label{eq:Bogeq}
\begin{equation}
\left(T+V_{\rm
tr}-\mu+\frac{8\pi\hbar^2a}{M}|\Psi|^2\right)u_j-\frac{4\pi\hbar^2a}{M}\Psi^2\,v
_j
= E_ju_j,
\end{equation}
\begin{equation}
\left(T+V_{\rm
tr}-\mu+\frac{8\pi\hbar^2a}{M}|\Psi|^2\right)v_j-\frac{4\pi\hbar^2a}{M}
\left(\Psi^*\right)^2u_j
=- E_jv_j
\end{equation}
\end{mathletters}
that determine the pair of quantum eigenamplitudes (in effect, they are wave
functions) $u_j({\bf r})$ and
$v_j({\bf r})$ along with the associated energy eigenvalue $E_j$ for the $j$th
eigenstate.

 This set of equations  can be rewritten in matrix form as ${\cal
M}{\cal U}_j = E_j\sigma_z{\cal U}_j$, where ${\cal M}$ is a
$2\times 2$  matrix differential operator,
${\cal U}_j$ is a two-component vector  with elements
$u_j$ and
$v_j$, and $\sigma_z$ (the usual Pauli matrix)  acts as a metric tensor in
 the scalar product $\int d^3 r\,{\cal U}_k^\dagger \sigma_z{\cal U}_j$.
 The diagonal elements ${\cal
M}_{11} = {\cal M}_{22} = T +V_{\rm tr} -\mu + 2V_H$ of the  matrix $\cal
M$ consist
of  hermitian operators (in the usual quantum-mechanical sense),   and the
off-diagonal elements
${\cal M}_{21}^* = {\cal M}_{12} = -(4\pi \hbar^2 a/M)\Psi^2$ are
complex-conjugate
functions, so that
${\cal M}^\dagger = {\cal M}$. With  suitable boundary conditions, an
integration by
parts shows that
$\int d^3r\,{\cal U}_k^\dagger{\cal M}{\cal U}_j =\int d^3r\,({\cal M} {\cal
U}_k)^\dagger\,{\cal U}_j$.  As in the usual Sturm-Liouville theory, it
follows~\cite{AP} that the eigenvalues
$E_j$ are real and that eigenfunctions for distinct positive energies
$E_k\neq E_j$
are ``orthogonal'' with
$\int d^3 r\,(u_k^*u_j-v_k^*v_j) = 0$. For $E_j>0$ [see eq.~(\ref{eq:hatK'qp})
below], the proper choice of normalization
\begin{equation}\int
d^3r\left(|u_j|^2-|v_j|^2\right)=1\label{eq:normj}\end{equation}
 is analogous to the condition
$u_k^2-v_k^2=1$ in eq.~(\ref{eq:boseqp}) for the uniform system;  it
ensures that the quasiparticle operators obey the usual bosonic commutation
relations.  Although the Bogoliubov equations describe the excited states, the
condensate wave function $\Psi$ itself acts as a ``zero-mode''
solution~\cite{AP}; its conjugate operator  completes the set of states and is
important in the quantum dynamics of the phase of the
condensate~\cite{Lewen,Vill}.

In addition,
for every solution $u_j$ and $v_j$ with energy $E_j$, there exists another
solution $v_j^*$ and $u_j^*$ with energy $-E_j$ and normalization
$\int
d^3r\left(|v_j|^2-|u_j|^2\right) = -1$;
this second set corresponds to interchanging
$\alpha_j^\dagger$
 and
$\alpha_j$, in effect exchanging quasiparticles and quasiholes. Although the
Bogoliubov equations describe bosonic amplitudes, their structure is very
reminiscent of the Dirac equation, where the coupled amplitudes again include
particles and holes that have positive and negative energies\cite{BD} (the
qualitative differences appear only when each  is reinterpreted as a
quantum field
theory with distinct  bosonic or fermionic commutation relations).

Equation (\ref{eq:nonconddens}) shows that the total non-condensate number $N'$
is the ground-state expectation value
$\int d^3r\langle\hat\phi^\dagger\hat\phi\rangle$.  Since the quasiparticle
operators represent non-interacting bosons, use of the  normal-mode expansion
(\ref{eq:Bogexpan}) yields $N'={\sum_j}\,'\,N_j'$, where $N_j' =
\int d^3r|v_j|^2$ is the ground-state occupation of the $j$th eigenstate.  In
addition, direct substitution of eq.\ (\ref{eq:Bogexpan})  into  eq.\
(\ref{eq:K'}) for the quasiparticle modified hamiltonian $\hat K'$ and use of
eq.~(\ref{eq:Bogeq})  gives the simple and intuitive expression

\begin{equation}
\hat K' = -{\sum_j}'E_j \,N_j' +
{\sum_j}'E_j\alpha_j^\dagger\alpha_j,\label{eq:hatK'qp}
\end{equation}
which is a direct generalization of the Bogoliubov quasiparticle
hamiltonian (\ref{eq:hqp}) for a uniform system.  The first term is the
ground-state thermodynamic potential
$K_g =
\langle \hat K'\rangle = -{\sum_j}'E_j \,N_j'$, and the second shows that
the excited states consist of various numbers of non-interacting bosons with
energies $E_j$ and  occupation-number operators $\alpha_j^\dagger\alpha_j$.
Since
these operators have non-negative integers as their eigenvalues, stability
of the
ground state requires that $E_j\ge 0$ for all eigenstates $j$ that obey the
proper
normalization in eq.\ (\ref{eq:normj}).

At low but finite temperature in the grand canonical ensemble, the appropriate
(unnormalized) weight factor for the excited states is $\exp(-\beta\hat K')$;
the average of any operator $\hat{\cal O}$ is simply the
trace $\langle
\hat{\cal O}\rangle = {\rm Tr}[\hat{\cal O}\exp(-\beta\hat
K')]/{\rm Tr}[\exp(-\beta\hat K')]$ with the same  weight factor
(normalized by the
grand partition function
${\cal Z} = {\rm Tr}[\exp(-\beta\hat K')]$). For example, the
non-condensate density
is

\begin{equation}
n'({\bf r}) = {\sum_j}'\left\{f(E_j)|u_j({\bf r})|^2 + \left[1+f(E_j)\right]
|v_j({\bf r})|^2\right\},\label{eq:noncond}
\end{equation}
where $f(E_j) = \left[\exp(\beta E_j)-1\right]^{-1}$ is the Bose-Einstein
distribution function for the $j$th excited state.  At zero temperature, the
spatial integral gives the preceding result for the total
non-condensate number.

It is instructive to specialize  the present  Bogoliubov quasiparticles to a
uniform system.  In sec.~\ref{sec:I}, the condensate number $N_0$ was
eliminated in
favor of the total number $N$, leading to a number-conserving description
(with no
chemical potential $\mu$ in the hamiltonian);  in contrast, the  theory here
relies on the grand canonical ensemble at fixed temperature $T$ and chemical
potential $\mu$, leading to  a modified hamiltonian
$\hat K =
\hat H-\mu
\hat N$.  In this latter formalism, the ground-state thermodynamic
potential per
unit volume  for the uniform condensate is $K_0/V= \frac{1}{2}gn_0^2-\mu n_0$,
where
$n_0=N_0/V$ is the condensate density (the kinetic energy vanishes for a
uniform condensate).  In equilibrium at fixed chemical potential and
temperature (here, $T=0$), the system adjusts all free parameters to
minimize the
thermodynamic potential.  In the present case where the non-condensate is
neglected,
only the condensate density
$n_0$ remains undetermined, and the condition $(\partial K_0/\partial
n_0)_\mu=0$
immediately gives the expected result $\mu = gn_0$, which is the Hartree energy
for adding one particle.  The corresponding solutions of the
Bogoliubov equations are plane waves $u_{\bf k}({\bf r}) = u_k\exp(i{\bf k\cdot
r})$ and $v_{\bf k}({\bf r}) = v_k\exp(i{\bf k\cdot r})$.  Use of the
preceding
expression for $\mu$ in the Bogoliubov equations (\ref{eq:Bogeq}) readily
reproduces all of previous Bogoliubov  results, for example,
$E_k  = \sqrt{2gn_0\epsilon_k^0+(\epsilon_k^0)^2}$.

More generally, the Bogoliubov equations  provide a flexible formalism to
study the effect of a non-uniform condensate wave function.  A particularly
interesting  example is a uniform system in which the  condensate moves with
velocity ${\bf v} = \hbar{\bf q}/M$ (discussed briefly at the end of
 sec.~\ref{sec:I}).  The condensate wave function has the form $\Psi \propto
e^{i{\bf q\cdot r}}$, and the GP equation then shows that the chemical
potential
becomes $\mu = \epsilon_q^0+ gn_0 = \frac{1}{2}Mv^2 + gn_0$. The structure
of the
two Bogoliubov eqs.\ (\ref{eq:Bogeq}) implies that the pair of amplitudes
has the
form

\begin{equation}
\pmatrix{u({\bf r}) \cr v({\bf r})\cr} = \pmatrix{u_k\,e^{i\bf q\cdot
r}\,e^{i\bf k\cdot r}\cr v_k\,e^{-i\bf q\cdot
r}\,e^{i\bf k\cdot r}\cr},
\end{equation}
and the resulting eigenvalue is given by $E_{\bf k} = \hbar{\bf k\cdot v} \pm
E_k^0$, where $E_k^0$ is the Bogoliubov energy for the stationary condensate.
Substitution back into the Bogoliubov equations shows that the $\pm$ sign here
corresponds to the  normalization  $u_k^2-v_k^2 = \pm 1$, so that the bosonic
commutation relations require the choice $E_{\bf k} = \hbar{\bf k\cdot v}
+E_k^0$ (as also follows by requiring continuity with  the small-$v$
limit).  For
a detailed discussion of these solutions, along with  the associated mass
current
and superfluid density (as well as the generalization to  a mixture of  two
distinct  interacting species), see, for example, refs.\
\cite{AP,CF}.

  Consider an
anisotropic harmonic trap with arbitrary frequencies $\omega_\alpha$ ($\alpha =
x, y, z$),  and let $\Psi$ denote {\it any\/} exact solution of the GP
equation.  In general, the Bogoliubov equations are difficult  to solve, but
they do
 have  {\it three exact solutions\/}  (the ``dipole'' modes)   in which this
general  condensate oscillates rigidly in each orthogonal direction with
frequency
$\omega_\alpha$.  To construct these explicit solutions \cite{FR}, define the
familiar raising and lowering operators
\begin{mathletters}\label{eq:dipole}
\begin{equation}
a_\alpha^\dagger = \frac{1}{\sqrt
2}\left(\frac{x_\alpha}{d_\alpha}-id_\alpha\frac{ p_\alpha}{\hbar} \right) =
\frac{1}{\sqrt 2}\left(\frac{x_\alpha}{d_\alpha}-d_\alpha\frac{
\partial}{\partial x_\alpha}
\right),
\end{equation}
\begin{equation}
a_\alpha = \frac{1}{\sqrt
2}\left(\frac{x_\alpha}{d_\alpha}+id_\alpha\frac{ p_\alpha}{\hbar} \right) =
\frac{1}{\sqrt 2}\left(\frac{x_\alpha}{d_\alpha}+d_\alpha\frac{
\partial}{\partial x_\alpha}
\right).
\end{equation}
\end{mathletters}
It is not difficult to prove that the corresponding dipole solutions have the
 explicit form
\begin{equation}\label{eq:dipolesoln}
\pmatrix{u_\alpha({\bf r})\cr v_\alpha({\bf r})}= \pmatrix{a_\alpha^\dagger
\Psi({\bf r})\cr a_\alpha \Psi^*({\bf r})}.
\end{equation}
For the non-interacting Bose gas, the solution of the GP equation is just
the Gaussian ground state $\Psi_g= \sqrt{N}\psi_g$ of the trap given in eq.\
(\ref{eq:psig0});  the associated dipole excitations are then characterized by
$u_\alpha
\propto a_\alpha^\dagger\psi_g$ (namely, the first excited states of the
oscillator) and
$v_\alpha=0$ (because
$a_\alpha$ annihilates the ground state $\psi_g$).

It is worth noting that the Bogoliubov equations (\ref{eq:Bogeq})
have an equivalent derivation \cite{EDCB,Rup1,ME} based directly on the
linear response of the time-dependent GP equation (although this approach
leads more directly to  the final equations, it relies only on
wave functions and does not treat the field operators explicitly).  Add a weak
sinusoidal perturbation  to the GP equation
\begin{equation}
i\hbar\,\frac{\partial \Psi}{\partial t} = \left(T+ V_{\rm tr} +
g|\Psi|^2\right)\Psi + \left(f_+e^{-i\omega t} +f_-e^{i\omega t}\right) \Psi,
\end{equation}
and seek a solution in the form

\begin{equation}
\Psi({\bf r},t) = e^{-i\mu t/\hbar}\left[\Psi({\bf r}) +u({\bf r}) e^{-i\omega
t} + v({\bf r})e^{i\omega t}\right].
\end{equation}
Linearizing in the small amplitudes $u$ and $v$ yields an inhomogeneous form of
the Bogoliubov equations  with driving terms proportional to
$f_\pm({\bf r})\Psi$.  A normal-mode expansion in terms of eigenfunctions
$u_j$ and $v_j$ gives precisely the same coupled equations as in
eq.\ (\ref{eq:Bogeq}), identifying the frequencies $\omega_j = E_j/\hbar$ as
resonances in the linear response.

\section{Hydrodynamic Description of a Dilute Trapped Bose Gas}\label{sec:III}

Although the Bogoliubov equations, in principle, characterize all the excited
states of the condensate, they are difficult to solve and their physical
meaning
is not always transparent.  Thus, it is important to consider alternative
formulations, and the present section concentrates on the hydrodynamic
approach that
emphasizes the familiar concepts of density and current fluctuations.

\subsection{Uniform dilute Bose gas}

It is instructive to review the theory of density fluctuations in a uniform
Bose
gas, where the field operator has the familiar form

\begin{equation}
\hat\psi({\bf r} ) = \sum_{\bf k} \frac{1}{\sqrt V}\,e^{i{\bf k\cdot
r}}\,a_{\bf
k}.
\end{equation}
The corresponding density operator  $\hat n({\bf r})= \hat\psi^\dagger({\bf
r})\hat\psi({\bf r})$ has a Fourier expansion

\begin{equation}
\hat n({\bf r}) =\sum_{\bf q} e^{i{\bf q\cdot r}}\,\hat n_{\bf q},
\end{equation}
where

\begin{equation}
\hat n_{\bf q}=\sum_{\bf k} a_{\bf k-q}^\dagger a_{\bf k}\label{eq:densfluct}
\end{equation}
is an operator that removes one quantum of density fluctuation with wave number
$\bf q$.   It is easy to see that the associated creation operator is given by
$\hat n^\dagger_{\bf q} = \hat n_{-\bf q}$. For an ideal  Fermi gas at $T=0$,
the   operator $\hat n_{\bf q}$ in
eq.\ (\ref{eq:densfluct}) removes a particle with wave number $\bf k$
(which must
lie inside the Fermi sea with $|{\bf k}|\le k_F$) and simultaneously creates a
particle with wave number
$\bf k-q$ (which  must lie outside the Fermi sea with $|{\bf k-q}|\ge k_F$);
 the final operator $\hat n_{\bf q}$ is a sum over all possible $\bf k$
consistent with  the above restrictions.  Thus the density-fluctuation
operator $\hat n_{\bf q}$ is really a particle-hole operator that conserves
particle
number,  and a similar characterization applies to both a general
interacting Fermi
gas and
 a non-condensed Bose gas.  Typically,  density fluctuations in a Fermi gas
have
a well-defined dispersion relation $\omega_{\bf k}$, and they are often called
``collective modes'' to distinguish them from the ``quasiparticles'' that are
associated with single-particle excitations (those resulting from the
removal or
addition of one particle).

The situation is qualitatively different for a condensed Bose gas, because the
condensate itself plays a special role.  For simplicity, consider a
 dilute uniform Bose gas with  stationary condensate at low temperature.  In
this case, the  two terms arising from the Bogoliubov prescription dominate
the
general sum  in eq.\ (\ref{eq:densfluct}), leading to the simple
density-fluctuation creation operator

\begin{eqnarray}
\hat n_{\bf q}^\dagger &\approx &\sqrt{N_0}\left(a_{\bf q}^\dagger+a_{-\bf
q}\right) \nonumber\\
&\approx &\sqrt{N_0}\left(u_q-v_q\right)\left(\alpha_{\bf
q}^\dagger+\alpha_{-\bf q}\right)\label{eq:Bosedens}
\end{eqnarray}
that is a coherent linear
combination of a quasiparticle and a quasihole.
When this operator acts on the Bogoliubov ground state $|\Phi\rangle$, it gives

\begin{equation}\hat n_{\bf q}^\dagger |\Phi\rangle =
\sqrt{N_0}\left(u_q-v_q\right)\alpha_{\bf q}^\dagger|\Phi\rangle
\end{equation}
 because, by definition,
$\alpha_{-\bf q}|\Phi\rangle $ vanishes.  Apart from an overall factor,
the action of  the  density-fluctuation operator $\hat n_{\bf q}^\dagger$
on the
Bogoliubov  ground state is the same as that of the quasiparticle operator
$\alpha_{\bf q}^\dagger$.  Consequently, the excitation spectrum associated
with a
density fluctuation in a dilute Bose gas is necessarily identical with that
for a
quasiparticle;  as discussed above, this situation is completely different from
that for a dilute Fermi gas.

  The current-density
operator has a similar expansion

\begin{equation}
\hat {\bf j}({\bf r}) = \frac{\hbar}{2Mi} \left[\hat\psi^\dagger ({\bf r})
{\nabla}\hat\psi({\bf r}) -\left({\nabla}\hat\psi^\dagger({\bf
r})\right)\hat\psi({\bf r})\right]
= \sum_{\bf q}\,\hat {\bf j}_{\bf q}\, e^{i\bf q\cdot r},
\end{equation}
with
\begin{equation}
\hat {\bf j}_{\bf q}=\frac{\hbar}{M}\sum_{\bf k} \left({\bf k} -
\case{1}{2} {\bf q}\right) a_{\bf k-q}^\dagger a_{\bf k}.
\end{equation}
In the Bogoliubov approximation, two terms again predominate with

\begin{eqnarray}
\hat {\bf j}_{\bf q}&\approx &\frac{\hbar \,{\bf
q}}{2M}\sqrt{N_0}\left(a_{\bf q}
-a_{-\bf q}^\dagger\right)\nonumber\\
&\approx &\frac{\hbar \,{\bf
q}}{2M}\sqrt{N_0}\left(u_q+v_q\right)\left(\alpha_{\bf
q} -\alpha_{-\bf q} ^\dagger\right);
\end{eqnarray}
as expected for a dilute gas, this operator is manifestly longitudinal (namely,
along
$\bf q$).

For time-dependent (Heisenberg) operators, conservation of particle number
requires that

\begin{equation}
\frac{\partial \hat n_{\bf q}(t)}{\partial t} +i{\bf q }\cdot \hat {\bf j}_{\bf
q}(t) = 0.\label{eq:consN}
\end{equation}
It is clear from eq.~(\ref{eq:Bosedens}) that the density-fluctuation
operator $\hat
n_{\bf q}(t)$  in a dilute Bose gas oscillates at the Bogoliubov frequency
$\omega_q
=E_q/\hbar$.

In a bulk uniform system, the ensemble average of $\hat n_{\bf q}^\dagger\hat
n_{\bf q}$ is called the ``static structure function,'' with the
precise definition
$S(q) = N^{-1}\langle \hat n_{\bf q}^\dagger\hat
n_{\bf q}\rangle$. For a dilute Bose gas at low temperature,  this
quantity is readily evaluated with eq.\ (\ref{eq:Bosedens}) to give

\begin{equation}
S(q)\approx \left(u_q-v_q\right)^2 \langle \alpha_{\bf q}\alpha_{\bf q}^\dagger
+ \alpha_{\bf q}^\dagger\alpha_{\bf q}\rangle =
\frac{\epsilon_q^0}{E_q}\coth\left(\frac{E_q}{2k_BT}\right),\quad\hbox{in
the Bogoliubov approximation,}\label{eq:Sq}
\end{equation}
since   off-diagonal  ensemble averages like $\langle\alpha_{\bf
q}\alpha_{-\bf q}\rangle $ vanish. At
$T= 0$, the thermal factor is simply 1, giving the zero-temperature limits

\begin{equation}
S(q) =\frac{\epsilon_q^0}{E_q}\approx \cases {\hbar q/2Ms,& for $q\xi \ll
1,$\cr
\noalign{\vspace{.1cm}}1,& for $q\xi \gg 1$,\cr}
\end{equation}
where $s$ is the speed of sound.  More generally, the Bogoliubov approximation
at $T=0$ reproduces the form of Feynman's variational approximation for the
zero-temperature density excitation spectrum of superfluid
${}^4$He
\cite{RPF}

\begin{equation}
E_q \approx \frac{\hbar^2q^2}{2MS(q)},
\end{equation}
with $S(q)$ the {\it measured\/} structure function.  For low but
finite temperature, in contrast, the static structure function for a dilute
Bose gas has the limiting long-wavelength form

\begin{equation}
S(q) \approx \frac{k_BT}{Ms^2} \quad\hbox{as $q\to 0$}.
\end{equation}

\subsection{Sum rules for a uniform dilute Bose gas}

For both Fermi and Bose systems, sum rules provide a powerful and exact
approach
to the description of collective modes \cite{PN,NP}.  Consider the
response of a uniform system to a weak scalar field $\Phi({\bf r},t)$ that
couples to the density with a perturbation hamiltonian (here expressed in
Fourier components)

\begin{equation}
\hat H_{\rm int}(t) = \sum_{\bf q}\int_{-\infty}^\infty \frac{d\omega}{2\pi}\,
\hat n_{\bf q}^\dagger\, \Phi({\bf q},\omega) \, e^{-i\omega t}
\end{equation}
The general theory of linear response shows that the induced change in the
density occurs at the same wave vector $\bf q$ and frequency $\omega$, with the
explicit form

\begin{equation}
\delta\langle\hat n({\bf q},\omega)\rangle = -\chi({\bf q},\omega)\,\Phi({\bf
q},\omega).
\end{equation}
Here, the coefficient is known as the ``dynamic susceptibility,''  and the
explicit minus sign is added here for later convenience;  $\chi$  is really the
Fourier transform of a density-density correlation function.  Linear-response
theory
\cite{PN} yields the explicit zero-temperature form

\begin{equation}
\chi({\bf q},\omega) = -\frac{1}{N}\sum_{j\neq 0} |\langle j\,|\hat n_{\bf
q}^\dagger\,|0\,\rangle |^2\bigg[\frac{1}{\omega-\omega_{j0}+i\eta} -
\frac{1}{\omega+\omega_{j0}+i\eta}\bigg],\label{eq:chi}
\end{equation}
where the limit $\eta\to 0^+$ is taken at the end of the analysis.  Here,
$|j\,\rangle$ is an exact eigenstate of the interacting but unperturbed
many-particle system, and $\omega_{j0} =(E_j-E_0)/\hbar$ is the exact
excitation
frequency of this state relative to the ground state.

It is clear by inspection that $\chi({\bf q},\omega)$ has poles in the complex
$\omega$ plane at the points $\omega = \pm \omega_{j0}-i\eta$.  Use of the
formal
identity

\begin{equation}
\lim_{\eta\to 0^+} \frac{1}{x-a+i\eta} = {\cal P}\frac{1}{x-a} -i\pi
\delta(x-a),
\end{equation}
where ${\cal P}$ denotes the Cauchy principal value of an integral, yields the
explicit expression for the imaginary part

\begin{equation}
{\rm Im} \chi({\bf q},\omega)= \frac{\pi}{N}\sum_{j\neq 0} |\langle
j\,|\hat n_{\bf
q}^\dagger\,|0\,\rangle
|^2\left[\delta(\omega-\omega_{j0})-\delta(\omega+\omega_{j0})\right].
\label{eq:Imchi}
\end{equation}
Note that ${\rm Im} \chi({\bf q},\omega) $ is an odd function of $\omega$ and
therefore vanishes at $\omega=0$.  Furthermore, a combination of eqs.\
(\ref{eq:chi}) and (\ref{eq:Imchi}) yields the spectral representation for the
susceptibility (the response function) at wave vector $\bf q$ and frequency
$\omega$

\begin{equation}
\chi({\bf q},\omega) = \int_{-\infty}^\infty \frac{d\omega'}{\pi}\, \frac{{\rm
Im}\chi({\bf q},\omega')}{\omega'-\omega-i\eta}.
\end{equation}
In particular, the limit $\omega\to 0$ is simply the static susceptibility

\begin{equation}
\chi_{\bf q} \equiv \chi({\bf q},0) = \int_{-\infty}^\infty
\frac{d\omega'}{\pi}\,\frac{{\rm Im}\chi({\bf
q},\omega')}{\omega'}\label{eq:sum-1}
\end{equation}
 that characterizes the response to a static plane-wave distortion with wave
vector
$\bf q$ [the convergence factor $i\eta$ is now unnecessary because
${\rm Im} \chi({\bf q},0)=0 $].

This result can be generalized by defining the $i$th moment of ${\rm Im}\chi$

\begin{equation}
{\cal M}_i \equiv \int_{-\infty}^{\infty} \frac{d\omega}{\pi} \,\omega^i\,{\rm
Im}\chi({\bf q},\omega) = \frac{1}{N}\sum_{j\neq 0} |\langle j\,|\hat
n_{\bf q}^\dagger\,|0\,\rangle |^2
\left(\omega_{j0}\right)^i\left[1-(-1)^i\right];
\end{equation}
>From this perspective, the preceding eq.\ (\ref{eq:sum-1}) is the $i= -1$
moment

\begin{equation}
{\cal M}_{-1} = \frac{2}{\pi}\int_0^\infty
\frac{d\omega}{\omega}\,{\rm Im}\chi({\bf q},\omega) =\chi_{\bf q}.
\end{equation}
In the limit ${ \bf q\to 0}$, this quantity is related to the bulk
thermodynamic
compressibility because $\chi_{ \bf q} $ is the response to a static plane wave
$\propto e^{i\bf q\cdot r}$;   a detailed analysis  \cite{PN} shows that

\begin{equation}
\lim_{\bf q\to 0}\, \chi_{\bf q} = \frac{\hbar}{Ms^2},\label{eq:compsum}
\end{equation}
giving the limit of ${\cal M}_{-1}$ as $q\to 0$ (this result is known as the
``compressibility sum rule'').

It is clear that the even moments vanish.  The next odd moment

\begin{eqnarray}
{\cal M}_1 &=& \frac{2}{\pi}\int_0^\infty d\omega\,\omega\,{\rm Im}\chi({\bf
q},\omega)\nonumber\\
&=&\frac{2}{N}\sum_{j\neq 0} |\langle j\,|\hat
n_{\bf q}^\dagger\,|0\,\rangle |^2 \,\omega_{j0}
\end{eqnarray}
can be related  \cite{PN} to the ground-state average of a double commutator
$\langle 0\,|\left[\hat n_{\bf q},[\hat H,\hat n^\dagger_{\bf
q}]\right]|0\,\rangle$.  The inner commutator follows from the dynamical
eq.\ (\ref{eq:consN}) that expresses the conservation of particles since $[\hat
H,\hat n_{\bf q}^\dagger] = i\hbar\,\partial \hat n_{\bf
q}^\dagger/\partial t =
-{\bf q\cdot \hat j_q^\dagger}$, leading to the final result

\begin{equation}
{\cal M}_1 = \frac{\hbar q^2}{M},\label{eq:sum1}
\end{equation}
which is known as the ``$f$-sum rule'' \cite{PN}.

It is also valuable to consider the third moment \cite{RDP,SS1,AJL}

\begin{eqnarray}
{\cal M}_3 &=& \frac{2}{\pi}\int_0^\infty d\omega\,\omega^3\,{\rm Im}\chi({\bf
q},\omega)\nonumber\\
&=&\frac{2}{N}\sum_{j\neq 0} |\langle j\,|\hat
n_{\bf q}^\dagger\,|0\,\rangle |^2 \,\left(\omega_{j0}\right)^3,
\end{eqnarray}
which can be related to the more complicated matrix element
 $$\langle 0\,|\left[[\hat n_{\bf q},\hat H],\left[\hat H,[\hat H,\hat
n^\dagger_{\bf q}]\right]\right]|0\,\rangle\propto \langle 0\,|\left[{\bf
q\cdot
\hat j_q},[\hat H,{\bf q\cdot j_q}^\dagger]\right]|0\,\rangle.$$  For a dilute
uniform Bose gas, a detailed analysis \cite{AJL} that neglects a small
kinetic-energy part of order $\sqrt{na^3}$ yields

\begin{equation}
{\cal M}_3 = \frac{q^2}{\hbar M}\epsilon_q^0\left(\epsilon_q^0+2ng\right) =
\frac{q^2}{\hbar M}\,E_q^2,\label{eq:sum3}
\end{equation}
where $E_q$ is seen to be  the Bogoliubov energy spectrum in eq.\
(\ref{eq:Esubk}).

These various sum rules can serve to characterize  the detailed structure of
${\rm Im}\chi({\bf q},\omega)$ \cite{RDP,SS1,AJL} for a  dilute uniform
Bose gas.
Assume that an excitation with wave vector $\bf q$ constitutes a {\it single\/}
long-lived collective mode with frequency
$\omega_{ q}$. In this case, the spectral weight has only one
$\delta$-function,  with ${\rm Im}\chi({\bf q},\omega) = A_{\bf q} \delta
(\omega-\omega_q)$ for positive $\omega$. Direct substitution shows that

\begin{mathletters}
\begin{equation}
{\cal M}_1  =  \frac{\hbar q^2}{M} = \frac{2}{\pi} A_{\bf q}\, \omega_{\bf q},
\end{equation}
\begin{equation}
{\cal M}_3  =  \frac{q^2}{\hbar M} E_q^2= \frac{2}{\pi} A_{\bf q} \,\omega_{\bf
q}^3.
\end{equation}
\end{mathletters}
The ratio of these results shows that $\omega_{\bf q}$ is precisely the
Bogoliubov frequency $E_q/\hbar$, with the weight factor  $A_{\bf q} =
\pi \epsilon_q^0 /E_q$.  In addition, substitution of the single-mode
approximation into the compressibility sum rule (\ref{eq:sum-1}) gives the
static susceptibility

\begin{equation}
\chi_{\bf q} = \frac{\hbar q^2}{M\omega_{\bf q}^2} \quad\to
\frac{\hbar}{Ms^2}\quad\hbox{as $q\to 0$},
\end{equation}
in agreement with  the general result in eq.\ (\ref{eq:compsum}).

Finite temperature requires an ensemble average, weighted with the Gibbs factor
$e^{-\beta E_i}$ for the general initial state $|i\rangle$, along with  the
partition function $Z\equiv \sum_j e^{-\beta E_j}$.  The ``dynamical structure
factor''
$S({\bf q},\omega)$ is defined as  \cite{PN}

\begin{equation}
S({\bf q},\omega) = \frac{1}{NZ}\sum_{fi} |\langle f\,|\hat n_{\bf
q}^\dagger|i\,\rangle|^2\,\delta\left(\omega- \omega_{fi}\right),
\end{equation}
where $\omega_{fi} = (E_f-E_i)/\hbar$ can now be both positive and negative.
For liquid ${}^4$He, neutron scattering can measure $S({\bf q},\omega)$
directly.  In the Bogoliubov approximation for a dilute uniform Bose gas, it is
not difficult to show that

\begin{equation}
S(q,\omega) \approx
\frac{\epsilon_q^0}{E_q}\left\{\left[1+f(E_q)\right]\delta\left(\omega-\frac{E_q
}
{\hbar}
\right)
+
f(E_q)\,\delta\left(\omega+\frac{E_q}{\hbar}\right)\right\},\label{eq:Sqomega}
\end{equation}
where $f(E_q)$ is the Bose-Einstein distribution function.  The first term
(representing emission of a quasiparticle with energy $E_q$) is known as a
``Stokes'' process;  it persists even at zero temperature.  The second term
(representing the absorption of a quasiparticle with energy $E_q$) is an
``anti-Stokes process'' that requires the presence of a previously excited
quasiparticle and thus vanishes as $T\to 0$.  On general grounds \cite{PN}, the
dynamical structure factor satisfies several important relations; for example,
its zeroth moment is the static structure function  $S(q)$, which here gives

\begin{equation}
S(q) = \int_{-\infty}^\infty d\omega\,S(q,\omega) =
\frac{\epsilon_q^0}{E_q}\coth\left(\frac{E_q}{2k_BT}\right),
\end{equation}
in agreement with the previous  eq.\ (\ref{eq:Sq});  furthermore, the first
moment is just the $f$-sum rule at finite temperature

\begin{equation}
\int_{-\infty}^\infty d\omega\,\omega\,S(q,\omega) = \frac{\hbar
q^2}{2M};
\end{equation}
finally, $S(q,\omega)$ obeys a detailed-balance condition $S(q,-\omega) =
e^{-\beta
\hbar\omega}S(q,\omega)$.   These relations  provide  non-trivial checks on the
Bogoliubov approximation to $S(q,\omega)$ in eq.\ (\ref{eq:Sqomega}).

\subsection{Hydrodynamic description of non-uniform dilute Bose gas}

The condensate wave function $\Psi$   can be rewritten as

\begin{equation}
\Psi = |\Psi|e^{iS},\label{eq:modphase}
\end{equation}
where the absolute value is related to the condensate density $|\Psi({\bf
r})| =
\sqrt{n_0({\bf r})}$.  Furthermore, the condensate particle current

\begin{equation}
{\bf j}_0 = \frac{\hbar}{2Mi}\left[\Psi^*\nabla\Psi -
\left(\nabla\Psi^*\right )\Psi\right]= \frac{\hbar}{M}\,n_0\nabla S
\end{equation}
 identifies the
condensate velocity as

\begin{equation}
{\bf v}_0({\bf r}) = \frac{\hbar}{M}\nabla S({\bf r}).\label{eq:condvel}
\end{equation}
The local superfluid velocity ${\bf v}_0$ is proportional to the  gradient
of the phase of the condensate wave function, implying that $\nabla\times {\bf
v}_0 = 0$.  Hence the condensate flow is longitudinal (or irrotational), as
proposed by Landau \cite{LDL} for the superfluid velocity ${\bf v}_s$ in
${}^4$He.

Substitution of eq.\ (\ref{eq:modphase}) into the   GP equation
(\ref{eq:GP}) provides an intuitive hydrodynamic picture of the condensate. The
real part gives a generalized Bernoulli's equation \cite{EPG,LPP}

\begin{equation}
\mu = V_{\rm tr} +\case{1}{2}Mv_0^2 + \frac{4\pi\hbar^2 a}{M} n_0
-\frac{\hbar^2}{2M\sqrt{n_0}}\nabla^2\sqrt{n_0},\label{eq:Bern}
\end{equation}
where the last term is a quantum pressure associated with spatial variations in
the condensate amplitude $|\Psi|$ (and thus the condensate density);
the imaginary part is the static  continuity equation for the condensate

\begin{equation}
\nabla\cdot(n_0{\bf v}_0) = \nabla\cdot{\bf j}_0 = 0.
\label{eq:cont}\end{equation}

What determines  the dynamics of small time-dependent
perturbations about the static solution $\Psi({\bf r})$?  Two equivalent
approaches yield the same final results.

1.  Write the time-dependent condensate wave function as $\Psi({\bf r},t) =
\sqrt{n({\bf r},t)}\exp\left[iS({\bf r},t)\right]$ and substitute into the
time-dependent GP equation (\ref{eq:TDGP}).  The real and imaginary parts yield
time-dependent generalizations of eqs.\ (\ref{eq:Bern}) and (\ref{eq:cont}),
which can then be linearized around the static condensate contributions
\cite{SS}.

2.  Retain the operators $\hat\phi$ and $\hat\phi^\dagger$ in the expansion of
the field operator $\hat\psi = \Psi +\hat\phi$ from eq.\ (\ref{eq:fluct1}).
This
latter method  allows a transparent treatment of the operator aspects of the
hydrodynamic variables
\cite{ALF} and will be used here.

In a dilute Bose gas at low temperature, the total number-density operator

\begin{equation}
\hat n=\hat\psi^\dagger\hat\psi \approx |\Psi|^2 + \Psi^*\hat\phi +
\Psi\hat\phi^\dagger
\end{equation}
naturally separates into a large condensate part $n_0 = |\Psi|^2$ and a small
non-condensate part

\begin{equation}
\hat n' \approx \Psi^*\hat\phi +\Psi\hat\phi^\dagger
\label{eq:densfl}
\end{equation}
that characterizes the local time-dependent particle-density fluctuations.
Similarly, the local time-dependent fluctuations in the current-density
operator have the form

\begin{equation}
\hat{\bf j}' = \hat n'{\bf v}_0+ n_0\nabla \hat\Phi',\label{eq:currfl}
\end{equation}
where
\begin{equation}
\hat\Phi' = \frac{\hbar}{2Mi|\Psi|^2}\left(\Psi^*\hat\phi -
\Psi\hat\phi^\dagger\right)\label{eq:velpot}
\end{equation}
is the perturbation in the
velocity potential operator (this quantity is  effectively the perturbation in
the phase operator). Since both $\hat n'$ and $\hat\Phi'$ are  linear functions
of $\hat\phi$ and $\hat\phi^\dagger$, it follows that   $\langle \hat
n'\rangle $
and
$\langle \hat\Phi'\rangle $ vanish in the ensemble discussed below
eq.~(\ref{eq:fluct1}), although  correlation functions like $\langle \hat
n'({\bf
r},t)\hat n'({\bf r'},t')\rangle$ are finite.  Note that $\hat n'$ is
distinct from
the operator $\hat\phi^\dagger\hat\phi$ whose ensemble average is the
non-condensate density in eq.~(\ref{eq:nonconddens}).

It is straightforward to verify from the equations of motion
(\ref{eq:fieldeq}) for $\hat\phi$ and $\hat\phi^\dagger$ that these
hydrodynamic operators $\hat n'$ and $\hat\Phi'$  obey the linearized
conservation law

\begin{equation}
\frac{\partial \hat n'}{\partial t}+\nabla\cdot\hat{\bf j}' = \frac{\partial
\hat n'}{\partial t}+\nabla\cdot\left(\hat n'{\bf v}_0+ n_0\nabla
\hat\Phi'\right)= 0\label{eq:consop}
\end{equation}
and the linearized Bernoulli's equation
\begin{equation}
\frac{\partial\hat\Phi'}{\partial t} + {\bf v}_0\cdot\nabla\hat\Phi' +
\frac{4\pi\hbar^2 a}{M^2}\hat
n'-\frac{\hbar^2}{4M^2n_0}\nabla\cdot\bigg[n_0\nabla\left(\frac{\hat
n'}{n_0}\right)\bigg]=0 \label{eq:Bernop}
\end{equation}
that is a direct quantum generalization of Bernoulli's equation for
irrotational
 compressible flow \cite{ALF} at constant entropy.  Here, the first two
terms constitute the total (or hydrodynamic) derivative (see, for example,
ref.\
\cite{FW1}, sec.\ 48), and  the last term  is again the quantum
(kinetic-energy)
contribution.  This eq.\ (\ref{eq:Bernop}) has one significant advantage over
the original field equations (\ref{eq:fieldeq}), for the chemical potential no
longer appears explicitly ($\mu$ does, of course, affect the condensate wave
function and condensate density).

These coupled operator equations can be rewritten with the normal-mode
expansions from eq.\ (\ref{eq:Bogexpan}).  In this way, the hydrodynamic
operators themselves have normal-mode expansions

\begin{mathletters}\label{eq:hydexpan}
\begin{equation}
\hat n' = {\sum_j}'\left(n_j'\alpha_j + n_j'^*\alpha_j^\dagger\right),
\end{equation}
\begin{equation}
\hat \Phi' = {\sum_j}'\left(\Phi_j'\alpha_j + \Phi_j'^*\alpha_j^\dagger\right),
\end{equation}
\end{mathletters}
where the amplitudes $n_j'$ and $\Phi_j'$ obey the same coupled eqs.\
(\ref{eq:consop}) and (\ref{eq:Bernop}) as the operators.  Comparison with the
Bogoliubov eqs.\ (\ref{eq:Bogeq}) shows that

\begin{mathletters}\label{eq:linear}
\begin{equation}
n_j' = \Psi^* u_j-\Psi v_j,
\end{equation}
\begin{equation}
\Phi_j' = \frac{\hbar}{2Mi|\Psi|^2}\left(\Psi^* u_j+\Psi v_j\right),
\end{equation}
\end{mathletters}
showing that the Bogoliubov amplitudes $u_j$ and $v_j$ are {\it linearly
related\/} to the hydrodynamic amplitudes $n_j'$ and $\Phi_j'$
\cite{ALF,WG}.  Thus, any solution for $n_j'$ and $\Phi_j'$ gives a
corresponding solution $u_j$ and $v_j$, and {\it vice versa\/}.

For example, eq.~(\ref{eq:dipolesoln}) shows that the Bogoliubov equations have
exact dipole-mode  solutions
$u_\alpha~=~ a_\alpha^\dagger\Psi$ and
$v_\alpha~=~a_\alpha\Psi^*$,  in which the
condensate oscillates
 rigidly  with frequency $\omega_\alpha$ along the
three orthogonal directions (this holds for any solution $\Psi$ of the GP
equation).  Use of the explicit form of the raising and lowering operators
$a_\alpha^\dagger$ and $a_\alpha$ in eqs.\ (\ref{eq:dipole}) and the linear
relations in eqs.\ (\ref{eq:linear}) readily yields the corresponding density
perturbation

\begin{equation}
n_\alpha' = -\frac{d_\alpha}{\sqrt 2}\frac{\partial |\Psi|^2}{\partial
x_\alpha},
\end{equation}
which is a rigid shift of the condensate density $|\Psi|^2 = n_0$, and
\begin{equation}
\Phi_\alpha' =
\frac{\hbar}{\sqrt{2}Mi}\left(\frac{x_\alpha}{d_\alpha}-id_\alpha\frac{\partial
S}{\partial x_\alpha}\right).
\end{equation}
The last term involving the phase $S$ is absent if $\Psi$ is  real, and the
velocity
perturbation is then a spatial constant, moving $90^\circ$ out of phase
with the
density oscillation.  More generally, the last term reflects the change
in the local  velocity arising from the rigid motion of the  condensate
velocity.

For a nearly ideal Bose gas, the Bogoliubov equations for $u_j$ and $v_j$ often
provide a simpler description than the hydrodynamic variables; in contrast,
the
hydrodynamic amplitudes
$n_j'$ and $\Phi_j'$ are usually preferable in the large-condensate (TF) limit,
where Stringari (see below) has constructed explicit solutions  for the
low-lying
normal modes\cite{SS}.  As
 an application  of the linear relation between the two descriptions,
these TF  hydrodynamic solutions immediately give the corresponding
Bogoliubov
amplitudes $u_j$ and $v_j$ in the TF limit.  The spatial integral of
$|v_j|^2$ is
the zero-temperature occupation number $N_j'(0)$ of the $j$th mode, and a
detailed
evaluation for a large spherical condensate shows that they are large (of order
$R_0^2/d_0^2$)
\cite{FR}.

\subsection{Hydrodynamics in the Thomas-Fermi limit}

Assume a large stationary condensate with ${\bf v}_0 = 0$ and
condensate density  $n_0 =
(M/4\pi\hbar^2a)(\mu-V_{\rm tr})\Theta(\mu-V_{\rm tr})$.  In this TF limit, it
is consistent to omit the quantum contribution in Bernoulli's equation,  in
which case the  hydrodynamic amplitudes $n_j'({\bf r}) e^{-i\omega_j t}$ and
$\Phi_j'({\bf r}) e^{-i\omega_j t}$ obey two coupled linear equations

\begin{mathletters}
\begin{equation}
-i\omega_jn_j' + \nabla\cdot(n_0\nabla\Phi_j')=0,
\end{equation}
\begin{equation}
-i\omega\Phi_j' + \frac{4\pi\hbar^2a}{M^2}n_j'=0.
\end{equation}
\end{mathletters}
They can be combined into a single second-order equation for the density
perturbation \cite{SS}

\begin{equation}
\omega_j^2n_j' +\frac{1}{M}\nabla\cdot\left[\left(\mu-V_{\rm
tr}\right)\nabla n_j'\right] = 0\label{eq:TFhyd}
\end{equation}
that  has numerous explicit solutions.

\subsubsection{spherical trap}

In the TF limit, the chemical potential becomes $\mu \approx
\frac{1}{2}\hbar\omega_0 R_0^2/d_0^2 = \frac{1}{2} M\omega_0^2R_0^2$, so that
eq.\ (\ref{eq:TFhyd}) becomes

\begin{equation}
\omega_j^2\,n_j' +
\case{1}{2}\omega_0^2\nabla\cdot\left[\left(R_0^2-r^2\right)\nabla
n_j'\right] = 0.
\end{equation}
The spherical symmetry allows solutions of the form
$n_{nl}'(r)Y_{lm}(\theta,\phi)$, where $Y_{lm}$ is a spherical harmonic and
$n_{nl}'$ is the corresponding $n$th radial eigenfunction.
Solutions with different
$m$ are degenerate, and the TF frequencies \cite{SS}

\begin{equation}
\omega_{nl}^2 = \omega_0^2\left[l + n\left(2n +2l
+3\right)\right],\label{eq:SSfreq}
\end{equation}
 can be compared with
the frequencies $\omega_{nl} = \omega_0(2n+l)$ of an ideal gas in an
isotropic harmonic potential.  For the lowest radial modes with $n=0$, the TF
eigenfrequencies $\omega_0\sqrt l$ differ from those for the harmonic
oscillator
$\omega_0 l$ except  for the lowest dipole mode ($l=1$).

The associated eigenfunctions are polynomials of the form $n_{nl}(r) =
r^lP_{nl}(r^2/R_0^2)$, where $P_{nl}(u)$ is a polynomial of order $n$ (a
type of
Jacobi polynomial), for example, $P_{0l}(u) =1$ and $P_{1l}(u) =
1-(2l+5)u/(2l+3)$.  The radial eigenfunctions are bounded at both  the
origin and
 the TF surface $r=R_0$.

\subsubsection{anisotropic  axisymmetric trap}

The preceding treatment can be generalized to an axisymmetric but anisotropic
trap with $V_{\rm tr}= \frac{1}{2} M\omega_\perp^2(r_\perp^2 + \lambda^2z^2)$,
where $\lambda = \omega_z/\omega_\perp$ is the asymmetry parameter.  The
condensate density  $n_0$ is proportional to $ 1-r_\perp^2/R_\perp^2 -
z^2/R_z^2$, where $R_\perp$ and $R_z$ are the radial and axial semiaxes of the
ellipsoidal condensate with  anisotropy ratio $R_z/R_\perp= 1/\lambda$.
Equation (\ref{eq:TFhyd}) now has eigenfunctions $n'$ proportional to $
e^{im\phi}$
and eigenvalues that depend explicitly on the azimuthal quantum number $m$.
They
fall into two distinct classes
\cite{SS,Fl,Ohb}

a.  $n'\propto r_\perp^{|m|} e^{im\phi}\times$ a polynomial in
$(r_\perp^2,z^2)$
(these modes are {\it even} in $z$);

b.    $n'\propto z\,r_\perp^{|m|} e^{im\phi}\times$  a
polynomial in $(r_\perp^2,z^2)$ (these modes are {\it odd} in $z$).

Numerous special cases are of interest, and it is simplest to start with the
choice that the polynomial equals one.

 a. The solutions of type a take the form
$n'\propto r^lY_{lm} $ with $ m=\pm l$, and the associated frequency is
$\omega^2_{m=\pm l}= l\omega_\perp^2$, independent of the asymmetry parameter
$\lambda$.  This mode is a generalization of that   with frequency
$\omega_{0l}^2
= l\omega_0^2$ in a spherical trap.  For $l = 1$, these two  modes are
the  circularly polarized radial dipole oscillations with frequency
$\omega_\perp$ and $m =\pm 1$.

 b. The solutions of type b take the form $n'\propto
r^lY_{lm} $ with $m=\pm(l-1)$, and the associated frequency is
$\omega_{m=
\pm(l-1)}^2 = (l-1)\omega_\perp^2 + \omega_z^2$.  For $l=1$, this mode is the
axial dipole oscillation with $m=0$ and frequency $\omega_z$.

The next simplest class takes a polynomial of the form $A +Br_\perp^2 + Cz^2$.
For $m=0$, those of  type a  have the two frequencies

\begin{equation}
\frac{\omega_{m=0}^2}{\omega_\perp^2} = 2
+\case{3}{2}\lambda^2\pm\sqrt{(2-\case{3}{2}
\lambda^2)^2+2\lambda^2}.
\end{equation}
The two eigenstates involve a coupled motion of a monopole mode with $n=1$,
$l=0$
and a quadrupole mode with $n=0$, $l=2$, $m=0$.  For a spherical trap
($\lambda =
1$), the frequencies  reproduce both  the  results given in eq.\
(\ref{eq:SSfreq}).  For a  disk-shape trap with $\lambda = \sqrt 8$, this
expression gives a good fit to the observations on ${}^{87}$Rb \cite{Jin}.  For
a cigar-shape trap with $\lambda\ll 1$, the two frequencies reduce to
$\sqrt{\case{5}{2}}\,\omega_z
$ and $2\omega_\perp$;  the first fits the observations on ${}^{23}$Na
\cite{MOM}  within 1\%.  More generally, theoretical eigenvalues and
eigenfunctions  are readily determined for
 larger values of
$|m|$ and for the solutions  of class b.

\subsection{Sum-rules for a non-uniform Bose condensate}

Stringari \cite{SS} has modified the sum rules for a uniform medium  to obtain
accurate estimates of the first few lowest  frequencies of the collective modes
in a trap for intermediate values of the interaction parameter $Na/d_0$,  where
neither the nearly ideal gas nor the TF approximation is accurate.
Consider a general operator $\hat F$ that excites the system from its ground
state $|0\,\rangle$ to an excited state $|j\,\rangle$.  Here,  $\hat
H|j\,\rangle =
E_j|j\,\rangle$  and $\hat H|0\,\rangle = E_0|0\,\rangle$, where $\hat H$
is the
interacting many-particle Hamiltonian and $E_j$ and
$E_0$ are the corresponding exact  energies.

In analogy with eq.\ (\ref{eq:Imchi}), define a ``spectral-density function''

\begin{equation}
S(\omega) = \sum_{j\neq 0} \big|\langle j|\hat
F|0\,\rangle\big|^2\delta\left(\omega-\omega_{j0}\right),
\end{equation}
where $\omega_{j0} = (E_j-E_0)/\hbar$.  In the following analysis, the matrix
elements of $\hat F$ are assumed to have the following time-reversal  behavior

\begin{equation}
\big|\langle j|\hat F|0\,\rangle\big|^2
 = \big|\langle j|\hat F^\dagger|0\,\rangle\big|^2=\big|\langle 0|\hat
F|j\,\rangle\big|^2,\end{equation}
which restricts the allowed class of excitation operators.  The various moments
of the spectral-density function are
\begin{equation}
{\cal M}_i \equiv \int_0^\infty d\omega \,\omega^i\, S(\omega).
\end{equation}

For example, the first moment is

\begin{eqnarray}
{\cal M}_1 &= &\sum_{j\neq 0} \big|\langle j|\hat
F|0\,\rangle\big|^2\,\omega_{j0}\nonumber\\
&=&\case{1}{2}\sum_{j\neq 0}\left(\langle 0|\hat F^\dagger|j\,\rangle
\,\omega_{j0}\,\langle j|\hat F|0\,\rangle +\langle 0|\hat
F|j\,\rangle \,\omega_{j0}\,\langle j|\hat
F^\dagger|0\,\rangle\right)\nonumber\\
&=&\frac{1}{2\hbar}\sum_{j\neq 0}\left(\langle 0|\hat F^\dagger|j\,\rangle
\langle j|[\hat H,\hat F]|0\,\rangle -\langle 0|[\hat H,\hat
F]|j\,\rangle \langle j|\hat F^\dagger|0\,\rangle\right).
\end{eqnarray}
The sum over the excited states can now be performed with the completeness
relation
[this is essentially the method used to prove the $f$-sum rule in eq.\
(\ref{eq:sum1})], leading to the expression

\begin{equation}
{\cal M}_1 = \frac{1}{2\hbar} \langle 0|\left[\hat F^\dagger,[\hat H,\hat
F]\right]|0\,\rangle,
\end{equation}
expressed in terms of a ground-state expectation value of a double commutator.
Similarly,
\begin{equation}
{\cal M}_3 = \frac{1}{2\hbar^3}  \langle 0|\left[\big[\hat F^\dagger,\hat
H\big],\big[\hat H,[\hat H,\hat F]\,\big]\right]|0\,\rangle.
\end{equation}

\subsubsection{surface modes}
It is  simplest to  treat a spherical trap with a
spherical condensate density.
 Consider the excitation operator
\begin{mathletters}
\begin{equation}
 F = \sum_{i=1}^N r_i^l\,Y_{lm}(\theta_i,\phi_i) \quad\hbox{in first-quantized
notation, or}
\end{equation}
\begin{equation}
\hat F  =  \int d^3 r\,\hat\psi^\dagger({\bf r})\,
r^lY_{lm}(\theta,\phi)\hat\psi({\bf r})\quad\hbox{in second-quantized
notation}.
\end{equation}
\end{mathletters}
This operator has no radial node and thus
predominantly excites
 modes with zero radial quantum number $n=0$ (called ``surface modes''
\cite{SS}). A direct evaluation of the two moments for the ground-state
condensate
wave function (the solution of the GP equation) yields the ratio ${\cal
M}_3/{\cal
M}_1$, which is an upper bound for the squared frequency of the transition
from the
ground state to the first state excited  by the operator
$\hat F$.  In fact, this ratio should provide a good approximation to the exact
lowest  squared frequency since
these surface modes are orthogonal to the higher
radial modes.

For the special case  of the  quadrupole  surface mode ($n=0$ and $l=2$), this
ratio gives the  expression

\begin{equation}
\omega_{02}^2 = 2\omega_0^2 \left(1+\frac{\langle T\rangle_0}{\langle V_{\rm
tr}\rangle_0}\right),
\end{equation}
where the angular brackets are the ground-state expectation values of the
kinetic
and trap potential energies evaluated with the condensate wave function
$\Psi$.
This expression holds for all values of the interaction parameter $Na/d_0$,
providing a uniform interpolation between the ideal gas and the TF limit.

a.  For an ideal gas, the familiar properties of the harmonic oscillator show
that the kinetic energy and trap potential energy are equal, reproducing the
expected result that $\omega_{02} = 2\omega_0$ for the excitation to the lowest
quadrupole state of the oscillator.

b.  In  contrast, the kinetic energy is negligible in the TF limit,
reproducing
the previous TF  result from eq. (\ref{eq:SSfreq}) that $\omega_{02} = \sqrt 2
\omega_0$.

\subsubsection{lowest compressional mode}
The operator
\begin{mathletters}
\begin{equation}
 F = \sum_{i=1}^N r_i^2,\quad\hbox{or, equivalently,}
\end{equation}
\begin{equation}
\hat F = \int d^3 r\,\hat\psi^\dagger({\bf r})\, r^2\hat\psi({\bf
r})
\end{equation}
\end{mathletters}
excites the lowest monopole mode with $n=1$ and $l=0$.  Direct evaluation
of the
moments ${\cal M}_1$ and ${\cal M}_3$, along with the virial theorem from eq.\
(\ref{eq:vir}) gives the result
\begin{equation}
\omega_{10}^2 = \omega_0^2 \left(5-\frac{\langle T\rangle_0}{\langle V_{\rm
tr}\rangle_0}\right).
\end{equation}

a.  In the ideal-gas limit, the last ratio is $1$, yielding the expected
expression
$\omega_{10} = 2\omega_0$ [from the general expression $\omega_{nl} =
\omega_0(2n+l)$ for an isotropic oscillator].

b.  In the TF limit, the last ratio is negligible, yielding $\omega_{10} =
\sqrt 5
\omega_0$, in agreement with the previous TF result from eq. (\ref{eq:SSfreq}).

\section{Vortices in a dilute trapped Bose gas}
The prediction and detection of quantized vortices in superfluid ${}^4$He were
early milestones in  understanding the role of the condensate and the
associated macroscopic wave function (see, for example, refs.\
\cite{BK,RJD}). Thus
it is natural to consider similar vortex structures in the dilute trapped Bose
gases, but the analysis involves several new and subtle features.

\subsection{Review of classical vortices}

Consider an incompressible nonviscous fluid with uniform number density
$\bar n$.
A long straight vortex line at the origin has the velocity field

\begin{equation}
{\bf v(r)} = \frac{\kappa}{2\pi r_\perp}\hat \phi,\label{eq:vortex}
\end{equation}
where $\hat\phi$ is a unit tangential vector in cylindrical polar coordinates.
The flow velocity consists of circular streamlines, and its magnitude  diverges
near the  symmetry axis (as $r_\perp\to 0$).  It is obvious from the flow
pattern
that $\nabla\cdot {\bf v} = 0$, so that the flow is transverse (or
solenoidal).  In
addition, it is not difficult to verify that $\nabla\times {\bf v} =0$  except
on the axis, and use of Stokes's theorem shows that

\begin{equation}
\nabla\times {\bf v} = \kappa \,\hat z \,\delta^{(2)}({\bf
r}_\perp).\label{eq:vorticity}
\end{equation}
Thus the flow is longitudinal (or irrotational) {\it nearly everywhere\/},
but  it
is singular on the axis (analogous to  the magnetic field of a long straight
current-carrying filament).  Finally, the line integral of $\bf v$ is known
as the
``circulation;''  for any  contour $C$  that encircles the vortex once, the
circulation

\begin{equation}
\oint_C d\,{\bf l\cdot v} = \kappa,\label{eq:circ}
\end{equation}
  is independent of $C$, with the constant $\kappa$ in eq.\
(\ref{eq:vortex}) fixing
the strength of the vortex.

The  energy is purely  kinetic in an incompressible fluid.  In the present
case
of a long vortex filament, it has the value (per unit length)

\begin{eqnarray}
E_v&=& \case{1}{2} M\bar n\int d^2r_\perp\,v^2\nonumber \\
&=&\frac{M\bar n\kappa^2}{4\pi} \int \frac{dr_\perp}{r_\perp} \approx
\frac{M\bar
n\kappa^2}{4\pi}\ln\left(\frac{R_0}{\xi}\right).\label{eq:Evortex}
\end{eqnarray}
Here, the integral must be cut off at an upper limit $R_0$, which can
be interpreted as either the size of the container or the intervortex
separation,
and a lower limit
$\xi$, which can be interpreted as the size of the vortex core.

\subsubsection{quantized circulation}
The idea of a quantized   vortex in superfluid ${}^4$He  can be
motivated by recalling the Bohr-Sommerfeld quantization condition $\oint
d\,{\bf
l\cdot p} = h$ for the periodic motion of a particle in phase space.  Since
${\bf
p} = M{\bf v} $ for a particle in the vortex, this result immediately
suggests that
the circulation in superfluid ${}^4$He should be quantized in units of
$h/M$.  Note
that $\kappa$ has the dimensions of length${}^2$/time (like a diffusion
constant or
a thermal conductivity or a  kinematic viscosity).   Onsager \cite{Ons} first
publicly proposed the concept of quantized circulation     in 1949
(sec.\ 2.3 of  ref.\
\cite{RJD} contains a brief history of Onsager's unpublished work).  London
\cite{London} documented one of   Onsager's earlier remarks and
 made a similar proposal \cite{Lon} concerning quantized magnetic flux in
superconductors  in 1950.  Independently, Feynman developed a more complete
description of quantized vortex lines in superfluid ${}^4$He~\cite{Feyn},
based on
the identification of the velocity with the gradient of the phase of the
appropriate  many-body wave function [compare eq.\ (\ref{eq:condvel}) for the
condensate velocity].

A   classical vortex in  an incompressible fluid
 requires  a specific model for  the  core radius~$\xi$.  In fact, all real
fluids
have a non-zero compressibility, implying a finite speed of sound~$s$.
If the hydrodynamic flow speed $|v|$ is small compared to $s$, then the
fluid is
essentially   incompressible.  This picture necessarily fails near the
vortex core,
where eq.\ (\ref{eq:vortex}) shows the the flow speed  diverges; hence a
singularity
develops when the magnitude of the circulating velocity becomes comparable
with the
speed of sound.  Setting
$v(\xi) =
\kappa/2\pi\xi \approx s$ gives the resulting core radius

\begin{equation}
\xi \approx \frac{\kappa}{2\pi s}.
\end{equation}
If the vortex is quantized with circulation $h/M$, this result implies that the
vortex core has a radius
\begin{equation}
\xi \approx \frac{\hbar}{Ms}.\label{eq:quantxi}
\end{equation}
As seen in eq.\ (\ref{eq:speedTF}), the same qualitative expression  holds for
the healing length
$\xi =\hbar/({\sqrt 2} Ms)$ in a dilute trapped Bose gas in the TF limit,
suggesting that $\xi$  also characterizes the  vortex core radius in such a
condensate.

\subsubsection{classical vortex dynamics}

The dynamics of a vortex in an incompressible fluid follows from the
celebrated ``circulation'' theorem of classical hydrodynamics (see, for
example,
sec.\ 48 of ref.\ \cite{FW1});  it states  that    the circulation around any
contour that moves with the fluid is a strict constant of the motion.  If the
contour initially contains a vortex with circulation $\kappa$, the vortex
can never
escape  the  moving contour.  In the particular case of a single long straight
vortex line in otherwise stationary fluid,    this result implies that the
vortex
itself remains at rest.  Similarly, in the presence of a uniform  external flow
${\bf v}_{\rm ex}$, the vortex line moves rigidly with the same velocity (this
conclusion also follows with a Galilean transformation to a moving coordinate
system).

A more interesting case is two long straight parallel vortices a distance $d$
apart.   Since neither vortex moves under its own influence, the motion arises
solely from its neighbor.   Assume first that they have the {\it same\/}
sense of
circulation, as shown in fig.\ 5a.  In this case, it is easy to see from the
circulating velocity fields that they each execute a circular trajectory at
fixed
separation
$d$.  Since the interaction energy depends only on $d$ (and the individual
circulation
$\kappa$),  the constancy of the
 separation $d$ follows from the conservation of energy in a nonviscous
fluid.  In
the presence of   weak dissipation,  the system will act to lower the energy;
here, the two vortices will slowly spiral apart because the total energy of two
identical vortex lines is lowest at infinite separation (for $d\to 0$, the
vortices
fuse into  a single vortex with circulation $2\kappa $ and four times the
energy
of a single vortex line, whereas, for $d\to \infty$, the energy reduces to two
times the energy of a single vortex line).

If the parallel vortices have {\it opposite\/} sense of circulation, they
are known
as a ``vortex pair''  (shown in fig.\ 5b). The mutual influence of the
circulating velocity fields means that they move perpendicular to the line
joining
them in the same sense as the local fluid between them, with fixed  separation
$d$  (again because of conservation of energy).  This configuration is the
two-dimensional analog of a vortex ring, which moves perpendicular to its plane
maintaining a fixed radius.  In the presence of  weak dissipation, the two
vortices
slowly drift together and annihilate when the separation $d$ is comparable
with the
vortex core size $\xi$.

\subsubsection{effect of rigid boundaries in classical hydrodynamics}

As a simple example, consider a long circular cylinder with rigid walls of
radius
$R_0$.  If a vortex line is located on the symmetry axis, its circulating flow
velocity is everywhere parallel to the boundary, so that the flow field is
unperturbed by the presence of the cylinder. The situation is more complicated,
however,  if the vortex is shifted rigidly a distance $r_0 <R_0$
off-axis.  In
this case, the flow of the vortex by itself no longer matches the boundary
condition
that the normal component of the flow velocity must vanish.  This
boundary-value
problem has an elementary solution consisting of a single opposite image
located
along the same ray at a distance $R_0^2/r_0$  outside the cylinder (see
fig.\ 6).
Comparison with fig.\ 5b shows that the vortex and its image constitute a
vortex
pair;  in the absence of dissipation, the original vortex will therefore
execute a
circular orbit at fixed
$r_0$ under the influence of its image (which moves synchronously with the
vortex).  If the system is weakly dissipative, the vortex will slowly
spiral outward
and eventually annihilate with the image vortex when $r_0 \approx R_0-\xi$.

\subsubsection{effect of rotation}

One of the original motivations for the introduction of quantized vortices
was the
observed behavior of a rotating superfluid.  For example, a bucket of ${}^4$He
rotating at an angular velocity $\Omega$ was known \cite{RJD} to retain its
classical parabolic  meniscus
$z =
\frac{1}{2}
\Omega^2r^2/g$ far  below the superfluid transition temperature
$T_\lambda$.  This observation appeared to contradict  Landau's  assertion
\cite{LDL} that   the superfluid velocity is irrotational with
$\nabla\times {\bf v}_s =0$
(implying that it could not rotate) and that $\rho_s/\rho \approx 1$ (so
that the
normal fluid played a negligible role).  To explain this apparent paradox,
Feynman
\cite{Feyn} relaxed the irrotational condition of strictly zero vorticity,
 allowing the vorticity to be singular at the cores of the quantized
vortices [compare eq.\ (\ref{eq:vorticity})];   in particular, he suggested
that a
rotating   superfluid contains  an array of quantized vortex lines with areal
density
$n_v = 2\Omega/\kappa= M\Omega/(\pi\hbar) $ and mean vorticity $n_v\kappa =
2\Omega$, thereby mimicking the solid-body rotation required to explain the
curved
meniscus (see sec.\ 2.4 of ref.\
\cite{RJD}).

>From this perspective, it becomes interesting to consider the effect of
rotating a
physical
 container at an angular velocity ${\bf \Omega } = \Omega\hat z$.  In
the laboratory (non-rotating) frame, the moving walls represent time-dependent
potentials that  do work on the system, precluding a description in terms of
thermodynamic equilibrium.  In the rotating frame, however, the walls are
stationary, so that the hamiltonian becomes time-independent, and the usual
thermal
Gibbs distribution remains valid when expressed in terms of the energy levels
appropriate for the rotating frame. In the rotating frame, the new
hamiltonian is
known to be  $H-{\bf \Omega\cdot L}= H-\Omega L_z$ (see, for example, sec.\ 5
of ref.\ \cite{LDL} and sec.\ 34 of ref.\ \cite{LP}).

 At zero temperature, a system rotating at an angular velocity
$\Omega$ will adjust any free parameters to minimize the quantity $F=
\langle H - \Omega L_z\rangle = E-\Omega L_z$, which can be considered a ``free
energy'' that depends on the parameter
$\Omega$.  As a simple example, consider a long cylinder of radius $R_0$.  If
the fluid does not contain a vortex,
 the velocity is everywhere zero,  so that the energy $E_0$ and angular
momentum
$L_0$ per unit length both vanish, with $F_0= E_0-\Omega L_0=0$. When a
vortex is
placed at the center of the cylinder, however, the energy per unit length is
\begin{equation} E_v = \frac{M\bar
n\kappa^2}{4\pi} \ln\left(\frac{R_0}{\xi}\right)
\end{equation}
from eq.\ (\ref{eq:Evortex}),  and an elementary calculation shows that the
corresponding angular momentum per unit  length is $L_v = \frac{1}{2} M\bar n
\kappa R_0^2$. Thus the free energy per unit length of the fluid containing one
vortex is given by
\begin{equation}
F_v = E_v-\Omega L_v =\frac{M\bar
n\kappa^2}{4\pi} \ln\left(\frac{R_0}{\xi}\right)-\case{1}{2} \Omega M\bar
n\kappa
R_0^2.\label{eq:F1}
\end{equation}
 This quantity is positive for sufficiently small $\Omega$, but it decreases
linearly with increasing angular velocity and becomes negative (and hence
less than
$F_0$ for the no-vortex state) at a critical angular velocity $\Omega_{c1} =
(\kappa/2\pi R_0^2)\ln(R_0/\xi)$ for the creation of a single vortex line in a
long cylinder of radius $R_0$. Use of the quantized circulation $\kappa
= h/M$  yields the expression  (see, for example,  sec.\ 2.4 of ref.\
\cite{RJD})

\begin{equation}
\Omega_{c1} \approx \frac{\hbar}{M R_0^2}\ln\left(\frac{R_0}{\xi}\right)
\quad\hbox{for a quantized vortex in uniform fluid}.\label{eq:omegac1}
\end{equation}

This elementary analysis can be generalized to the case of a long straight
vortex a
distance $r_0$ from the center of the cylinder.  The free energy per unit
length
remains zero if there is no vortex, and a detailed calculation
\cite{Hess,PS} gives
the free energy per unit length of a  vortex displaced a fractional distance
$x_0=r_0/R_0$ from the center

\begin{equation}
F_v(x_0) = \frac{M\bar n\kappa^2}{4\pi} \bigg[\ln\left(\frac{R_0}{\xi}\right) +
\ln(1-x_0^2) -\frac{\Omega}{\Omega_0}(1-x_0^2)\bigg],\label{eq:delF1}
\end{equation}
where  $\Omega_0 = \kappa/(2\pi R_0^2) = \hbar/(MR_0^2)$ is a characteristic
angular velocity.  Figure 7 shows this quantity as a function of $x_0$ for
various
fixed values of $\Omega$.  If $\Omega = 0$, the free energy decreases
monotonically
with increasing $x_0$, indicating that the vortex is unstable in the
presence of a
weak dissipation.  This behavior persists  to the value $\Omega =
\Omega_0$, when
the curvature at the origin $x_0=0$ vanishes.  For
$\Omega_0<\Omega<\Omega_{c1}$,
the free energy develops a local minimum at $x_0=0$, although  $F_v(0)\ge
F_v(1)$,
with an intermediate barrier that hinders the vortex  from reaching the
outer wall
at
$x_0=1$. This behavior suggests that $\Omega_m= \Omega_0$ represents the
onset of
{\it metastability} in the presence of weak dissipation, which  persists up
to the
critical angular velocity
$\Omega_{c1}$, when the free energy $F_v(0)$ for a vortex at the origin becomes
equal to  that $F_v(1)$ for a vortex at the wall.  For
$\Omega_{c1}<\Omega$, the
vortex  becomes  stable relative to the no-vortex state.  Experiments on
rotating
superfluid
${}^4$He
\cite{PS} have verified these predictions in considerable detail.

\subsection{Vortices in a dilute uniform Bose gas}

The original application of the GP equation \cite{EPG,LPP} was the
description of a
long straight vortex in an otherwise uniform dilute Bose gas.  Recall
eq.\ (\ref{eq:condvel}) for the condensate
velocity ${\bf v}_0 = (\hbar/M)\nabla S$, where $S$ is the phase of the
condensate
wave function, and consider the circulation $\Gamma = \oint_C d\,{\bf
l\cdot v}_0$
around an arbitrary closed contour $C$.  A combination of  these expressions
gives

\begin{equation}
\Gamma = \frac{\hbar}{M}\oint_C d\,{\bf l\cdot}\nabla S = \frac{\hbar}{M}
\Delta
S\big|_C,
\end{equation}
where $\Delta S|_C$ is the change in the phase on once going around $C$.  If
the condensate wave function is single valued, this change must be an
integer times
$2\pi$, implying that the circulation in any condensate is quantized in
units of
$h/M$.

In general, the GP equation for a bulk dilute Bose gas is

\begin{equation}
\left( T+ g|\Psi|^2 -\mu\right) \Psi = 0,
\end{equation}
since $V_{\rm tr} $ is absent.  Assume that the condensate contains a singly
quantized vortex line at $r_\perp=0$,  oriented along $\hat z$, with bulk
condensate
density $n_0$ far from the vortex ($r_\perp\to \infty$).  The condensate wave
function has the form \cite{EPG,LPP}

\begin{equation}
\Psi = \sqrt{n_0}\,e^{i\phi} f(r_\perp),
\end{equation}
where $f$ is real and approaches 1 as $r_\perp\to \infty$. Here, the phase is
given by $S=\phi$, so that the condensate velocity ${\bf v}_0 =
(\hbar/M)\nabla\phi = (\hbar/Mr_\perp)\hat\phi$ is precisely that for a vortex
with circulation $h/M$.  The condensate density $|\Psi|^2 = n_0(r_\perp) =
n_0f(r_\perp)^2$ is cylindrically symmetric. The radial amplitude obeys the
nonlinear  GP equation

\begin{equation}
\left(-\frac{\hbar^2}{2M}\frac{1}{r_\perp}\frac{d}{dr_\perp}\,r_\perp\frac{d}{dr
_\perp}
+ \frac{\hbar^2}{2Mr_\perp^2} + gn_0f^2 -\mu\right)f = 0,\label{eq:GPvor}
\end{equation}
where the asymptotic behavior of $f$ fixes the chemical potential $\mu =
gn_0$ at
the value for the uniform system.

In a dilute uniform Bose gas, the   healing length  is defined by $\xi^2  =
\hbar^2/(2Mgn_0) = 1/(8\pi an_0)$,  balancing the kinetic energy and the
interaction energy. With the dimensionless variable $x = r_\perp/\xi$, the
radial
GP equation becomes

\begin{equation}
\left(-\frac{1}{x}\frac{d}{dx}\,x\,\frac{d}{dx} + \frac{1}{x^2} + f^2 -
1\right)f =
0,\label{eq:radialGP}
\end{equation}
where the term $1/x^2$ is the centrifugal barrier associated with the
circulating
velocity field.  For small $x\ll 1$, this barrier forces the amplitude to
vanish
linearly inside the vortex core $x\lesssim 1$, and a detailed analysis
shows that

\begin{equation}
f(x) \approx \cases{0.583\,x,&for $x\ll 1$,\cr
1-(2x^2)^{-1},&for $x\gg 1$.\cr}
\end{equation}
Figure 8 shows the form of this function, indicating that this quantum vortex
{\it automatically} forms its own core with radius $\approx \xi =
\hbar/(\sqrt 2 M
s)$, where $s$ is the Bogoliubov speed of sound [compare eq.\
(\ref{eq:quantxi})].
The condensate density $n_0(r_\perp)$ varies like $r_\perp^2$ near the
origin, and
the corresponding  condensate current
${\bf j}_0 (r_\perp)~=~n_0(r_\perp){\bf v}_0(r_\perp)$ reaches a maximum
near the
core and then vanishes linearly as $r_\perp\to 0$, even though the velocity
itself
diverges.  Numerical analysis yields the GP vortex energy per unit length
\cite{Ginz}

\begin{equation}
E_v\approx \frac{\pi\hbar^2 n_0}{M} \ln\left(1.46\frac{R_0}{\xi}\right)
\approx
\frac{M n_0\kappa^2}{4\pi} \ln\left(1.46\frac{R_0}{\xi}\right),\label{eq:Ev}
\end{equation}
where $R_0$ is an outer cutoff;
this result has precisely the same form as the classical expression in eq.\
(\ref{eq:Evortex}).

The GP description of a vortex in a bulk dilute Bose gas can be generalized to
study the dynamics of a set of widely separated parallel vortex lines
\cite{ALF2}
located at
$\{{\bf r}_{\perp j}\}$, assuming that $|{\bf r}_{\perp i}-{\bf r}_{\perp
j}|\gg
\xi$ for all pairs.  An approximate GP condensate wave function can be
constructed
as a product of the individual radial functions $f(|{\bf r}_{\perp}-{\bf
r}_{\perp
j}|)$, each of which approaches 1 far from the vortex cores; the phase is
given as
the sum of the azimuthal angles
$S_j$ measured  relative to ${\bf r}_{\perp j}$ as origin, where $\tan S_j =
(y-y_j)/(x-x_j)$, so that

\begin{equation}
\Psi({\bf r}_\perp, t) = \sqrt{n_0} \,e^{-i\mu
t/\hbar}\prod_j\left[e^{iS_j}f(|{\bf
r}_{\perp}-{\bf r}_{\perp j}|)\right].
\end{equation}
This state changes with
time both because of the overall chemical potential and because each vortex
follows
its own dynamical trajectory.  Substitution of this condensate wave function
into the time-dependent GP eq.\ (\ref{eq:TDGP}) determines the subsequent
motion of
the individual  vortices; they obey the classical hydrodynamic prescription
that
each vortex moves with the local velocity induced at its position  by all
the other
ones.  This result should  not be  very surprising, because the GP equation
itself
can be recast in  a hydrodynamic form.

\subsection{A vortex in a dilute  trapped Bose condensate}

Assume a singly quantized vortex in an axisymmetric trap, where the
condensate wave
function takes the form

\begin{equation}
\Psi({\bf r}) = \sqrt{N_0}\,e^{i\phi}f(r_\perp,z).
\end{equation}
Like the case of a vortex in a uniform condensate, the condensate velocity
field
here is just
 ${\bf v}_0 = (\hbar/Mr_\perp)\hat\phi$, representing a vortex with circulation
$\kappa = h/M$.    The amplitude  function $f$ satisfies the GP equation
and can be
determined numerically \cite{DS}.  This function vanishes linearly as
$r_\perp\to
0$, giving
 a node on the symmetry axis.  Consequently, the condensate density for a
vortex in
a trapped condensate is  toroidal (in contrast to the non-vortex
condensate, where
the density decreases monotonically away from the center).

\subsubsection{critical angular velocity $\Omega_{c1}$}

Given the condensate wave function for the vortex state and that for no
vortex, it
is straightforward to evaluate the additional energy $\Delta E$ associated
with the
formation of the vortex, along with the angular momentum $L_z$.  If the system
rotates with angular velocity $\Omega$, a vortex line on the
symmetry axis becomes stable when $\Delta F = \Delta E-\Omega L_z$ becomes
negative, so that the critical angular velocity is $\Omega_{c1} = \Delta
E/L_z$.

This quantity can be evaluated analytically in two limits.  If the system
is nearly
ideal, with $Na/d_0\ll 1$, then the condensate wave function for the vortex
is one
with all
$N_0$ particles in the single-particle state

\begin{equation}
\chi_{10}({\bf r}_\perp)\psi_0(z)\propto r_\perp e^{i\phi}
\exp\left(-\frac{r_\perp^2}{2d_\perp^2}-\frac{z^2}{2d_z^2}\right),
\label{eq:vorideal}
\end{equation}
where $\chi_{n_+ n_-}$ is a two-dimensional oscillator state with $n_+$
and $n_-$
 right- and left-circular quanta and $\psi_n$ is the usual one-dimensional
oscillator wave function \cite{CT}. The increased energy is essentially
 $N_0$  times the increased single-particle energy $\hbar\omega_\perp$
associated with the transition from the two-dimensional ground state
$\chi_{00} $ to
the vortex state $\chi_{10}$, and the associated angular momentum is just
$N_0\hbar$
because every particle has unit angular momentum.  Thus $\Omega_{c1}\to
\omega_\perp$ for an ideal gas in an axisymmetric trap.  In this ideal-gas
limit,
however, the transition is hugely degenerate, for the same $\Omega_{c1}$ also
characterizes a condensate with multiple quanta of circulation. With the GP
equation, it is not difficult to evaluate the first correction to this
value for a
singly quantized vortex in a nearly ideal Bose gas

\begin{equation}
\frac{\Omega_{c1}}{\omega_\perp} \approx 1
-\frac{1}{\sqrt{8\pi}}\,\frac{N_0a}{d_z}+\cdots,\label{eq:omegac1ideal}
\end{equation}
showing that the critical angular velocity  decreases linearly as the
interaction
parameter
$N_0a/d_z =
\sqrt{\lambda}\,N_0a/d_\perp$ initially increases  from $0$ (here, $\lambda=
\omega_z/\omega_\perp$ is the asymmetry parameter for the trap).

In the opposite limit of a large interaction parameter $Na/d_0 \gg 1$,  initial
estimates  \cite{BP} relied on classical hydrodynamics, using eq.\
(\ref{eq:omegac1}) to write
\begin{equation}
\frac{\Omega_{c1}}{\omega_\perp }\approx \frac{\hbar}{M\omega_\perp
R_\perp^2}\ln\left(\frac{R_\perp}{\xi}\right)=\frac{d_\perp^2}{R_\perp^2}
\ln \left(\frac{R_\perp}{\xi}\right),\label{eq:omegac1TF0}
\end{equation}
since
$d_\perp^2 = \hbar/M\omega_\perp$.  Note that this ratio is small in the TF
limit
where
$d_\perp^2/ R_\perp^2 \sim \xi/R_\perp\ll 1$;   the explicit TF relation

\begin{equation}
\frac{d_\perp^2}{R_\perp^2} =
\left(\frac{d_\perp}{15Na\lambda}\right)^{2/5}
\label{eq:TFsmall}
\end{equation}
shows precisely how this small parameter scales with $N$ and $\lambda$.

An improved TF estimate follows by writing the condensate wave function as
$\Psi =
e^{i\phi}|\Psi|$, and ignoring the gradients  of $|\Psi|$.  The new feature
is that
the TF density profile now has a centrifugal barrier \cite{Sinha,DR}

\begin{equation}
 gn_1\approx \left(\mu_1-\case{1}{2}M\omega_\perp^2r_\perp^2
-\case{1}{2}M\omega_z^2z^2-\frac{\hbar^2}{2Mr_\perp^2}\right)
\Theta\!\left(\mu_1-\case{1}{2}M\omega_\perp^2r_\perp^2
-\case{1}{2}M\omega_z^2z^2-\frac{\hbar^2}{2Mr_\perp^2}\right),
\end{equation}
where $n_1$ and $\mu_1$ are the density and chemical potential for the
state with
one quantum of circulation.  This density can be compared with that for the
no-vortex state in eq.\ (\ref{eq:TFdens})

\begin{equation}
 gn_0\approx \left(\mu_0-\case{1}{2}M\omega_\perp^2r_\perp^2
-\case{1}{2}M\omega_z^2z^2\right)
\Theta\!\left(\mu_0-\case{1}{2}M\omega_\perp^2r_\perp^2
-\case{1}{2}M\omega_z^2z^2\right).
\end{equation}
The fractional change in the chemical potential caused by the vortex
$(\mu_1-\mu_0)
/\mu_0$ can be shown \cite{SF} to be small, of order $(d_\perp^4/R_\perp^4)
\ln(R_\perp^2/d_\perp^2)$.
 As a result,  the TF density profile for the condensate with a singly
quantized
vortex
 has the simple form

\begin{equation}
 n_1(r_\perp,z) \approx n_0(0)\left(1-\frac{r_\perp^2
}{R_\perp^2}-\frac{z^2}{R_z^2}-\frac{\xi^2}{r_\perp^2}\right)
\Theta\!\left(1-\frac{r_\perp^2
}{R_\perp^2}-\frac{z^2}{R_z^2}-\frac{\xi^2}{r_\perp^2}\right),\label{eq:TFvorden
s}
\end{equation}
where $n_0(0)$ is the central density for the no-vortex condensate.  The
important
new qualitative feature of a vortex in the TF limit  is the appearance of  a
{\it small} flared hole of radius
$\sim
\xi$ \cite{DS,Sinha,DR}, but the remainder of the condensate density is
essentially unchanged;  in practice, it is usually sufficient to retain the
no-vortex density and simply cut off any divergent  radial integrals at the
core
size $\xi$.

A detailed solution of the GP equation for an axisymmetric trap in the TF limit
\cite{LPS,Svid} gives the more careful estimate

\begin{equation}
\Omega_{c1} \approx \frac{5}{2}\,
\frac{\hbar}{MR_\perp^2}\ln\left(\frac{R_\perp}{\xi}\right)\quad\hbox{or,
equivalently,}\quad  \frac{\Omega_{c1}}{\omega_\perp} \approx
\frac{5}{2}\, \frac{d_\perp^2}{R_\perp^2}\ln\left(\frac{R_\perp}{\xi}\right),
\label{eq:omegac1TF1}
\end{equation}
which holds with logarithmic accuracy;  it is larger than the previous naive
estimate in eq.\ (\ref{eq:omegac1TF0}) by a factor $\frac{5}{2}$ that
arises from
the reduced angular momentum for the non-uniform density relative to that for a
uniform density.  Reference
\cite{LPS} shows that  this asymptotic estimate agrees well with the
numerical work
of ref.\ \cite{DS} for the particular asymmetry $\lambda= \sqrt 8$,
appropriate for
the original experiments \cite{And} on ${}^{87}$Rb.

The TF description has also been extended \cite{Svid} to the case of  a
rotating
axisymmetric trap containing a straight vortex that is  displaced  laterally a
distance $r_0$ from the center.   The dependence  of the free energy on the
dimensionless parameter $x_0 = r_0/R_\perp$ for fixed $\Omega$ is qualitatively
similar to that in fig.\ 7, but the details differ considerably.  In
particular,
the vortex at the center is unstable for $\Omega \le \Omega_m = \frac{3}{5}
\Omega_{c1}$, where $\Omega_{c1}$ is given above in eq.\
(\ref{eq:omegac1TF1}), and
metastable for $\Omega_m\le \Omega\le\Omega_{c1}$. Thus the presence of the
trap
reduces (but does not eliminate)  the regime of metastability.  A similar
result
also holds for a simpler classical model  where the uniform density $\bar n$ is
replaced by a parabolic radial density profile \cite{ALFJLTP}.  In both these
cases, the reduced outer density in a trap lowers the free energy
$F_v(x_0)$ for
$x_0\gtrsim
\frac{1}{2}$ compared to that
for uniform density, raising the threshold for metastability relative to
that in
fig.~7.

\subsubsection{possible scenario for creation of a vortex}

A basic question is: how can the condensate be made to rotate?  For
 ${}^4$He in a rotating circular container, the microscopically rough
walls can
spin up the normal (viscous) fluid when the temperature is above the superfluid
transition.  If the fluid is then cooled  below the transition temperature, the
superfluid is created in a state of rotation for sufficiently large rotation
speeds.  This method  fails for a condensate in an axisymmetric trap,
because the
trap is simply a smooth potential.  Consequently, only a non-axisymmetric
trap can
spin up the condensate, for the asymmetry in the trap potential exerts a
torque on
the condensate, similar to the effect of moving walls in  classical
hydrodynamics.

It is helpful first to consider the classical example of a uniform fluid in
a long
rotating  elliptical cylinder with semi-major and semi-minor
axes  $a$ and $b$.  If $a>b$, the rotating walls push the fluid,
creating a classical velocity potential $\Phi_{\rm cl} = \Omega\, xy
\,(a^2-b^2)/(a^2+b^2)$
\cite{Lamb}, even in the absence of a vortex. The resulting  angular momentum
$L_{z0} = I_0\Omega$ and kinetic energy
$E_0 = \frac{1}{2}I_0\Omega^2$ have the expected dependence on the rotation
speed
$\Omega$,  with
 an effective moment of inertia per unit length $I_0$ that is less than the
classical solid-body value
$I_{\rm sb}$ by the ratio

\begin{equation}
\frac{I_0}{I_{\rm sb}}= \left(\frac{a^2-b^2}{a^2+b^2}\right)^{\!\!2}.
\end{equation}
This quantity vanishes for a circular cylinder ($a=b$), but it can approach
1 from
below   for extreme asymmetry ($b\ll a$).  Unlike the case
of a circular cylinder, the free energy  $F_0 = E_0-\Omega  L_{z0}=
-\frac{1}{2}I_0\Omega^2$ for a vortex-free state is now negative.
Note that the classical velocity ${\bf v}_{\rm cl}= \nabla\Phi_{\rm cl}$ is
everywhere irrotational with zero vorticity $\nabla\times {\bf v}_{\rm cl}=0$.
Nevertheless, the moment of inertia $I_0$ is non-zero;  indeed, for $b\ll
a$, it
can be comparable with that for solid-body rotation
${\bf v}_{\rm sb}= {\bf
\Omega\times r}$, whose vorticity is uniform with
$\nabla\times {\bf v}_{\rm sb}= 2{\bf
\Omega}$.

 The  critical
angular velocity $\Omega_{c1}$ for creating  a vortex  at the center
depends on the asymmetry ratio $b/a$ \cite{ALFaniso}, and   experiments
confirm the
 theoretical  predictions in considerable detail
\cite{DCP}.  In the extreme asymmetric limit
$b\ll a$, a detailed calculation yields

\begin{equation}
\Omega_{c1}\approx
\frac{\hbar}{2Mb^2}\ln\left(\frac{b}{\xi}\right),\label{eq:omegac1aniso}
\end{equation}
which holds with logarithmic accuracy  (here, the circulation is taken as
$h/M$).
Qualitatively, this value can be interpreted as that for a circular
cylinder with
the radius  equal to the largest inscribed circle that fits inside  the
elliptical cross section.

An approximate TF solution of the GP equation in a rotating triaxial (totally
asymmetric) trap with
$\omega_x\neq\omega_y\neq\omega_z$ bears out these qualitative conclusions
\cite{Svid}. As usual in the TF limit,  the chemical potential $\mu$
determines  the
condensate radii
$R_\alpha^2 = 2\mu/M\omega_\alpha^2$.
Even in the absence of a  vortex,  the rotating walls push the condensate,
inducing
a phase for the condensate wave function that is linear in the angular velocity
${\bf \Omega} = \Omega\hat z$

\begin{equation}
S_0(x, y) \approx
-\frac{M\Omega}{\hbar}\left(\frac{\omega_x^2-\omega_y^2}{\omega_x^2+\omega_y^2}
\right)
xy =
\frac{M\Omega}{\hbar}\left( \frac{R_x^2-R_y^2}{R_x^2+R_y^2} \right) xy.
\end{equation}
This expression has the correct angular symmetry  $\propto xy =
\frac{1}{2}r^2\sin2\phi$ and vanishes for an axisymmetric trap;  it is
the quantum analog of the classical velocity potential for a rotating elliptic
cylinder\cite{Lamb}.

 For a non-axisymmetric trap rotating about
the $\hat z$ axis, it is convenient to define the mean  radius

\begin{equation}
{\cal R}_\perp^2  = \frac{2R_x^2R_y^2}{R_x^2+R_y^2}, \quad\hbox{or,
equivalently,}\quad \frac{2}{{\cal R}_\perp^2} =
\frac{1}{R_x^2}+\frac{1}{R_y^2},
\end{equation}
so that ${\cal R}_\perp\to R_\perp= R_x=R_y$ for an axisymmetric trap.  A
calculation of the energy and angular momentum of a vortex on the axis of
the trap
yields the critical angular velocity \cite{Svid}

\begin{equation}
\Omega_{c1} \approx \frac{5}{2} \,\frac{\hbar}{M{\cal
R}_\perp^2}\ln\left(\frac{{\cal R}_\perp}{\xi}\right)
\end{equation}
for the creation of a vortex in an arbitrary non-axisymmetric trap in the TF
limit.  For an
axisymmetric trap (${\cal R}_\perp\to R_\perp$), the resulting
expression for $\Omega_{c1}$  reproduces the previous result in eq.\
(\ref{eq:omegac1TF1});   in contrast, it      becomes

\begin{equation}
\Omega_{c1} \approx \frac{5}{4} \,\frac{\hbar}{M
R_y^2}\ln\left(\frac{R_y}{\xi}\right)
\end{equation}
in the extreme anisotropic limit $R_y\ll R_x$, very similar to the classical
expression in eq.~(\ref{eq:omegac1aniso}).

One possible experimental
scenario for creating a vortex in a trap follows from the analogy with
superfluid
${}^4$He
\cite{PS}.  Cool a rotating trap with significant anisotropy in the $xy$
plane from
the normal state
 through
$T_c$ to some low temperature $T\ll T_c$.  If
$\Omega$ exceeds the critical angular velocity $\Omega_{c1}$,  the
resulting Bose
condensate should have a vortex.  Note the crucial
order of operations, with the condensate  created in a rotating state.
Although
reversing the order (cooling and then rotating) does often create a vortex in
superfluid ${}^4$He, such a procedure is hysteretic and displays
considerable metastability \cite{YGP,YP}.  Type-II superconductors exhibit
similar
metastability when a magnetic field is applied after the sample has been
cooled in
zero field.

\subsubsection{possible detection of a vortex line}

In principle, it may be possible to observe directly the reduced density in the
vortex core (somewhat like the optical detection of localized trapped
charge in the
core of vortex lines in superfluid ${}^4$He \cite{YGP,YP}), but the small
healing
length
$\xi$ in the TF limit may well suppress the effect.  An alternative and
promising
approach  considers the effect of a vortex on the collective modes of the
condensate.  The basic physical idea  is that a uniform rotation splits the
time-reversal degeneracy of the two modes with $\pm |m|$  that are
originally degenerate (like the Zeeman effect of a uniform magnetic field
on the
states with different azimuthal components of magnetic moment).

Two distinct  approaches  have been  developed (they both yield the same
results in
the TF limit). One  uses sum rules \cite{ZS} to generalize the earlier
study of the
low-lying collective modes  for a vortex-free condensate \cite{SS},
discussed in
the previous section.  The other uses the hydrodynamic equations in the TF
limit
\cite{SF}, where the vortex provides a weak perturbation (ref.\ \cite{Sinha}
also obtained some, but not all,  of the same results).

To estimate the magnitude
of the effect,  recall that a static velocity field
${\bf v}_0$ alters the usual time derivative $\partial/\partial t$  to the
hydrodynamic derivative $\partial/\partial t + {\bf v}_0\cdot \nabla$.  Thus
the original frequency $\omega$ of a wave is shifted to $\sim \omega\pm
v_0/R_0$,
where $R_0$ is the typical dimension of the condensate.   For a vortex, the
circulating velocity is of order $v_0\sim \hbar/(MR_0)$, and the fractional
change
in the frequency is $\delta \omega/\omega \sim \hbar/(MR_0^2\omega_\perp) \sim
d_\perp^2/R_\perp^2
\ll1$ in the TF limit.  Alternatively, the fractional change in the
frequency is
comparable to the ratio $v_0/s$, where $s$ is the speed of sound.  In a dilute
trapped Bose gas, this speed is of order $s\sim\hbar/(M\xi)$, which becomes
$s\sim\hbar R_0/(Md_0^2)$ in the TF limit and reproduces the
previous estimate for $\delta\omega/\omega$.   Similar ideas (see, for
example, pp.\
144-145 of ref.\
\cite{deG}) have served to study the effect of a quantized vortex line in a
type-II
superconductor on the BCS quasiparticles.

Consider  a large (TF) vortex-free  condensate in an axisymmetric trap, and
focus
on a particular collective mode with  degenerate  frequency
$\omega_{|m|}^0$. In
the presence of a vortex, the fractional splitting of the  modes with
$\pm |m|$  can be characterized by the
 ratio

\begin{equation}
\frac{\omega_{|m|}-\omega_{-|m|}}{\omega_{|m|}^0} = \Delta_{|m|}\,
\frac{d_\perp^2}{R_\perp^2} \approx  \Delta_{|m|} \left( \frac{d_\perp}{15
Na\lambda}\right)^{\!\!2/5},
\end{equation}
where $\Delta_{|m|}$ is a dimensionless number that depends on the specific
mode in
question.   For $|m|>1$, it is proportional to  the matrix element of
$1/r_\perp^2$, evaluated with the normalized unperturbed
(vortex-free) eigenfunctions, but the case of $|m|=1$ requires a more careful
analysis~\cite{SF}.

a. For the lowest quadrupole mode with $|m| = 2$, the unperturbed frequency is
$\omega_{2}^0 = \sqrt 2\omega_\perp$, and the dimensionless factor is
$\Delta_2 =
7/\sqrt 2$.

b.  For the lowest quadrupole mode with $|m|=1$, the unperturbed frequency is
$\omega_1^0 = \sqrt{1+\lambda^2}\,\omega_\perp=\sqrt{\omega_\perp^2
+\omega_z^2}$,
and the dimensionless factor is
$\Delta_1 = 7\lambda^2/(1+\lambda^2)^{3/2}$.

	\noindent  Typically, the predicted fractional splitting is of
order $10\%$ for the
low-lying modes.  Current measurements of frequencies for non-rotating
condensates
are accurate to  $\lesssim 1\%$, so that this vortex-induced splitting
should be
readily detectable.

\subsubsection{stability of a vortex in a trapped Bose condensate}

If the condensate rotates with an angular velocity $\Omega\gtrsim
\Omega_{c1}$, a
singly quantized vortex is believed to be the stable equilibrium
configuration.
If, however,  the angular velocity is  reduced so that
$\Omega<\Omega_{c1}$, does
the vortex remain stable (especially if $\Omega/\Omega_{c1} \ll 1)$? In the
present
case of a trapped condensate, the question can become even more intricate,
depending on whether the trap remains non-axisymmetric (so that the
reduced rotation  exerts a braking torque on the condensate) or is
first adiabatically transformed to axisymmetric (in which case,  the
reduced angular
velocity of the  trap potential cannot affect the rotation of the
condensate).

Rokhsar \cite{DR} has given a qualitative argument that a vortex in a
non-rotating
axisymmetric nearly ideal condensate should be unstable.  In this limit, the
single-particle vortex wave function has the form $\psi_v \approx
e^{i\phi}|\psi_v|$, with the radius
 of the vortex core  comparable to the oscillator length  and condensate
radius $d_0$ (note that the coherence length $\xi$ here is much larger than
$d_0$
and  therefore does not characterize the  vortex core radius in the nearly
ideal
limit).  This large  depleted region   of reduced repulsive  Hartree potential
energy favors the formation of  a localized bound core state,  with a real
single-particle wave function $\psi_b$.  The actual condensate wave function
becomes a linear combination $\Psi \approx
\sqrt{N_v}\psi _v+\sqrt{N_b}\psi_b$, where $N_v$ is the number of particles
in the
vortex condensate and $N_b = N-N_v$ is that in the   bound  core state.
For small
$N_b$, this complex wave function $\Psi \approx
\sqrt{N_v}e^{i\phi}|\psi _v|+\sqrt{N_b}|\psi_b|$ has a node away from the
axis of
symmetry, so that the vortex core shifts radially outward  from the origin.
This motion can occur only if there is some mechanism for the condensate to
lose
energy and angular momentum;   it is a quantum analog of the hydrodynamic
instability  below the threshold for metastability, discussed in connection
with
fig.\ 7.  As in that case, the time for a vortex to move outward depends on the
details of the dissipation.  In addition, the vortex core itself becomes
small in
the TF limit
 and cannot support a bound state for $Na/d_0\gtrsim 1$ \cite{DR}, so that this
argument is not applicable in the TF limit.

As an alternative physical picture in the non-interacting limit, note that the
vortex has all the particles   in the single-particle state $\chi_{10}
({\bf r}_\perp)\psi_0(z)$ given in eq.\ (\ref{eq:vorideal}), with excitation energy
$\hbar\omega_\perp$ and angular momentum
$\hbar$.  Clearly, the system can lower its
energy by transferring one particle from this (excited) condensate to the
original
ground state $\chi_{00}({\bf r}_\perp)\psi_0(z)$;  the same process can be
repeated many times, which is the physical source of the instability.

This argument can be sharpened somewhat by considering the Bogoliubov equations
for a singly quantized vortex in a nearly ideal condensate with $Na/d_z\ll 1$
\cite{ALFJLTP}. Among the many normal modes with positive norm $\int d^3
r(|u|^2
-|v|^2) = 1$, only one is anomalous and has a negative frequency

\begin{equation}
\frac{\omega_a}{\omega_\perp} \approx -1 +
\frac{1}{\sqrt{8\pi}}\frac{N_0a}{d_z} + \cdots.\label{eq:negfreq}
\end{equation}
This mode consists of a coherent linear combination of both $u$ and $v$
amplitudes with an angular dependence $m_a = -1$ relative to the vortex.
When re-expressed in terms of the particle density,  it represents a
displacement
of the vortex core relative to the center of mass of the condensate.

More generally,  Dodd {\it et al.} \cite{DBEC} have studied the normal
modes of a
singly quantized vortex numerically for small and intermediate values of the
dimensionless coupling parameter $Na/d_0$   (note that ref.~\cite{DBEC} uses
$m$  to denote the quantum number of the static vortex condensate, whereas
$m$
here denotes  the  angular dependence
$\propto e^{im\phi}$ of a density perturbation~\cite{SF}, in direct analogy
with
the same
$m$ dependence of a spherical harmonic $Y_{lm}$).  For the lowest-lying
dipole modes
with
$|m| = 1$ relative to the vortex,
Dodd {\it et al.} \cite{DBEC} find two dipole-sloshing modes with frequency
$\omega_\perp$ in which the condensate oscillates rigidly with right- and
left-circular helicity, independent of the interaction parameter (as
expected).   In
addition, they exhibit   an additional  (anomalous) dipole mode with $m =
-1$ that
has the frequency
$\omega_\perp$ for the ideal gas and decreases toward zero with increasing
coupling
strength.  Rokhsar~\cite{DR} argues that this mode  actually describes an
antiparticle;  it is the opposite-frequency and negative-norm  partner of a
physical mode with {\it negative} frequency and {\it positive\/} norm.
With this
reinterpretation,  the frequency of the  anomalous mode reduces to the expected
value
$-\omega_\perp$ for an ideal gas and grows toward zero from below with
increasing
$Na/d_z$.  In the context of the preceding discussion, this anomalous mode is
precisely the unstable (negative-frequency) mode found analytically in eq.\
(\ref{eq:negfreq}) from the Bogoliubov equations in the weak-coupling limit. At
present, little is known of its behavior in the TF limit, when $Na/d_z\gg 1$,
although the trend of the numerical data \cite{DBEC} suggests that the
frequency
remains negative.

As noted several times, an applied rotation can stabilize a vortex.
Specifically,
the frequency $\omega_j$ of a given normal mode in a non-rotating condensate is
shifted to $\omega_j-\Omega m_j$ when the system rotates with angular velocity
$\Omega$.  In particular, the anomalous mode with negative frequency
$\omega_a$ has
$m_a = -1$;  it therefore has an apparent frequency
$\omega_a +\Omega$ when observed in the rotating frame, indicating that a
positive rotation raises the (negative) frequency, tending to stabilize the
mode.
Recall  that eq.~(\ref{eq:omegac1ideal}) gives the thermodynamic critical
angular
velocity $\Omega_{c1} $ for the creation of a singly quantized vortex in the
near-ideal limit, and comparison with the independently computed negative
frequency
$\omega_a$ in eq.~(\ref{eq:negfreq})  shows that $\omega_a +
\Omega_{c1} $ vanishes identically for small $N_0a/d_z$.  Hence,  a rotation at
$\Omega_{c1}$ just suffices to stabilize this unstable anomalous mode in the
weak-coupling limit.  The corresponding situation in the TF limit remains
unknown,
where the metastable frequency $\Omega_m< \Omega_{c1}$ discussed in
connection with
fig.\ 7 may well be relevant.

In principle,  various other  methods can  stabilize a
vortex in a dilute trapped Bose condensate;  in
the present context, the most relevant proposal  is to
force the non-rotating condensate to assume a toroidal form \cite{DR2,JPY} (for
example,  piercing it with a blue-detuned narrowly focused laser beam that acts
as a repulsive localized force).    In this case, the externally imposed
hole in
the stationary condensate renders it multiply connected, even  in the
absence of a
vortex.  Strictly speaking, the excited states of interest here are  those with
irrotational flow and quantized circulation around the circumference of the
torus,
but they are frequently considered to represent the corresponding quantized
vortex
with its  core ``pinned'' in the central void.  Such states have been
studied in
detail for superfluid
${}^4$He in a rotating annulus (see, for example, secs.\ 2.6 and 5.3 of ref.\
\cite{RJD}), where a row of physical vortices eventually appears in the gap
between
the walls at sufficiently high angular velocity.

 \acknowledgments{I am grateful for the hospitality of  the Institute
for Theoretical Physics, University of California, Santa Barbara, where
part of this
work was carried out, and for
valuable discussions with M.\ Andrews, E.\ Cornell, P.\ Drummond, M.\
Edwards,  D.\
Feder, M.\ Girardeau, A.\ Griffin, M.\ Gunn, T.-L.\ Ho, A.\ Leggett, D.\
Rokhsar,
G.\ Shlyapnikov, S.\ Stringari, A.\ Svidzinsky,  B.\ Svistunov, L.\ You,
and P.\
Zoller.  I especially  thank A.\ Svidzinsky for assistance with the figures
and D.\
Feder for  comments on a preliminary draft.  This work was supported in
part by the
National Science Foundation, under Grant No.\  DMR 94-21888.}

\begin{figure}
\caption{Dimensionless excitation-energy spectrum $E/ng$ as a function of
 the dimensionless wavenumber $k\xi$: (a)  Bogoliubov spectrum in eq.\
(\ref{eq:Esubk}); (b)  free-particle spectrum.}
\label{fig1}
\end{figure}

\begin{figure}
\caption{Bogoliubov coherence factors  (a) $u^2$ and (b) $v^2$ from eq.\
(\ref{eq:uv}) as functions  of the dimensionless wavenumber $k\xi$.}
\label{fig2}
\end{figure}

\begin{figure}
\caption{Dimensionless squared Bogoliubov energy $E^2/n^2g^2$ in eq.\
(\ref{eq:unstable}) for a uniform system subject to an attractive
interaction with
$s$-wave scattering length
$-|a|$, as a function of the dimensionless wavenumber $k\xi$, where $8\pi n|a|
\xi^2 = 1$.  In these units, the unstable range is $0\le k\xi\le \sqrt 2$. }
\label{fig3}
\end{figure}

\begin{figure}
\caption{Dimensionless variational energy $2E/(N\hbar\omega_0)$  in eq.\
(\ref{eq:unstable1}) for a trapped condensate,
 as a function of the dimensionless parameter $\beta$ that characterizes the
actual radius $\beta d_0$ for various values of the interaction parameter:  (a)
$Na/d_0=0.5$ (locally stable), (b) $Na/d_0 = 0.67$ (onset of local
instability),
and (c)
$Na/d_0 = 0.84$ (locally unstable).}
\label{fig4}
\end{figure}

\begin{figure}
\caption{Induced local velocity for two parallel vortex lines a distance
$d$ apart
in unbounded incompressible dissipationless  fluid:  (a) Two parallel
vortex lines
with {\it same} sense of circulation; induced motion as shown yields circular
orbits at fixed separation
$d$;  (b) two antiparallel vortex lines with {\it opposite}
sense of circulation; induced motion as shown yields uniform motion
perpendicular
to line joining centers in direction of fluid at center,  maintaining   fixed
separation $d$.}
\label{fig5}
\end{figure}

\begin{figure}
\caption{Cylinder of radius $R_0$ containing a long straight vortex line
displaced
from the center by a distance $r_0<R_0$.  One image vortex with opposite
circulation on same ray at the exterior point with radius $R_0^2/r_0$
suffices to
satisfy the boundary condition that the normal component of velocity vanish
at the
boundary.}
\label{fig6}
\end{figure}

\begin{figure}
\caption{Dimensionless free energy per unit length $ MF_v/(\pi\bar n\hbar^2)
$ from eq.\
(\ref{eq:delF1}) for a vortex in a cylinder of radius $R_0$; the vortex is
displaced
radially a fractional distance
$x_0= r_0/R_0$ from the axis of symmetry, and the different curves
correspond to
 various fixed values of the angular velocity $\Omega$: (a) $\Omega = 0$
(unstable); (b)
$\Omega =
\Omega_m =\Omega_0\equiv
\hbar/(MR_0^2)$ (marginally metastable at origin); (c) $\Omega =
0.5\,\Omega_{c1}
$ (metastable at origin); (d) $\Omega = \Omega_{c1}$ (stable at origin), where
$\Omega_{c1}= \Omega_0\ln(R_0/\xi)\equiv (\hbar/MR_0^2)\ln(R_0/\xi)$ is
evaluated
for
$R_0/\xi \approx 100$.}
\label{fig7}
\end{figure}

\begin{figure}
\caption{Radial wave function $f(r_\perp/\xi)$ obtained by numerical
solution of
eq.\ (\ref{eq:radialGP}). }
\label{fig8}
\end{figure}


\begin{references}

\bibitem{Bog} Bogoliubov N.\ N., {\it J.\ Phys.\  (USSR)}, {\bf 11}
(1947) 23.

\bibitem{LP} See, for  example,  Lifshitz E.\ M.\ and Pitaevskii L.\
P., {\it Statistical Physics}, Third edition, Part 1 (Pergamon, Oxford)
1980, secs.\ 54, 56, and 62.

\bibitem{FW}  See, for example, Fetter A.\ L. and Walecka J.\ D., {\it
Quantum Theory of Many-Particle Systems} (McGraw-Hill, New York)
1971, secs.\ 11 and 35.

\bibitem{FW1} See, for example, Fetter A.\ L. and Walecka J.\ D., {\it
Theoretical Mechanics of Particles and Continua} (McGraw-Hill, New
York) 1980, sec.\ 50.


\bibitem{LHY}  Lee T.\ D., Huang K., and Yang C.\ N., {\it Phys.\
Rev.\/}, {\bf 106} (1957) 1135.

\bibitem{AG1}  Griffin A., {\it Excitations in a Bose-Condensed
Liquid\/} (Cambridge University Press, Cambridge) 1993, chaps.~3 and 5.

\bibitem{AG2} Griffin A., {\it Phys.\ Rev.\ B\/}, {\bf 53} (1996)
9341.

\bibitem{LY} Lee T.\ D., and Yang C.\ N., {\it Phys.\ Rev.\/}, {\bf
105} (1957) 1119.

\bibitem{GN}  Gavoret J.~and Nozi\`eres P., {\it Ann.\ Phys.\ (N.\
Y.)}, {\bf 28} (1964) 349.

\bibitem{HM}  Hohenberg P.\ C.\  and Martin P.\ C.,  {\it Ann.\ Phys.\ (N.\
Y.)}, {\bf 34} (1965) 291.

\bibitem{LDL}  Landau L.\ D., {\it J.\ Phys.\  (USSR)}, {\bf 5}
(1941) 71.

\bibitem{LP1}   Lifshitz E.\ M.\ and Pitaevskii L.\
P., {\it Statistical Physics},  Part 2 (Pergamon, Oxford)
1980, secs.\ 22 and 23.

\bibitem{Wilks}  See, for example, Wilks J., {\it The Properties of Liquid and
Solid Helium} (Oxford University Press, Oxford) 1967, chaps.\ 2-6.

\bibitem{And} Anderson M.~H., Ensher J.~R., Matthews M.~R., Wieman C.~E.,
and Cornell  E.~A., {\it Science\/}, {\bf 269} (1995)  198.

\bibitem{Dav} Davis  K.~B.,  Mewes  M.-O., Andrews  M.~R., van Druten  N.~J.,
 Durfee D.~S.,  Kurn D.~M., and  Ketterle W.,
{\it Phys.~Rev.~Lett.\/}, {\bf 75}  (1995) 3969.

\bibitem{Brad} Bradley C.~C., Sackett  C.~A., and Hulet R.~G.,
{\it Phys.~Rev.~Lett.\/}, {\bf 78}  (1997) 985;  see, also, Bradley C.\ C.,
Sackett
C.\ A., Tollet J.\ J., and Hulet R.\ G., {\it Phys.\ Rev.\ Lett.},  {\bf
75} (1995)
1687.

\bibitem{RMP}  For a more complete overview, see Dalfovo F., Giorgini S.,
Pitaevskii L.\ P., and Stringari S., {\it Rev.\  Mod.\ Phys.} (to be
published).

\bibitem{Mewes2} Mewes M.-O., Andrews M.~R., van Druten N.~J., Kurn D.~M.,
Durfee
D.~S., and Ketterle W., {\it Phys.~Rev.~Lett.\/}, {\bf 77} (1996) 416.

\bibitem{AP} Fetter A.~L., {\it Ann.~Phys.~(N.~Y.)}, {\bf 70}  (1972) 67.

\bibitem{EPG} Gross E.~P., {\it Nuovo Cimento}, {\bf 20} (1961) 454;
{\it J.~Math.~Phys.}, {\bf 4} (1963) 195.

\bibitem{LPP}  Pitaevskii L.~P., {\it Zh.~Eksp.~Teor.~Fiz.}, {\bf 40}
(1961) 646
 [{\it Sov.~Phys. JETP}, {\bf 13} (1961) 451].

\bibitem{DS}    Dalfovo F.\  and Stringari S., {\it
Phys.~Rev.~A\/}, {\bf 53} (1996) 2477.

\bibitem{Wu} Wu T.\ T., {\it J.\ Math.\ Phys.\/}, {\bf 2} (1961) 105.

\bibitem{HS}  Huse D.~A.\  and Siggia E.\ D., {\it J.\ Low Temp.\  Phys.\/},
{\bf 46} (1982) 137.

\bibitem{BP}  Baym G.\ and Pethick C.\ J., {\it Phys.~Rev.~Lett.\/}, {\bf 76}
(1996) 6.

\bibitem{Jin} Jin D.\ S., Ensher J.\ R., Matthews M.\ R.,
Wieman C.\ E., and Cornell E.\ A.,
 {\it Phys.\ Rev.\ Lett.\/}, {\bf 77}  (1996) 420.

\bibitem{MOM} Mewes M.-O., Andrews  M.\ R., van Druten N.\ J.,
Kurn D.\ M., Durfee D.\ S.,
Townsend C.\ G., and  Ketterle  W., {\it Phys.\ Rev.\ Lett.\/}, {\bf 77}
(1996) 988.

\bibitem{SS} Stringari S., {\it  Phys.\ Rev.\ Lett.\/}, {\bf 77} (1996)  2360.

\bibitem{ALF1}  Fetter A.\ L.\ (e-print cond-mat/9510037, unpublished).

\bibitem{Stoof} Stoof H.\ T.\ C., {\it J.\ Stat.\ Phys.}, {\bf 87} (1997) 1353.

\bibitem{UL} Ueda M.\  and Leggett A.\ J., {\it Phys.\ Rev.\ Lett.\/}, {\bf 80}
(1998) 1576.

\bibitem{Rup} Ruprecht P.~A., Holland M.\ J.,   Burnett  K.,  and
Edwards M.,  {\it Phys.\ Rev.\  A\/}, {\bf 51}  (1995) 4704.

\bibitem{DPS} Dalfovo F., Pitaevskii L.\ P., and Stringari S., {\it Phys.\
Rev.\
A\/}, {\bf 54} (1996) 4213.

\bibitem{LPS}  Lundh E., Pethick C.\ J., and Smith H., {\it Phys.\ Rev.\ A\/},
{\bf 55} (1997) 2126.

\bibitem{FF}  Fetter A.\ L.\ and Feder D., {\it Phys.\ Rev.\ A} (to be
published).

\bibitem{deG} The analogous
self-consistent equations for a superconductor are now generally known as
the Bogoliubov-de Gennes equations [see,
for example, de Gennes P.~G., {\it Superconductivity of Metals and
Alloys\/} (Benjamin, NY) 1966, Chap.~5].

\bibitem{Lewen}  Lewenstein M.\  and You L., {\it Phys.\ Rev.\ Lett.}, {\bf 77}
(1996) 3489.

\bibitem{Vill}  Villain P., Lewenstein L., Dum R., Castin Y., You L.,
Imamo\=glu A.,
and Kennedy T.\ A.\ B., {\it J.\ Mod.\ Opt.}, {\bf 44} (1997) 1775.

\bibitem{BD}  See, for example, Bjorken  J.~D., and Drell S.~D., {\it
Relativistic Quantum Mechanics\/} (McGraw-Hill, NY) 1964, chaps.\ 1-5.

\bibitem{CF}  Colson W.\ B.\  and Fetter A.\ L., {\it J.\ Low Temp.\
Phys.}, {\bf
33} (1978) 231.

\bibitem{FR} Fetter A.\ L.\ and Rokhsar D., {\it Phys.\ Rev.\ A}, {\bf 57}
(1998) 1191.

\bibitem{EDCB}   Edwards M., Dodd R.\ J.,
 Clark C.\ W., and  Burnett K., {\it J.~Res.~Natl.~Inst.~Stand.~Technol.},
{\bf 101}
(1996) 553.

\bibitem{Rup1} Ruprecht P.~A.,  Edwards M.,  Burnett  K.,  and
Clark C.~W., {\it Phys.\ Rev.\  A\/}, {\bf 54}  (1996) 4178.

\bibitem{ME} Edwards M., Ruprecht P.\ A., Burnett K., Dodd R.~J., and
Clark C.~W., {\it Phys.\ Rev.\ Lett.}, {\bf 77} (1996) 1671.

\bibitem{RPF}  Feynman R.\ P., {\it Phys.\ Rev.}, {\bf 94} (1954) 262.

\bibitem{PN} Pines D.\  and Nozi\`eres P., {\it The Theory of Quantum Liquids;
Volume I:  Normal Fermi Liquids} (Benjamin, New York) 1966, chaps.\ 1 and 2.

\bibitem{NP}   Nozi\`eres P. and Pines D., {\it The Theory of Quantum
Liquids; Volume II: Superfluid Bose Liquids} (Benjamin, New York) 1990, chaps.\
2, 7, and 9.

\bibitem{RDP}  Puff R.\ D., {\it Phys.\ Rev.\ A}, {\bf 137} (1965) 406.

\bibitem{SS1}  Stringari S., {\it Phys.\ Rev.\ B}, {\bf 46} (1992) 2974.

\bibitem{AJL}  Leggett A.\ J.,  lecture notes (1996) (unpublished).

\bibitem{ALF} Fetter A.~L., {\it Phys.~Rev.~A}, {\bf 53}  (1996) 4245.

\bibitem{WG} Wu W.-C., and Griffin A., {\it Phys.\ Rev.\ A}, {\bf 54} (1996)
4204.

\bibitem{Fl}  Fliesser M.,  Csord\'as A., Sz\'epfalusy P., and Graham R.,
{\it  Phys.~Rev.~A}, {\bf 56} (1997) R2533.

\bibitem{Ohb} \"Ohberg P., Surkov  E.~L., Tittonen I., Stenholm  S.,
Wilkens M.,
and Shlyapnikov G.~V., {\it Phys.~Rev.~A}, {\bf 56} (1997) R3346.

\bibitem{BK}  Fetter A.\ L., in {\it The Physics of Liquid and Solid
Helium}, Part
I, edited by K.\ H.\ Bennemann and J.\ B.\ Ketterson (John Wiley and Sons, New
York) 1976, chap.\ 3.

\bibitem{RJD}  Donnelly R.\ J., {\it Quantized Vortices in Helium II}
(Cambridge
University Press, Cambridge) 1991, chaps.\ 1 and 2.

\bibitem{Ons} Onsager L., {\it Nuovo Cimento\/}, {\bf 6}  Suppl.~2 (1949)
249.

\bibitem{London} London F., {\it Superfluids, vol.\ II: Macroscopic Theory of
Superfluid Helium} (Dover, New York) 1954, p. 151.

\bibitem{Lon}  London F., {\it Superfluids, vol.\ I:, Macroscopic Theory of
Superconductivity} (Dover, New York) 1964, p. 152.

\bibitem{Feyn} Feynman R.\ P., in {\it Progress in Low-Temperature Physics\/},
edited by C.\ J.\ Gorter (North-Holland, Amsterdam) 1955, vol.~1, p.~17.

\bibitem{Hess}  Hess G.\ B., {\it Phys.\ Rev.}, {\bf 161} (1967) 189.

\bibitem{PS}  Packard R.\ E.\  and Sanders T.\ M., Jr., {\it Phys.\ Rev.\
Lett.}, {\bf 22} (1969) 823; {\it Phys.\ Rev.\ A}, {\bf 6} (1972) 799.

\bibitem{Ginz}  Ginzburg V.\ L.\  and Pitaevskii L.\ P., {\it Zh.\ Eksp.\
Teor.\
Fiz.}, {\bf 34} (1958) 1240 [{\it Sov.\ Phys.--JETP}, {\bf 7} (1958) 858].

\bibitem{ALF2} Fetter A.\ L., {\it Phys.\ Rev.}, {\bf 151} (1966) 100.

\bibitem{CT}  Cohen-Tannoudji C., Diu B., and Lalo\"e F., {\it Quantum
Mechanics}
(Wiley, New York) 1977, pp.\ 500-502 and 733-738.

\bibitem{Sinha}  Sinha S., {\it Phys.\ Rev.\ A}, {\bf 55} (1997) 4325.

\bibitem{DR}  Rokhsar D.\ S., {\it Phys.\ Rev.\ Lett.}, { \bf 79} (1997) 2164.

\bibitem{SF}  Svidzinsky A.\ A.\  and Fetter A.\ L., {\it Phys.\ Rev.\ A}
(to be
published).

\bibitem{Svid}  Svidzinsky, A.\ A.\ (unpublished).

\bibitem{ALFJLTP} Fetter A.\ L., {\it J.\ Low Temp.\ Phys.} (to be published).

\bibitem{Lamb}  Lamb H., {\it Hydrodynamics} (Dover, New York) 1945, 6th
ed., pp.\
86-88.

\bibitem{ALFaniso}  Fetter A.\ L., {\it J.\ Low Temp.\ Phys.}, {\bf 16}
(1974) 533.

\bibitem{DCP}  DeConde K.\  and Packard R.\ E., {\it Phys.\ Rev.\ Lett.},
{\bf 35}
(1975) 732.

\bibitem{YGP}  Yarmchuk E.\ J., Gordon M.\ J.\ V., and Packard R.\ E., {\it
Phys.
Rev.\ Lett.}, {\bf 43} (1979) 214.

\bibitem{YP}    Yarmchuk E.\ J.\ and Packard R.\ E., {\it J.\ Low Temp.\
Phys.},
{\bf 46} (1982) 479.

\bibitem{ZS}  Zambelli F.\  and Stringari S., {\it Phys.\ Rev.\ Lett.} (to be
published).

\bibitem{DBEC} Dodd R.~J., Burnett K., Edwards  M., and Clark C.~W.,
{\it Phys.~Rev.~A}, {\bf 56} (1997)  587.

\bibitem{DR2}  Rokhsar D. (e-print cond-mat/9709212, unpublished).

\bibitem{JPY}  Javanainen J., Paik S.\ M., and Yoo S.\ M., {\it Phys.\
Rev.\ A},
{\bf 58} (1998) 580.


\end{references}
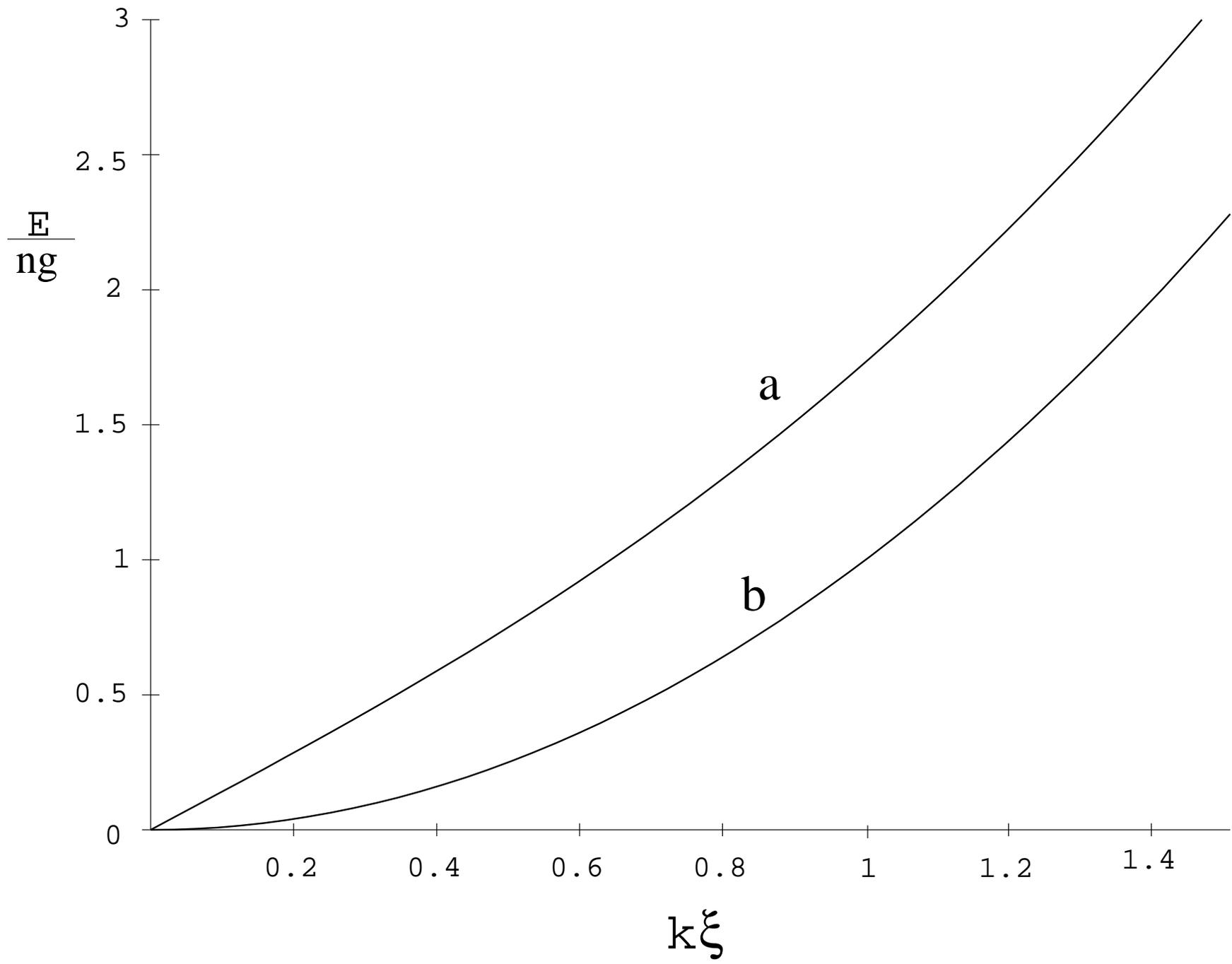
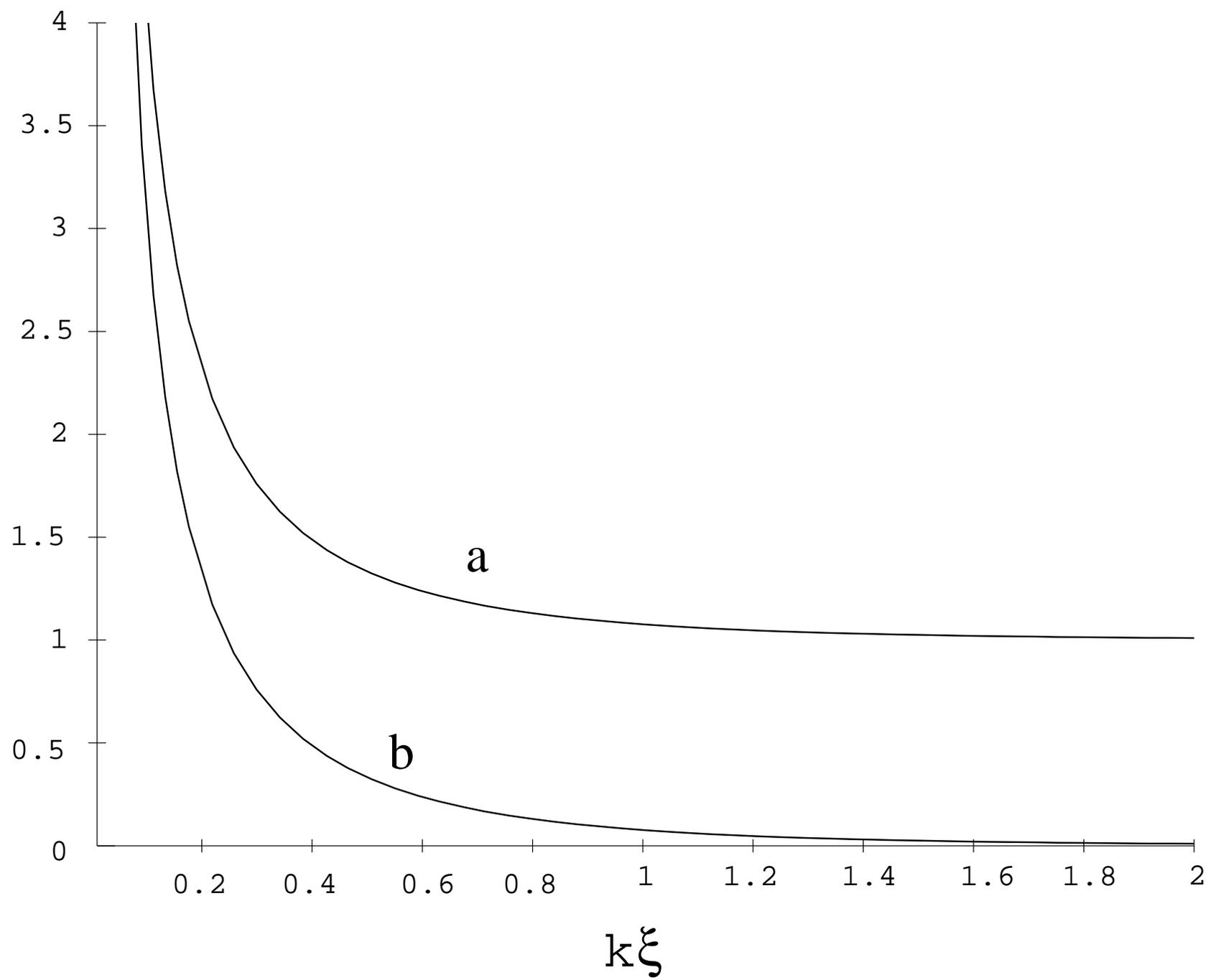
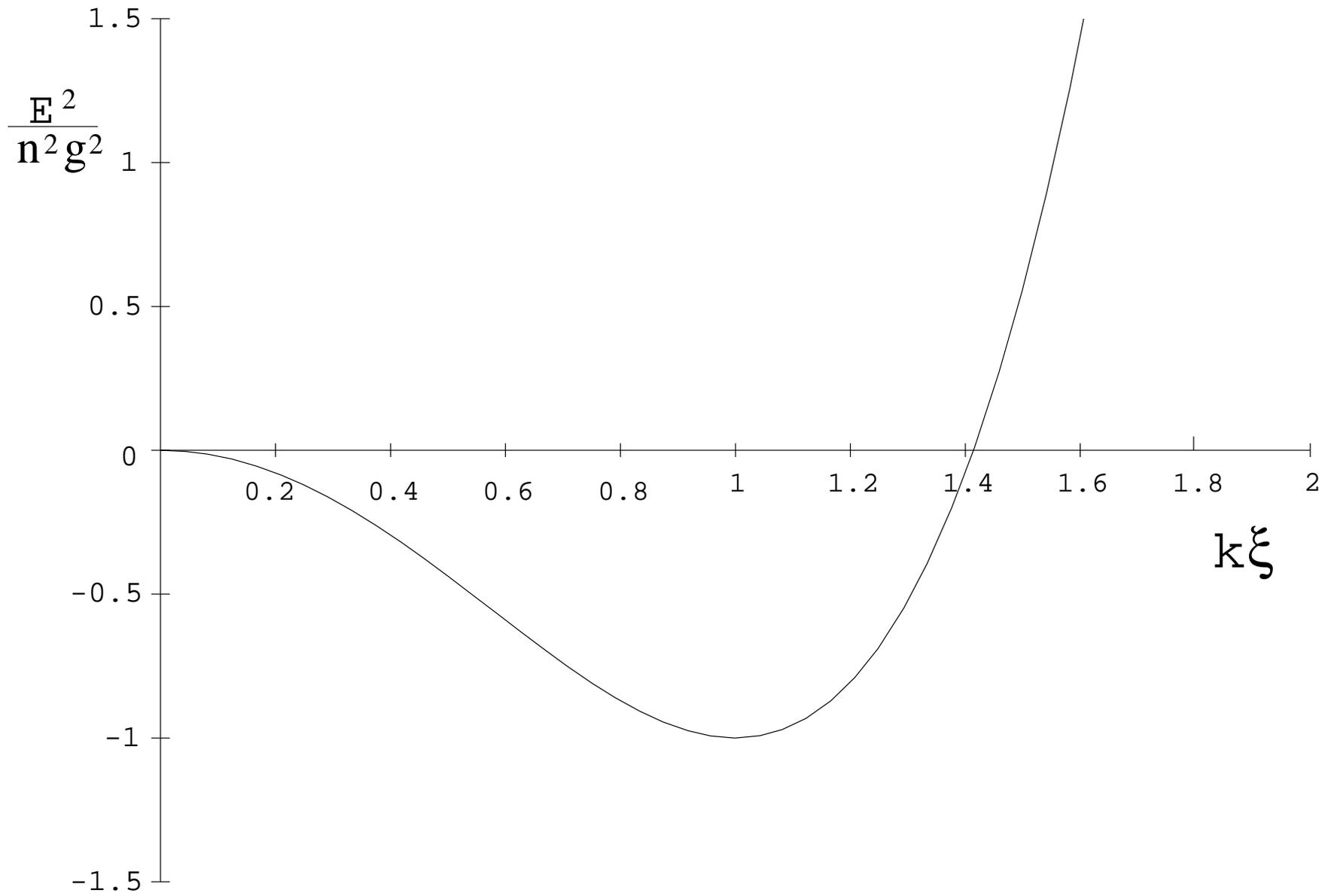
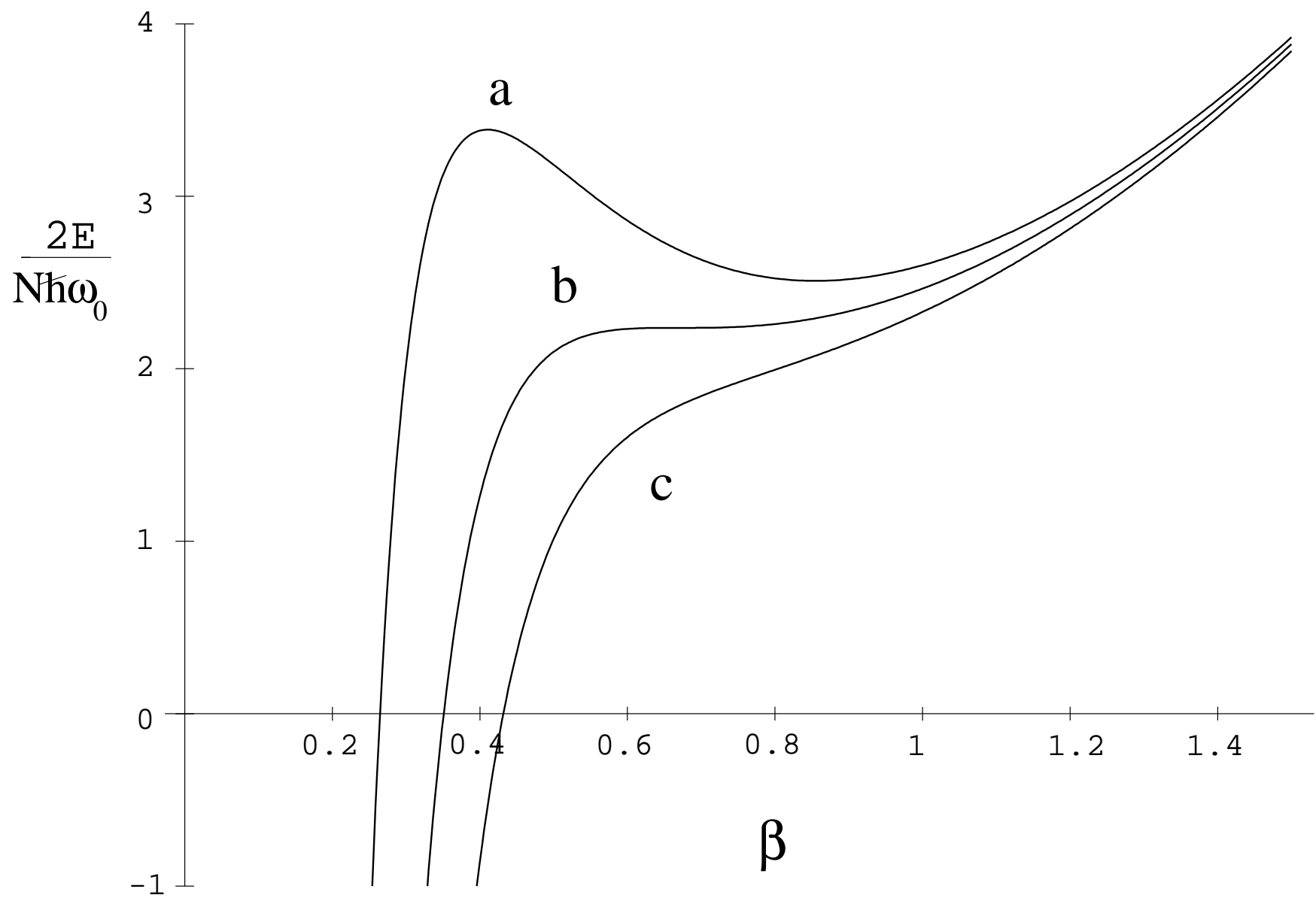
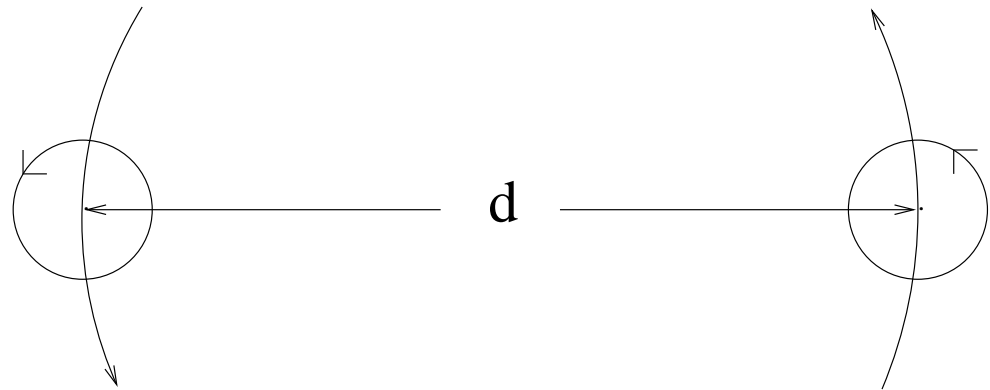
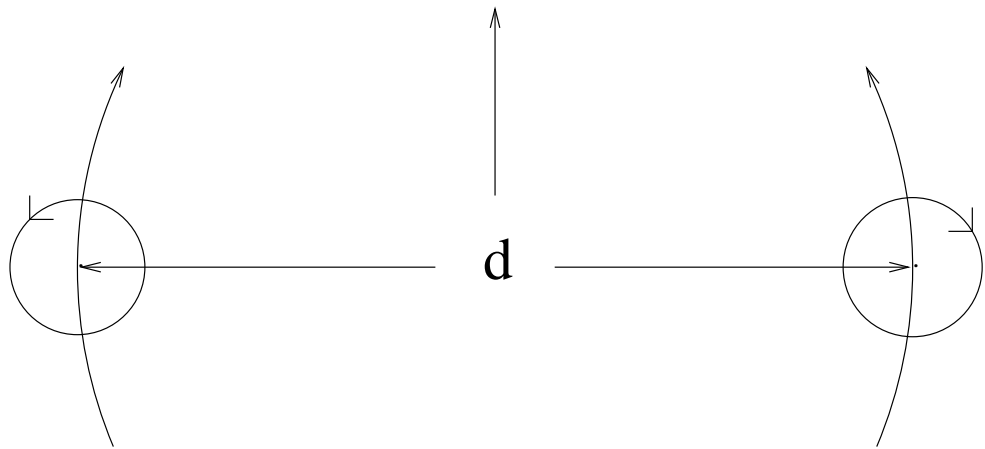
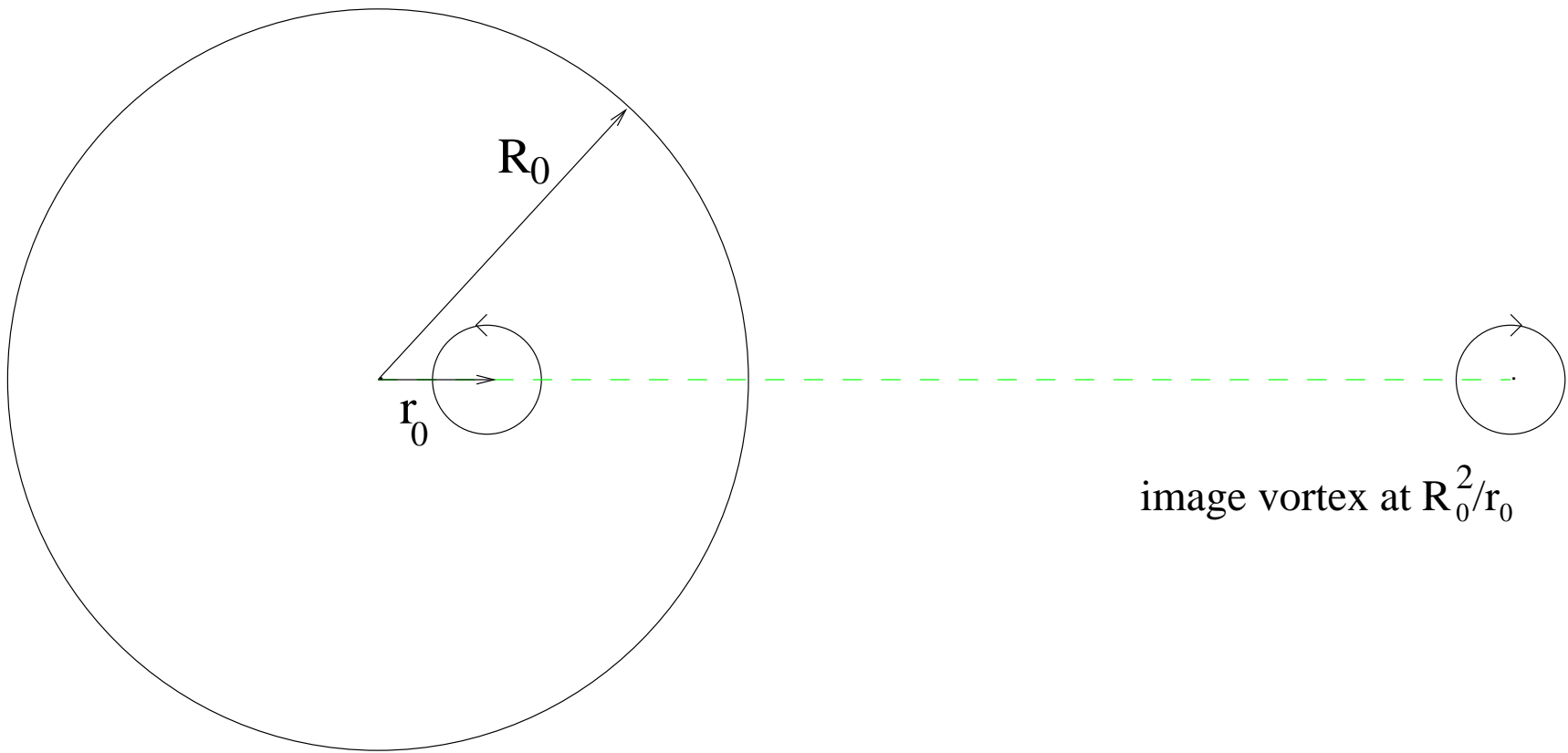
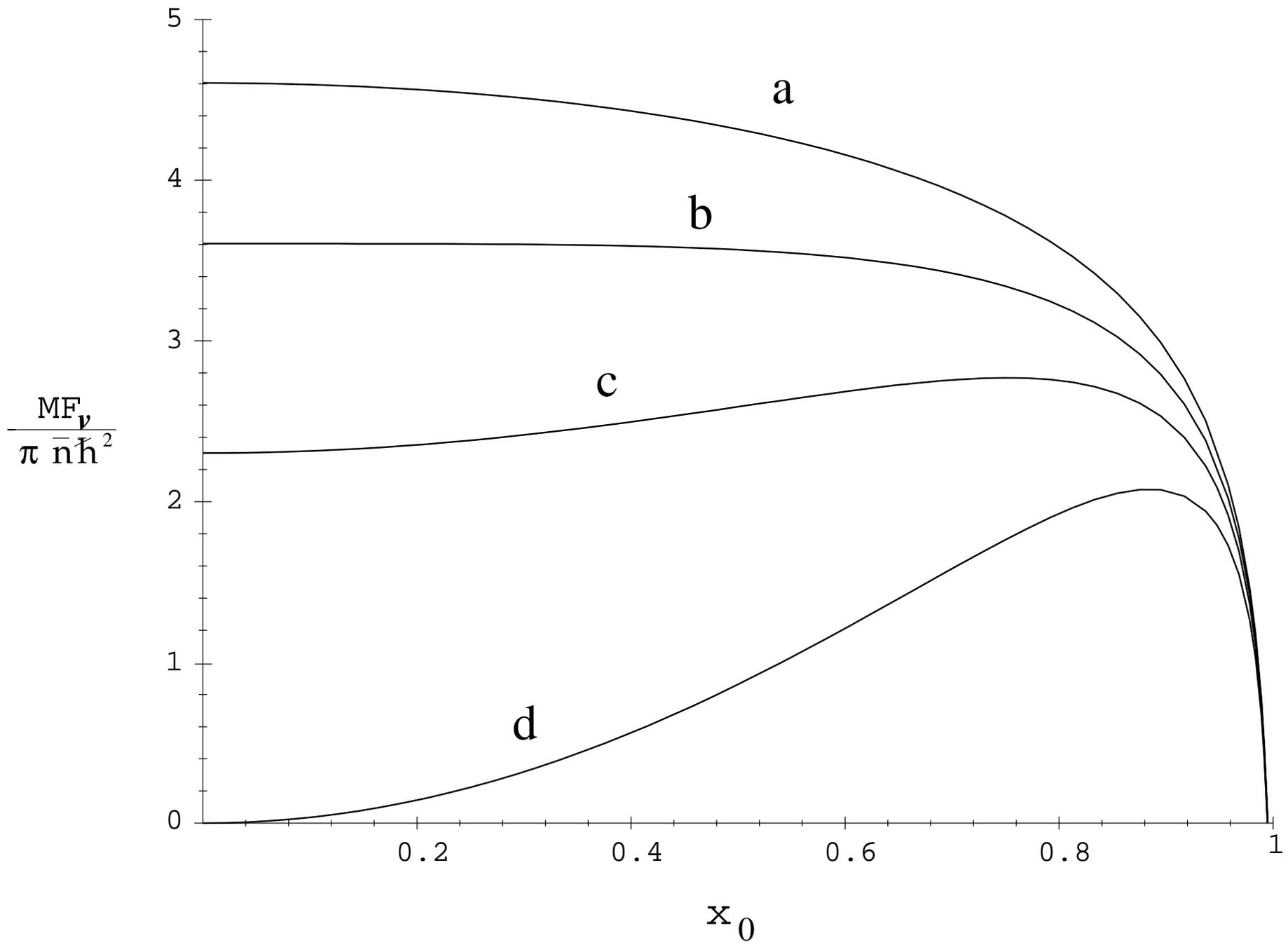
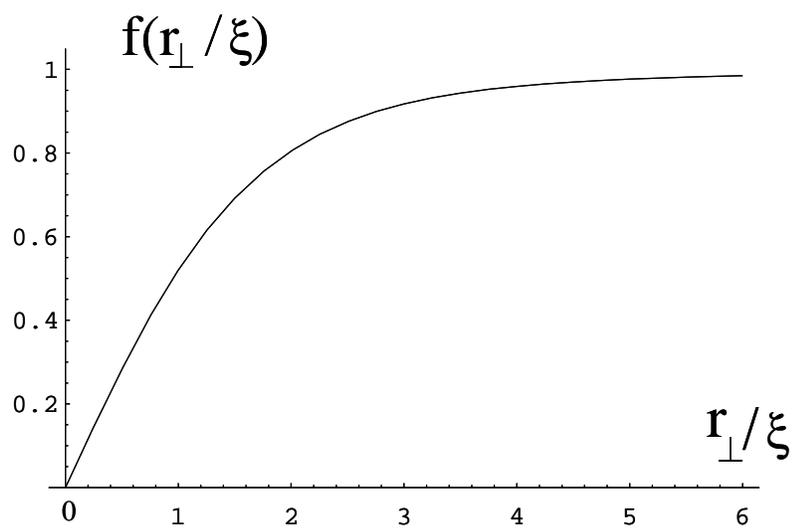
\end{document}